\documentclass[useAMS,referee]{mn2e}
\usepackage{amsmath,amssymb,graphicx,deluxetable}
\usepackage{subfigure}
\def\arcs{\hbox{$^{\prime\prime}$}}    
\def\lum{$\rm{ergs}~\rm{s}^{-1}$}       
\def\lx{$\rm{L}_{X}$}                 
\def\cts{counts s$^{-1}$}
\def\td{$\tau_{d}$}
\def\tr{$\tau_{r}$}

\title[A study of X-ray flares - I.  Active late-type dwarfs]{X-ray flares.  Active late-type dwarfs}
\author[J. C. Pandey and K. P. Singh]{J. C. Pandey$^1$\thanks{E-mail:jeewan@aries.ernet.in (JCP); singh@tifr.res.in (KPS)}\thanks{Present address: Aryabhatta Research Institute of Observational Sciences, Nainital -263 129, India} and K. P. Singh$^{1\star}$\\
$^{1}$Tata Institute of Fundamental Research, Mumbai - 400 005, India}

\begin{document}

\date{} 

\pagerange{\pageref{firstpage}--\pageref{lastpage}} \pubyear{2007}

\maketitle   \label{firstpage}

\begin{abstract}
We present  temporal and spectral characteristics of X-ray flares observed
from six late-type G-K active dwarfs (V368 Cep, XI Boo, IM Vir, V471 Tau,
CC Eri and EP Eri) using data from  observations with the XMM-Newton
observatory.  All the stars were found to be flaring frequently and altogether
a total of seventeen flares were detected above the ``quiescent''  state X-ray
emission which varied from 0.5 to 8.3 $\times 10^{29}$ \lum. The largest flare
was observed in a low activity dwarf XI Boo with a decay time of 10 ks and ratio of peak flare luminosity to ``quiescent'' state luminosity of 2.  We have studied the spectral changes during the flares by using colour-colour diagram and by detailed spectral analysis  during the temporal evolution of the flares. The exponential decay of the X-ray light curves, and time evolution of the plasma temperature and emission measure  are similar to those observed in compact solar flares.  We have derived the semiloop lengths of flares based on the hydrodynamic flare model. The size of the flaring loops is found to be less than the stellar radius. The hydrodynamic flare decay analysis indicates the presence of sustained heating during the decay of most flares.
\end{abstract}

\section{Introduction}
\label{sec:intro}

Flares are events in which a large amount of energy is released in a short
interval
of time. Such events take place at almost all frequencies of the electromagnetic
spectrum and have been observed in the Sun as well as in  many types of cool stars
(Garcia-Alvarez, et al. 2002). These stellar flares could be radiating
several orders of magnitude more energy
than a solar flare. Optical flares in UV Cet (dMe) type stars  are a
common phenomenon.
The flares produced by other stellar sources (e.g. RS CVn and BY Dra) are usually
detected only in the UV or X-rays. These UV and X-ray flares show extreme
luminosities and very hot temperatures ($\gtrsim 10 $MK). Even though, flares
in these stars present many analogies with the solar flares, there are also
significant differences, such as the amount of energy released. A general model
for stellar flares has emerged from numerous solar flare studies (G\"{u}del 2004).
The flare reconnection region, located somewhere at large coronal
heights, primarily accelerates electrons and ions upto MeV
energy (Dennis \& Schwartz  1989). The accelerated electrons participate
along the magnetic fields into chromosphere where they heat the cool plasma
to coronal flare temperatures, thus evaporating a part of the chromosphere
into the corona. Early X-ray observations with the EXOSAT observatory of flare stars
revealed examples
of two different type of flares: 1) Impulsive flares which are like
compact solar  flares, and 2) Long decay flares which are like two-ribbon
solar flares (Pallavicini, Tagliaferri \& Stella  1990). The compact flares are less
energetic ($\sim$ 10$^{30}$ ergs s$^{-1}$), short in duration ($< 1$ h),
and confined to a single loop while the  long-decay flares are more energetic
($\sim$ 10$^{32}$
ergs s$^{-1}$), of long duration ($\ge 1$ h), and release the energy in an
entire arcade of loops (Garcia-Alvarez 2000). The most likely flare process
relates to an opening up of magnetic fields and subsequent relaxation by closing
the open field lines.

Analysis of light curves during flares can provide us with insights into
the characteristics of the coronal structures and, therefore, of the magnetic
field (e.g., Schmitt \& Favata 1999; Favata, Micela \& Reale 2000a; Reale et al. 2004).
Even though stellar flares are spatially unresolved, a great deal of
information on the coronal heating and on the plasma structure morphology can
be inferred from a detailed modeling of stellar flares; for instance, if
sufficient data are available for moderately time-resolved spectral
analysis, a study of the complete evolution of a flare  can allow us to (a) infer
whether the flare occurs in closed coronal structures (loops), (b) determine
the size of the flaring structures, (c) determine whether continuous heating is present throughout the
flare, and (d) put constraints on the location and distribution of the heating (see Reale et al. 2004).

Active solar-type (G-K) dwarfs are known to possess magnetic fields on
their surfaces, and the field strengths are as large as several kilogauss, i.e.
stronger than the strongest field observed on the Sun (e.g., Saar 1990;
Johns-Krull \& Valenti 1996).
Cyclic behavior has been identified
in several late-type active stars (Baliunas et al. 1995, Olah \& Strassmeier 2002),
suggesting that dynamos similar to the solar dynamos are also operative in
solar-type stars.
These  solar-type  stars are
less  active than the dMe stars (so called flare stars).
However, they show more frequent flaring activity than the
late-type evolved stars (e.g. RS CVns; G{\"u}del 2004).
In this  paper, we present analysis of archival data obtained from
XMM-Newton observations of
six G-K dwarfs, namely V268 Cep, XI Boo, IM Vir, V471 Tau, EP Eri and CC Eri
with the aim to understand the spectral and temporal characteristics of X-ray flares in them.
In subsequent papers we will examine the characteristics of X-ray flares
in the subgiants and the giants.

The  paper is organized as follows: In \S \ref{sec:basic} we give the basic
parameters of stars in our sample, in \S \ref{sec:obs} we describe the
observational data sets that we have analysed and the methods of data reduction,
\S \ref{sec:analysis} contains our analysis and results, and in \S \ref{sec:discussion} we present our discussion and
conclusions.

\section{Basic parameters of sample stars}
\label{sec:basic}
The basic parameters of stars in our sample are given in Table 1.
V368 Cep (= HD 220140; V = 7.54 mag) is a G9V spectral-type chromospherically active star with photometric period of
2.74 d (Chugainov, Lovkaya \& Petrov 1991, Chugainov, Petrov \& Lovkaya 1993).
 XI Boo (= HD 131156; V = 4.55 mag) is a nearby visual binary, comprising a primary G8 dwarf and a secondary K4 dwarf
with an orbital period of 151 yrs (Hoffleit \& Jaschek 1982). In terms of the outer
atmospheric emission, UV and X-ray observations show that the primary  dominates
entirely over the secondary (Hartman et al. 1979, Ayres, Marstad \& Linsky 1981,
Schimtt 1997, Laming \& Drake 1999). XI Boo A (primary) is a slow rotator
with rotation period of 6.2 d (Gray et al. 1996).
A cool main sequence eclipsing  binary (G5V+KV), IM Vir (= HD 111487; V = 9.69 mag ) has  an orbital period of 1.31d (Malkov et al. 2006).
V471 Tau (= BD+16 516; V = 9.71) is a rapidly rotating  detached eclipsing binary first
observed by Nelson \& Young (1970). The system is composed of a cool main-sequence
chromospherically active K2 V star (Guinan \& Sion 1984) and a degenerate
hot white dwarf separated by 3.4 stellar radii.
CC Eri (= HD 16157, V = 8.8) is a spectroscopic binary with orbital period of
1.561 d. It consists of a K7Ve primary and a dM4 secondary, with mass ratio
of $\sim 2$ (Strassmeier et al. 1993). The primary star corotates with orbital motion
due to the tidal lock, and is one of the fastest rotating K dwarfs in the solar
neighborhood. EP Eri (=HD 17925, V = 6.0 mag) is very
nearby (10.3 pc), very young (high Li I abundances),
active K2-type dwarf with a rotation period of 6.85 days (Cayrel de Strobel
\& Cayrel 1989; Cutispoto 1992; Henry, Fekel \& Hall 1995). The presence of an
unresolved companion in this star has been
suggested by {\rm Henry et al. (1995)} based on the variable widths of the
photospheric absorption lines reported in the literature ($v\sin{i}$ ranges
from 3 to 8 km s$^{-1}$; see Fekel 1997).

\section{Observations and Data Reduction}
\label{sec:obs}
The  late-type dwarf stars in our sample were observed with the XMM-Newton satellite using varying detector set-ups.
The XMM-Newton satellite is composed of three co-aligned X-ray telescopes
(Jansen et al. 2001) which observe
a source simultaneously, accumulating photons in three CCD-based instruments: the twin
MOS 1 and MOS 2 and the PN (Turner et al. 2001; Str\"{u}der et al. 2001),
all three detectors constituting the EPIC (European Photon Imaging Camera) camera.
The EPIC instrument consists of three
CCD cameras with two different types of CCD, two MOS and one PN, providing the imaging
and spectroscopy in the energy range from 0.15 to 15 keV with a good angular
(PSF = 6 arcsec (FWHM)) and a moderate spectral resolution ($E/\Delta E \approx 20 - 50 $).
Exposure time for each star was in the range of 30-60 ks.
A log of observations is provided in Table 2.

The data were reduced  with standard XMM-Newton Science Analysis System (SAS)
software, version 7.0 with updated calibration files (Ehle et al. 2004).
The preliminary processing of raw EPIC Observation Data Files was done
using the {\it epchain} and {\it emchain} tasks which allow calibration both in energy and
astrometry of the events registered in each
CCD chip and to combine them in a single data file for MOS
and PN detectors.
The background contribution is particularly relevant at high energies
where coronal sources have very little flux and are often undetectable.
Therefore, for further analysis we have selected the energy range between
0.3 to 10.0 keV.
Event list file was extracted using the SAS task {\it evselect}.
The {\it epatplot} task was used for checking the existence of pile-up
affecting the inner region of all stars. Only the star CC Eri was found to be
affected by pile-up.
X-ray light curves and spectra of all target stars were generated
from on-source counts obtained from circular regions with
a radius $\sim$ 40-50\arcs ~around each source. However, for the star CC Eri,
X-ray light curves and spectra were extracted with events taken from an
annulus of 32\arcs with inner radius of 12\arcs to avoid pile-up effect in the inner region.
The background was taken from  several source free regions on the detectors
at nearly the same offset as the source and surrounding the source.

\section{Analysis and Results}
\label{sec:analysis}
\subsection{Multi-band X-ray light curves and characterization of flares}
\begin{figure}
\centering
\includegraphics[height=13.5cm, width=7.5cm,angle=-90]{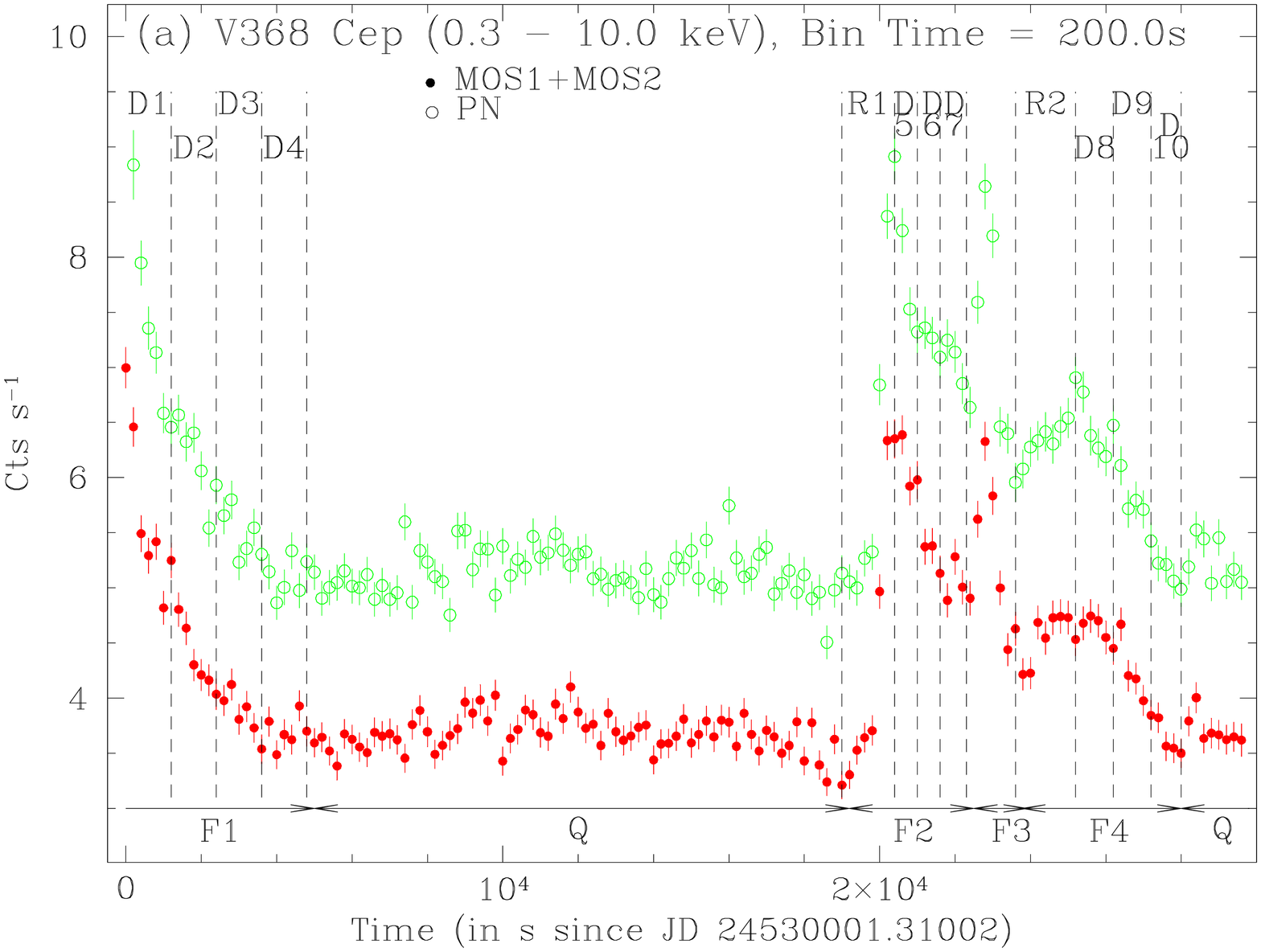}
\includegraphics[height=13.5cm, width=7.5cm,angle=-90]{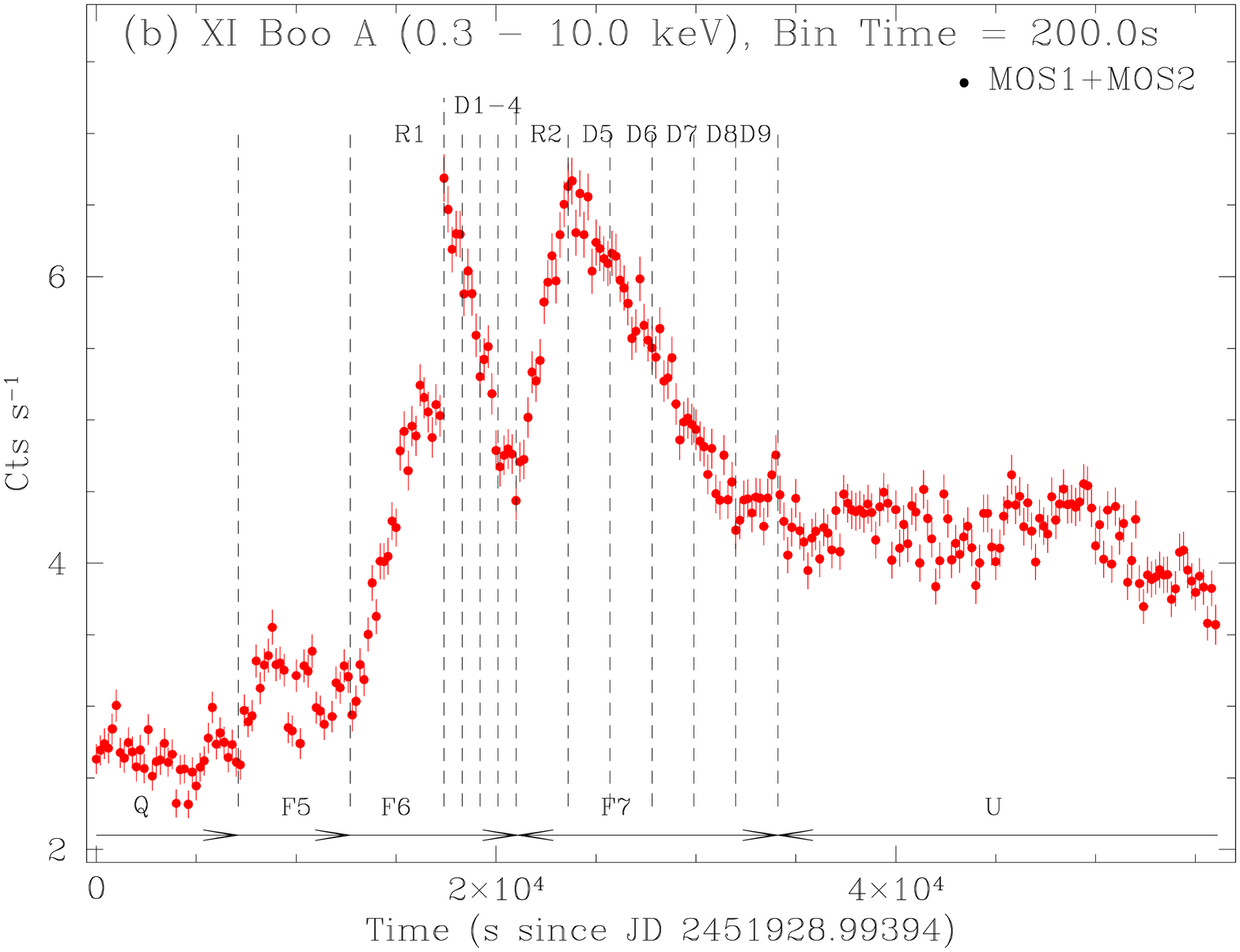}
\includegraphics[height=13.5cm, width=7.5cm,angle=-90]{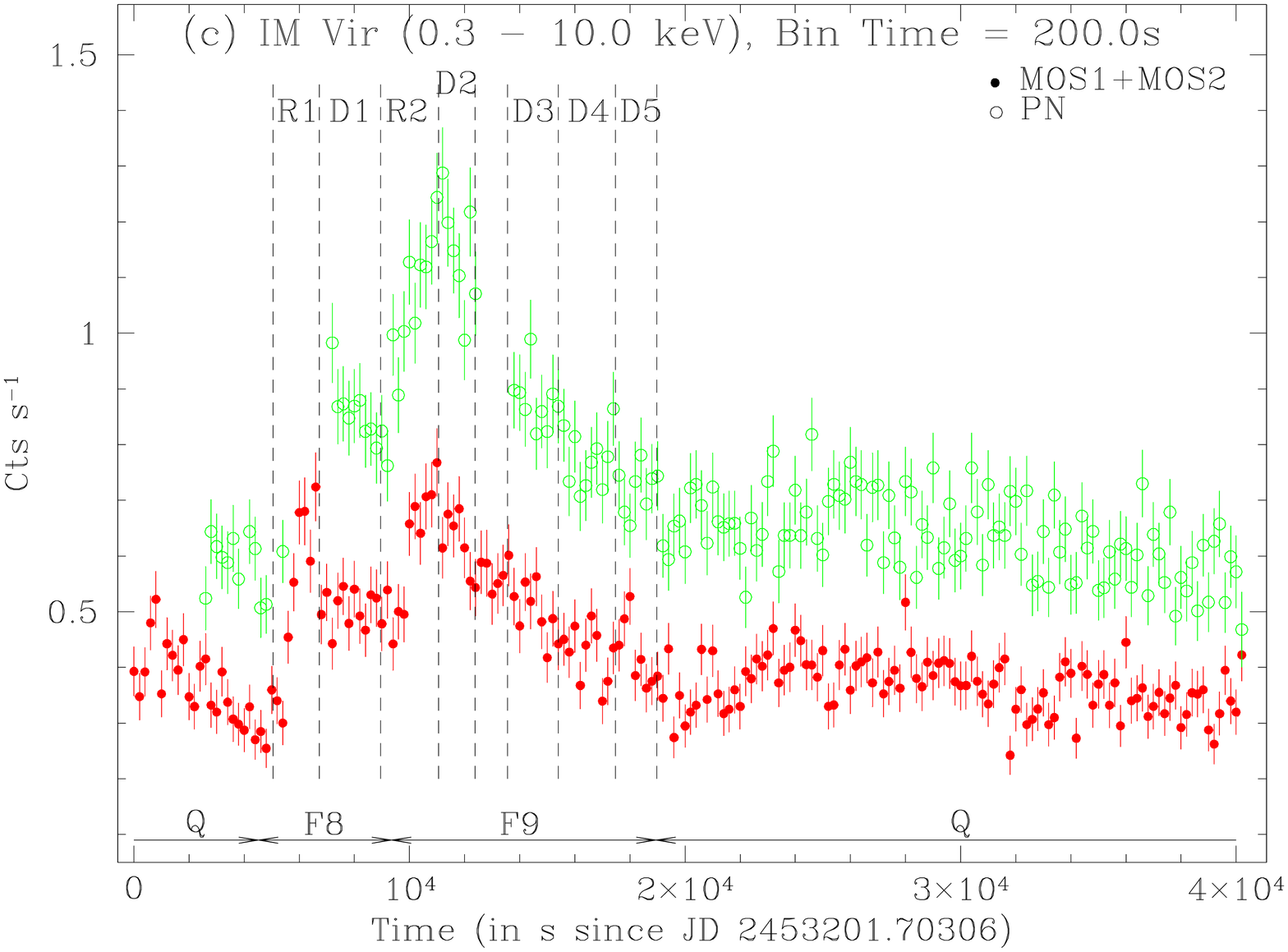}
\caption{MOS and PN Light curves of six G-K dwarfs}
\label{fig:mospnlc}
\end{figure}

The background subtracted X-ray light
curves of six stars viz. V356 Cep, IM Vir, V471 Tau, CC Eri and EP Eri,
as observed with the MOS and PN detectors are shown in Figures \ref{fig:mospnlc}(a) to
\ref{fig:mospnlc}(f), respectively. All light curves are in the energy band 0.3-10.0 keV.
The MOS and PN light curves
are represented by solid and open circles, respectively. The temporal
binning of light curves is 200 s for all sources.
The PN light curve of the star IM Vir was affected by proton flares twice
during the observations, first for 800 s after 10 ks from the beginning of the
observation, and second at the end, after 40 ks of the observations. However,
the MOS light curve was affected only at the end of the observation. For the
star EP Eri, the high proton flare background region in the light curve
is for a duration of 3 ks after  14.7 ks from the beginning of the observation.
We have removed  data during the proton flare background from the MOS and PN light curves of
both the stars IM Vir and EP Eri.  The star XI Boo
was not observed by the PN detector, and only the light curve obtained with
the MOS is presented here.

The light curves of all the individual stars show variability on a time
scale of ks, most of which resembles flaring activity.
The customary definition of a flare is a significant increase in intensity,
after which the initial or quiescent level of intensity is reached again.
Such patterns are observed in the light curves of the sample stars
shown in Figures \ref{fig:mospnlc}(a) to \ref{fig:mospnlc}(f), and
where  the
``flare regions'' are represented by arrows and marked by Fi, where i=1,2...17 refers to the flare number. The
mean quiescent level count rates of the sources are taken from
regions that are free from flares and are  marked by Q in the Figure \ref{fig:mospnlc}.
The start
time, end time, flare duration and flux during quiescent and  at flare peak for all the
17 flares observed are listed in Table 3. The peak and quiescent state
count rates  as given in Table 3 are converted into flux by using the WebPIMMS\footnote{http://heasarc.gsfc.nasa.gov/Tools/w3pimms.html},
where we assumed the plasma temperature of 1 keV.
Flares F6, F7, F9, F14 and F16 were found to be long lasting ($\geq 2.3$ h) flares.
 Infact, the flare  F7 of the XI Boo
is longest flare observed with total duration of $\sim 3.1$ h.
The peak flux in these long duration flares was found to be more than two times
than the quiescent state.  Other detected flares
were shorter than 1.3 h. The shortest duration flare was detected one
in V368 Cep (F3) and other in CC Eri (F12). In these short duration flares,
peak X-ray flux was 1.3 - 1.8 times more than that of the quiescent state.

To characterize the flares,
we have fitted the light curves of the flares with an exponential function

\begin{equation}
c(t) = A_{0}{\rm e}^{-\frac{t-t_0}{\tau_d}} + q
\label{eq:efold}
\end{equation}

where c(t) is the count rate as a function of time t, t$_0$ is the time of peak
count rate, q is the count rate in the quiescent state, $\tau_d$ is the decay
time of the flare, and $A_0$ is the count rate at flare peak. The best fitted
parameters for the flares are given in Table 4 for both the MOS and the PN data.
Five flares, F6 and F7 of XI Boo,
F9 of IM Vir, F14 of CC Eri and F16 of EP Eri the e-fold decay time  was
found to be more than 1 h.
Flare F7 of XI Boo was found to be the longest decay
flare ($\tau_d = 10$ ks) observed among all the seventeen flares.
The flare F16 shows a very different  structure  when
compared to the other flares, i.e. a sharp rise and a highly structured decay
as if more flares were present during the decay. Though it was tough to
locate the peak of flare F16 because of its complex structure, the decay
time was determined to be 4 ks. For
the remaining eleven flares the decay time was found to be  in the range of
8 - 47 minutes. The decay phase of the flare F17 was not observed. The rise times
of all flares were found to be less than 1 h. The rise and decay time of the
shortest flare F3 of V368 Cep was found to be 0.5 ks.
In XI Boo, after the flare F6
an active level U was identified, where average flux was 1.8 times more than
that of the Q state (see Figure \ref{fig:mospnlc}(b)), that was higher than
the peak of the flare F5  indicating a very high activity level.
Two similar highly active regions U1 and U2 were identified between the flares F13 and
F14.  The average fluxes during the
regions U1 and U2 were 1.3 and 1.4 times more than that of the quiescent
state.  Presence of
strong substructures in these active regions indicates superposition of a large
number of flaring regions.

\begin{figure}
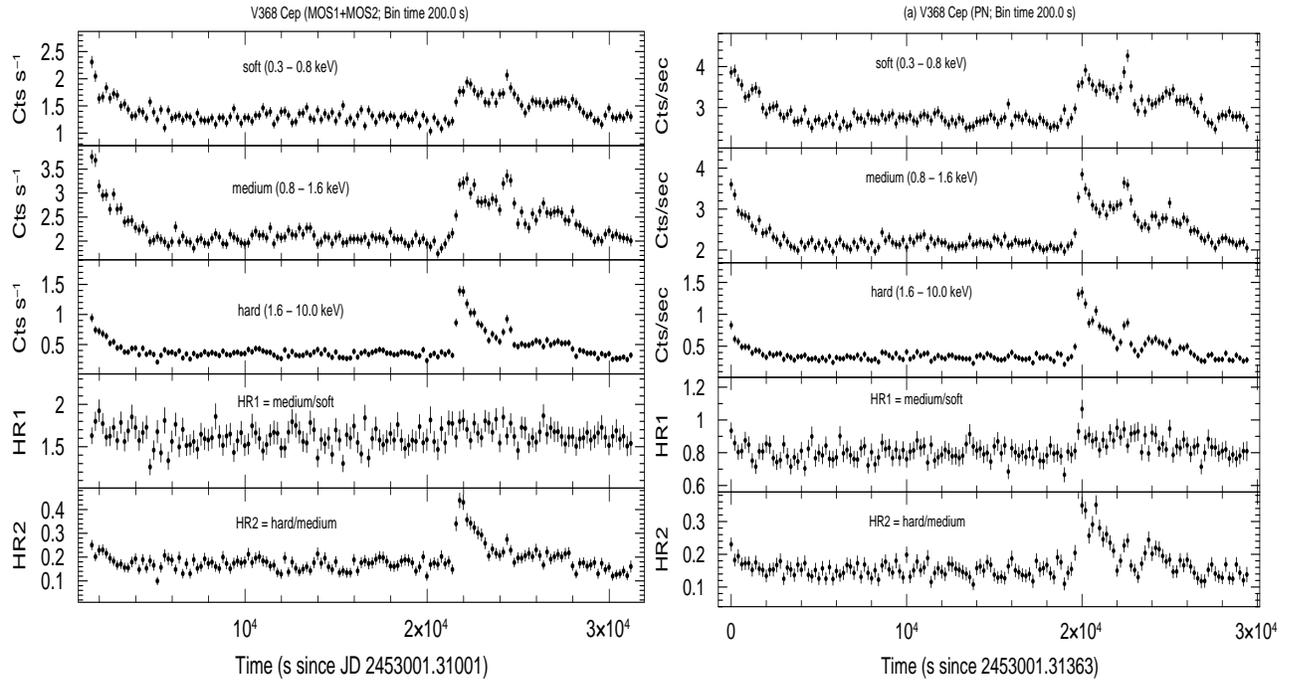

\hbox{
\includegraphics[height=8.5cm, width=9.0cm,angle=-90]{V368Cep_mos_b200.ps}
\includegraphics[height=8.5cm, width=9.0cm,angle=-90]{V368Cep_pn_b200.ps}
}
\caption{(a) MOS and PN light curves at three bands soft (0.3-0.8 keV), medium (0.8-1.6 keV), and
hard(1.6-10.0 keV) and hardness ratio HR1 and HR2 curve, where HR1=medium/soft
and HR2=hard/medium of V368 Cep.}
\label{fig:lchr}
\end{figure}

To investigate the behavior of the flares detected in different energy bands,
the light curves of the stars obtained with the MOS and PN
are divided into three energy bands namely soft (0.3 - 0.8 keV), medium (0.8 -
1.6 keV), and hard (1.6 - 10 keV). The boundaries of the selected energy bands are
chosen as the line free regions of the low resolution PN spectra.
The hardness ratio HR1 and HR2 are defined
by the ratio of medium to soft band, and hard to medium band count rate,
respectively.  The soft, medium, hard band intensity curves, and the hardness
ratio curves HR1 and HR2 as a function of time
are shown in the sub-panels running from top to bottom in left(MOS) and
right(PN) of Figure \ref{fig:lchr} (a) for the star V368 Cep.
Similar plots of intensity and hardness ratios observed from the
stars XI Boo, IM Vir, V471 Tau, CC Eri and EP Eri are shown in the
Figures \ref{fig:lchr}(b) to \ref{fig:lchr}(f), respectively.
The light curves in individual band also show significant variability in all sources.
We have determined the e-folding decay time of observed flares
in the three bands by fitting  equation (\ref{eq:efold}).
The decay times, rise times, count rates at flare peaks,
and quiescent state count rates at soft, medium and hard band
for PN data are
given in the Table 5. For most of the flares,  e-folding decay time in the soft
band was found to be more than  the decay time in the medium and the hard band.
However, for  the flare F3 of V368 Cep and F7 of XI Boo, it appears that
decay time in the medium band was more than in the soft band.
For the flares F6, F7 and F9 the decay time in the soft band was
found to be more than 3.0 ks than that of the hard band.
However, for other flare events the decay time in the
soft  band was  $\lesssim 2$ ks than that in the hard band.
Two smaller flare like structures were seen just after the peak
of flare F13 of CC Eri,  in the hard band but were not observed in the soft and medium band
light curves (see Figure \ref{fig:lchr}(e)). The rise time for all flares in all band
was found to be well within  $1\sigma$ level (Table 5).

The flare peak to the quiescent state flux ratio in the hard band was found in between
2-16, which is more than that of the similar ratios in the medium and the soft band (see also Table 5) for all
flares. In terms of the peak flux, the flares F6 and F7 of XI Boo, and F13 and F14 of CC Eri
were stronger in the medium band than in the soft and the hard band.
The variation in the  hardness ratio are indicative of changes in coronal temperature.
An X-ray flare is well defined by a rise in the temperature and subsequent decay.
For the flares F6, F7, F9, F12, F13 and F14, both HR1 and HR2 varied in the similar fashion
to their light curves (see Figure \ref{fig:lchr}). This implies that
an increase in the temperature at flare peak and subsequent cooling. The structures
observed during the decay phase of the flares F2 of V368 Cep and F13 of CC Eri in the hard
band were also
observed in the HR2 curve. This could probably be due to emergence of flares
during the decay phase of these flares, which were comparatively stronger in the hard band.
During the flares F2, F3 and F4 of V368 Cep, and  F16 of EP Eri the HR1 did not
show any variation but HR2 varied according to the flare intensity. However,  HR1 and HR2 did
not show any significant variation during the flares F1, F5, F8, F10, F11, F15 and F17.
The PN data have more count rates than the MOS and have similar characteristics
to that of MOS, therefore, for further analysis in this paper we use only
the PN data.

\subsection{Colour-colour diagrams}
\label{sec:ccd}
\begin{figure}
\centering
\includegraphics[height=11.0cm, width=7.5cm,angle=-90]{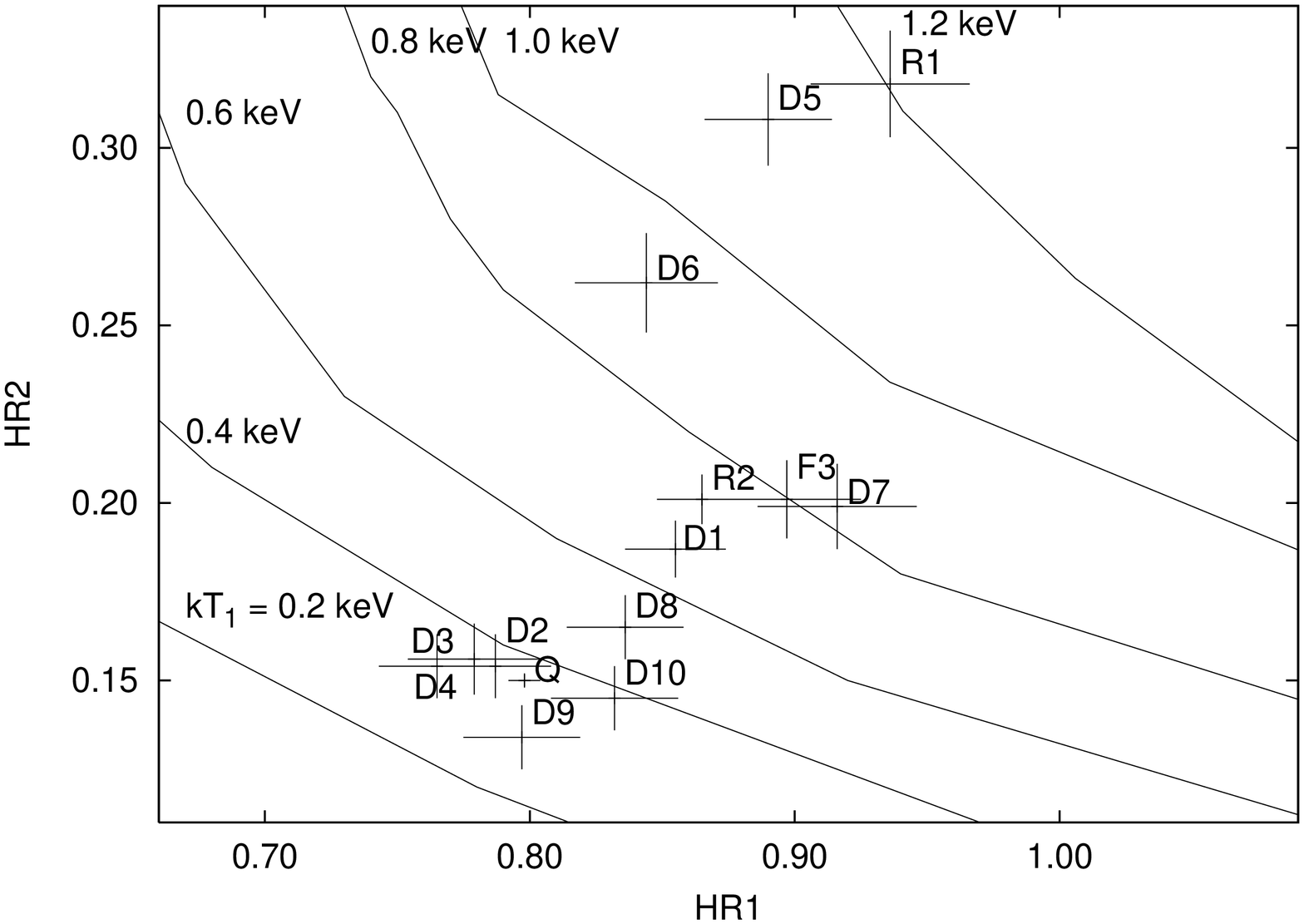}
\includegraphics[height=11.0cm, width=7.5cm,angle=-90]{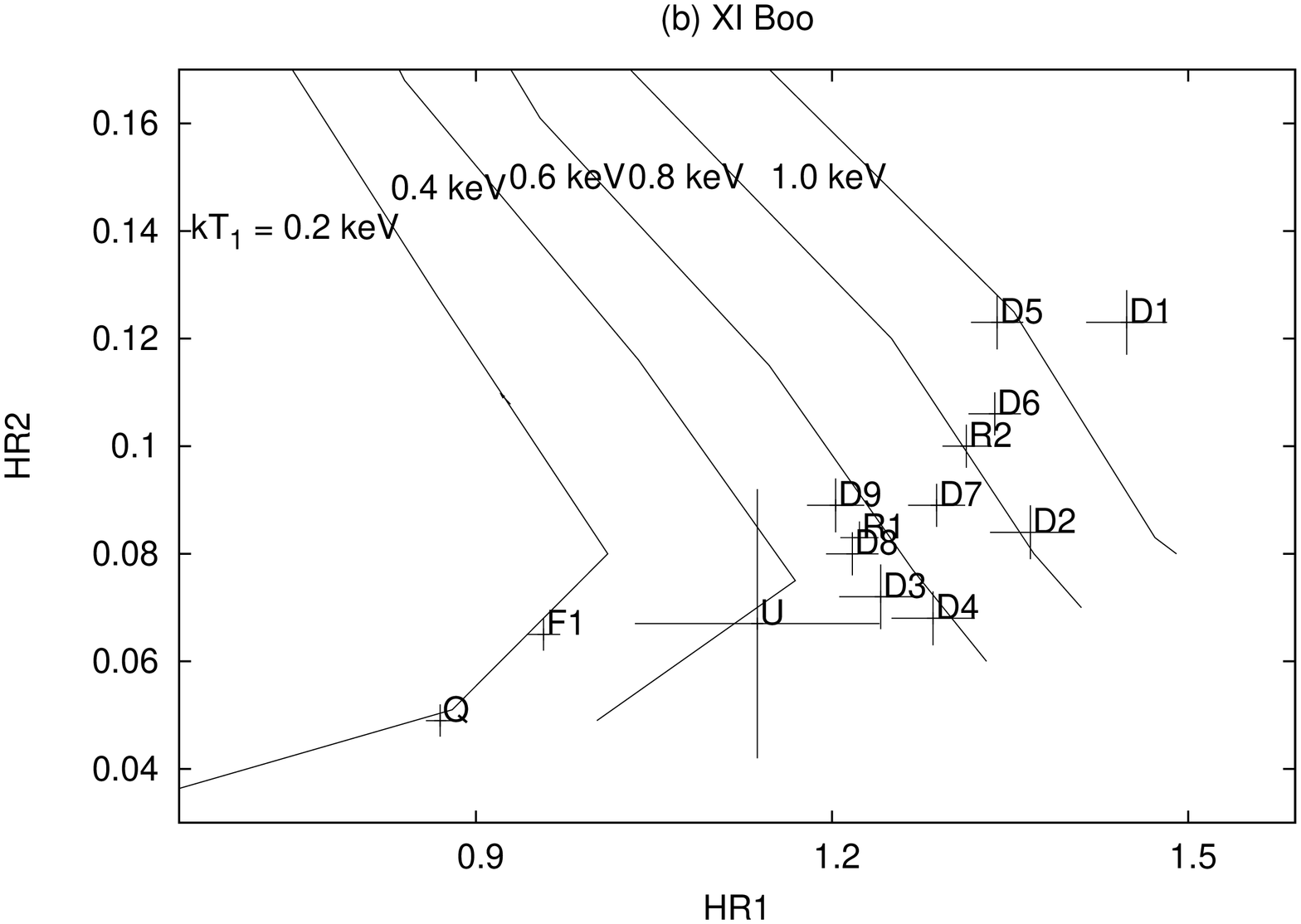}
\includegraphics[height=11.0cm, width=7.5cm,angle=-90]{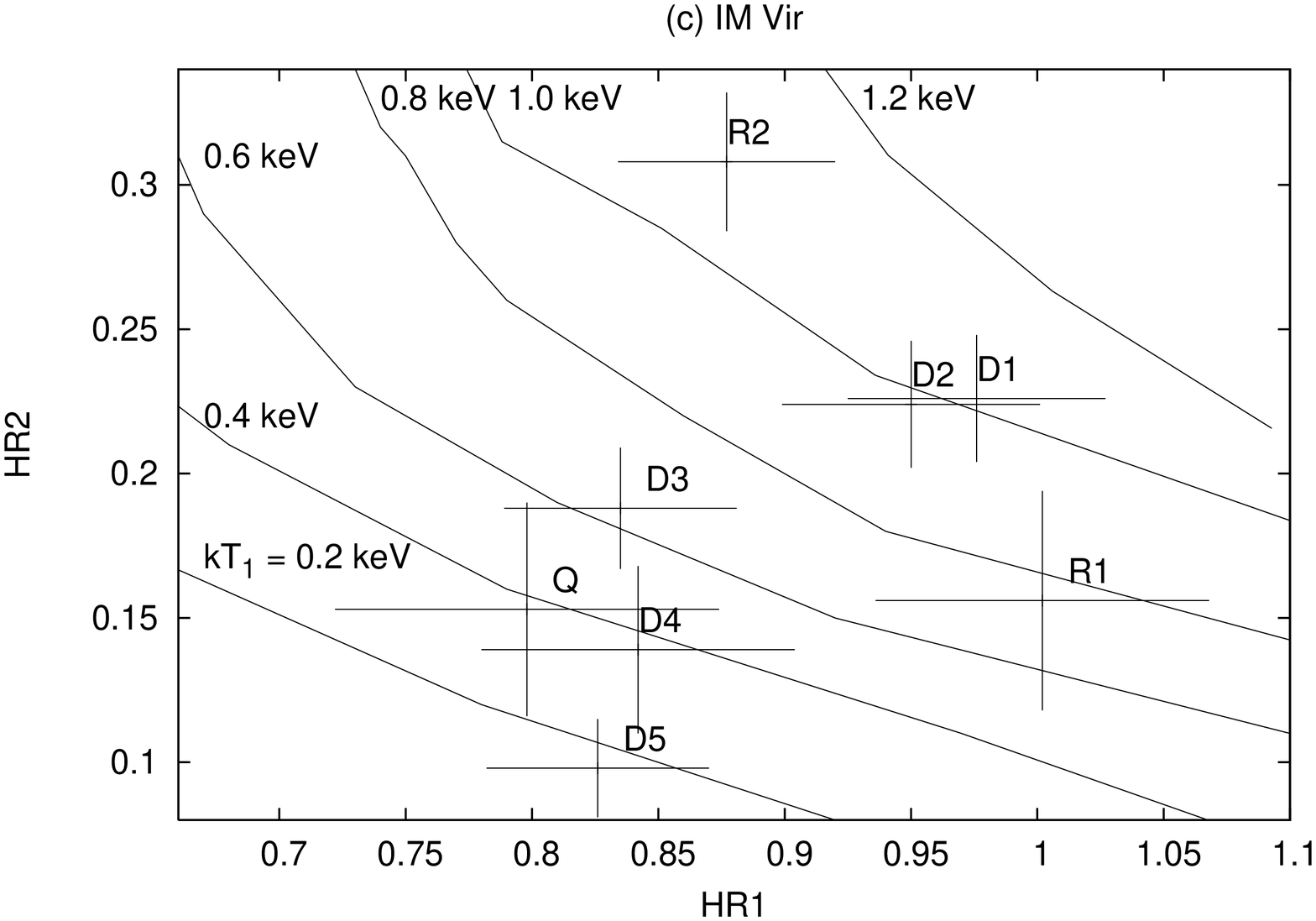}
\caption{Colour-colour diagram (HR1 versus HR2) for the  various time
segments (Q: quiescent state, Ri(i=1,2..): rise phase, Di(i=1,2..): decay phase) observed for each source.
The curves overplotted on the data are for model simulation where two temperature plasma
is assumed to predict the colors (see \S \ref{sec:ccd} for details).}
\label{fig:hr}
\end{figure}

Plots of hardness ratio in the form of colour-colour (CC) diagram can reveal
information about spectral variations and serve as a guide for a more
detailed spectral analysis. For this purpose,  the light curves and  the
hardness ratio curves of the flares were divided into different time segments
covering their rising and decaying phases as
shown in Figure \ref{fig:mospnlc}, where Ri(i=1,2..) represents the  rising phase, Di(i=1,2,..) represents the
decay phase and Q represents the quiescent state.
We determined the HR1 and HR2 from the PN data for each  of these
segments  and plotted the values in Figure \ref{fig:hr}.
To understand the observed behavior of HR1 and HR2 in terms of simple spectral models,
we generated the soft(0.3-0.8 keV), medium(0.8-1.6 keV) and
hard(1.6-10.0 keV) band count rate using the {\it ``fakeit''}
provision in the XSPEC,
using the most recent Canned Response Matrices downloaded from
http://xmm.vilspa.esa.es/external/xmm\_SW\_cal/calib/epic\_files.html, and
simple two temperature plasma models. The two temperature APEC  models
were mixed such that the emission measure were in the
ratio $EM_2/EM_1$ of 0.4, 0.6, 0.8, 1.0 and 1.2 (based on spectral
results in \S 4.3 and \S 4.4). The abundances (0.25)
and hydrogen column density ($10^{20}$ cm$^{-2}$) were kept fixed for each
set of
the $EM_2$/$EM_1$ (see \S \ref{sec:quiescent} and \S \ref{sec:flare}). Further by keeping  $kT_1$ fixed at different values
ranging from 0.2 to 1.2 keV in steps of 0.2 and varying $kT_2$ from 0.4 to 2.4 keV such that
$kT_2 > kT_1$, we
generated count rates in the  soft, medium and hard bands. These predicted count rates
were used to generate the hardness ratios.
The families of curve thus generated
are over-plotted on the data points in Figure \ref{fig:hr} to understand
the CC diagram. The kT$_2$ in Figure  \ref{fig:hr} increases from
 bottom to top of each curve.

Figure \ref{fig:hr} (a) shows the plot between HR1 and HR2 for a set of
$EM_2$/$EM_1$ = 1.0 for the star V368 Cep.
In the case of the flare F1, 'D1' intersects
the generated CC curves for which $kT_1$ is 0.6 keV, however,
 'D4'  intersects the curve generated for $kT_1 ~=$
0.2 keV. A similar pattern is seen for the other flares F2 and F4 of the stars
V368 Cep, where the top of the flare is located near the CC curve
for a high temperature component whereas the colours near the
end of the flare intersect the CC  curve corresponding to a
 lower temperature. The flare F2 shows a very high
temperature during its rise phase 'R1' and the decay phase 'D5', as both are
located between the CC curves of 1.0 and 1.2 keV. The decay phase D6
of the flare F2 is located  near the 0.8 keV CC curve.
The peak seen during the decay phase D6 of flare F2 (Fig \ref{fig:hr}(a)) can be
explained by a rise in the temperature of coronal plasma. It appears that
the quiescent state 'Q' of the star V368 Cep has a cool temperature between
0.2 to 0.4 keV.

The peak of the largest flare 'F9' observed in the star IM Vir appears to have a temperature
between 1.0 - 1.2 keV. As the flare decayed the temperature was also observed to decrease.
A similar trend is seen during the flares observed in the stars XI Boo (Figure \ref{fig:hr}(b)), V471 Tau
(Figure \ref{fig:hr}(d)), CC Eri (Figure \ref{fig:hr}(e)), and EP Eri (Figure \ref{fig:hr}(f)). The observed CC curves for the star IM Vir, V471 Tau and CC Eri
were well matched with the generated family of CC curves for a set of $EM_1/EM_2$
= 1.0. However, the observed CC diagram of EP Eri were well matched with
generated  CC curves for a set of $EM_1/EM_2 = 0.4$.
It appears that high temperatures are needed during
the flare peak, and the plasma temperatures decrease as flare decays.
To, further confirm and quantify the spectral changes during the
flare, we have also performed a detailed spectral analysis of the flares and
the quiescent states of the stars.
The results of these analyses are  presented  below.

\subsection{Quiescent state X-ray spectra}
\label{sec:quiescent}
\begin{figure}
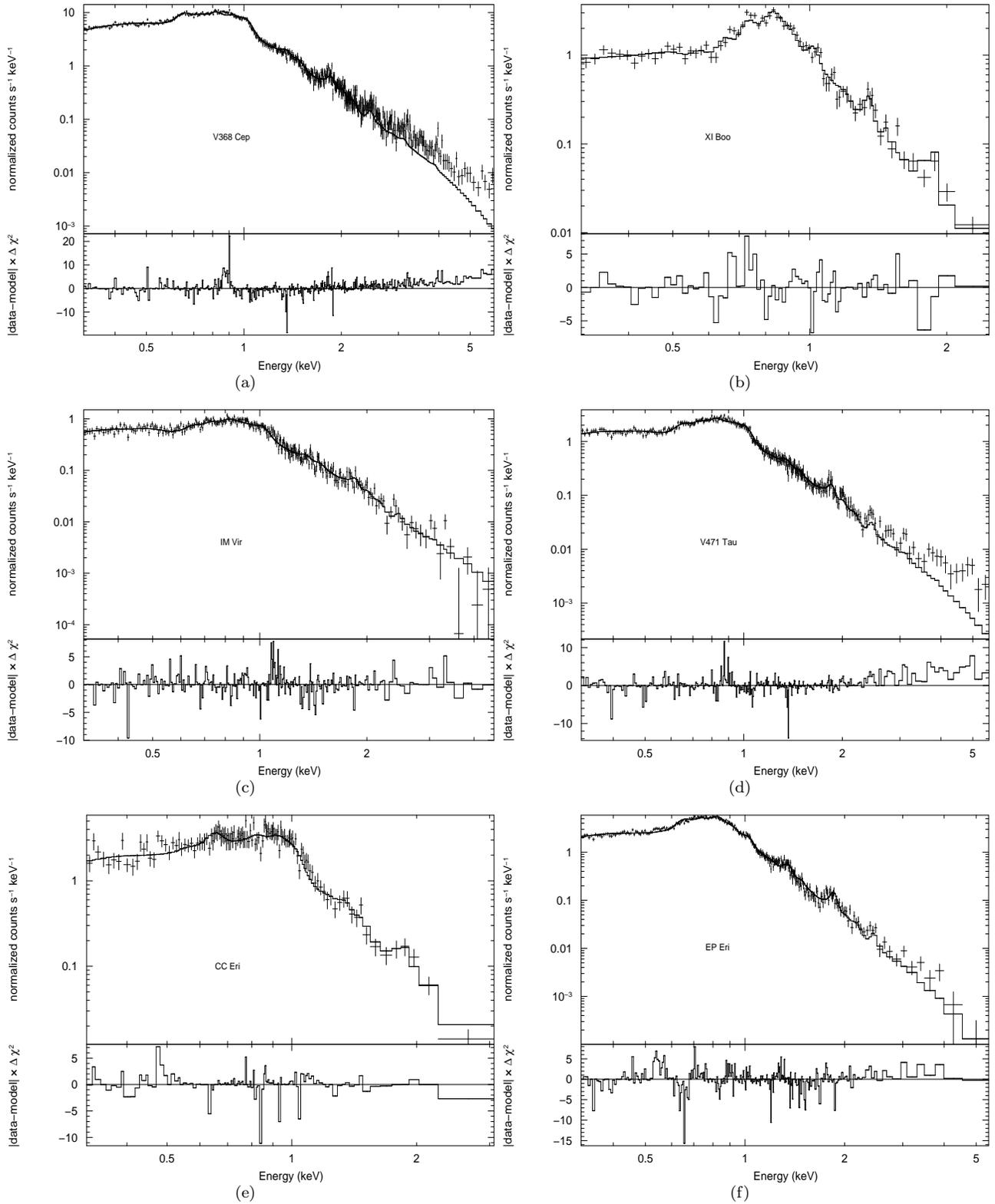

\subfigure[]{\includegraphics[height=8.5cm, width=6.5cm,angle=-90]{V368Cep_Q.ps}}
\subfigure[]{\includegraphics[height=8.5cm, width=6.5cm,angle=-90]{XIBoo_Q.ps}}
\subfigure[]{\includegraphics[height=8.5cm, width=6.5cm,angle=-90]{IMVir_Q.ps}}
\subfigure[]{\includegraphics[height=8.5cm, width=6.5cm,angle=-90]{V471Tau_Q.ps}}
\subfigure[]{\includegraphics[height=8.5cm, width=6.5cm,angle=-90]{CCEri_Q.ps}}
\subfigure[]{\includegraphics[height=8.5cm, width=6.5cm,angle=-90]{EPEri_Q.ps}}
\caption{Quiescent state spectra of PN data with APEC 2T model in upper
subpanels of each graph. $\chi^2$ contributions are given in the lower subpanel
of the each graph.}
\label{fig:qspec}
\end{figure}

X-ray spectra  for each star during the quiescent state were
analysed using  XSPEC version 12.3 (Arnaud 1996).
Data  were cleaned of proton flares by removing the affected time
periods for the stars IM Vir and  EP Eri.
Spectral analysis of EPIC data was performed in the energy band
between 0.3-10.0 keV for the star V368 Cep, XI Boo, V471 Tau and
CC Eri, while for the stars IM Vir and EP Eri spectral analysis
was performed between 0.3-5.1 keV.
Individual spectra were binned so as to have a minimum of 20 counts per energy bin.
The EPIC spectra of the stars were fitted with a single (1T) and two (2T)
 temperature
collisional plasma model known as APEC (Astrophysical Plasma Emission Code, Smith et al. 2001),
with variable elemental abundances.
Abundances for all the elements in the APEC were varied together.
The interstellar hydrogen
column density ($N_H$) was left free to vary.
For all the stars, no 1T or 2T
plasma models with solar photospheric (Anders \& Grevesse 1989)
abundances could fit the data, as unacceptably large values of
$\chi^{2}_\nu$ were obtained.
 Acceptable APEC 2T
fits were achieved only when the abundances were allowed to depart
from the solar values.
The best-fit 2T plasma models with sub-solar abundances along with the
significance of
the residuals in terms of $\Delta \chi^2$  are shown in Figures
\ref{fig:qspec}(a) to (f) for all the stars.
Table 6 summarizes the best-fit values obtained for the various
parameters along with the minimum $\chi_\nu^2 = \chi^2/\nu$
(where $\nu$  is degrees of freedom) and the
90\% confidence error bars estimated from the minimum $\chi^2+2.71$.
The derived values $N_H$ from the spectral analysis  were found to be lower than that of
the total galactic HI column density (Dicky \& Lockman 1990) towards the
direction of that star. The cool and hot temperatures at the quiescent state
of these stars were found in the range of 0.2 - 0.5 and 0.6 - 1.0 keV, respectively.
The corresponding  EM2/EM1 during the quiescent state of each star was found in the range of
0.8 - 1.2, except for the star EP Eri where it was found to be $\sim 0.4$.

\subsection{Spectral evolution of X-ray flares}
\label{sec:flare}
\begin{figure}
\subfigure[]{ \includegraphics[height=8.0cm, width=6.5cm,angle=-90]{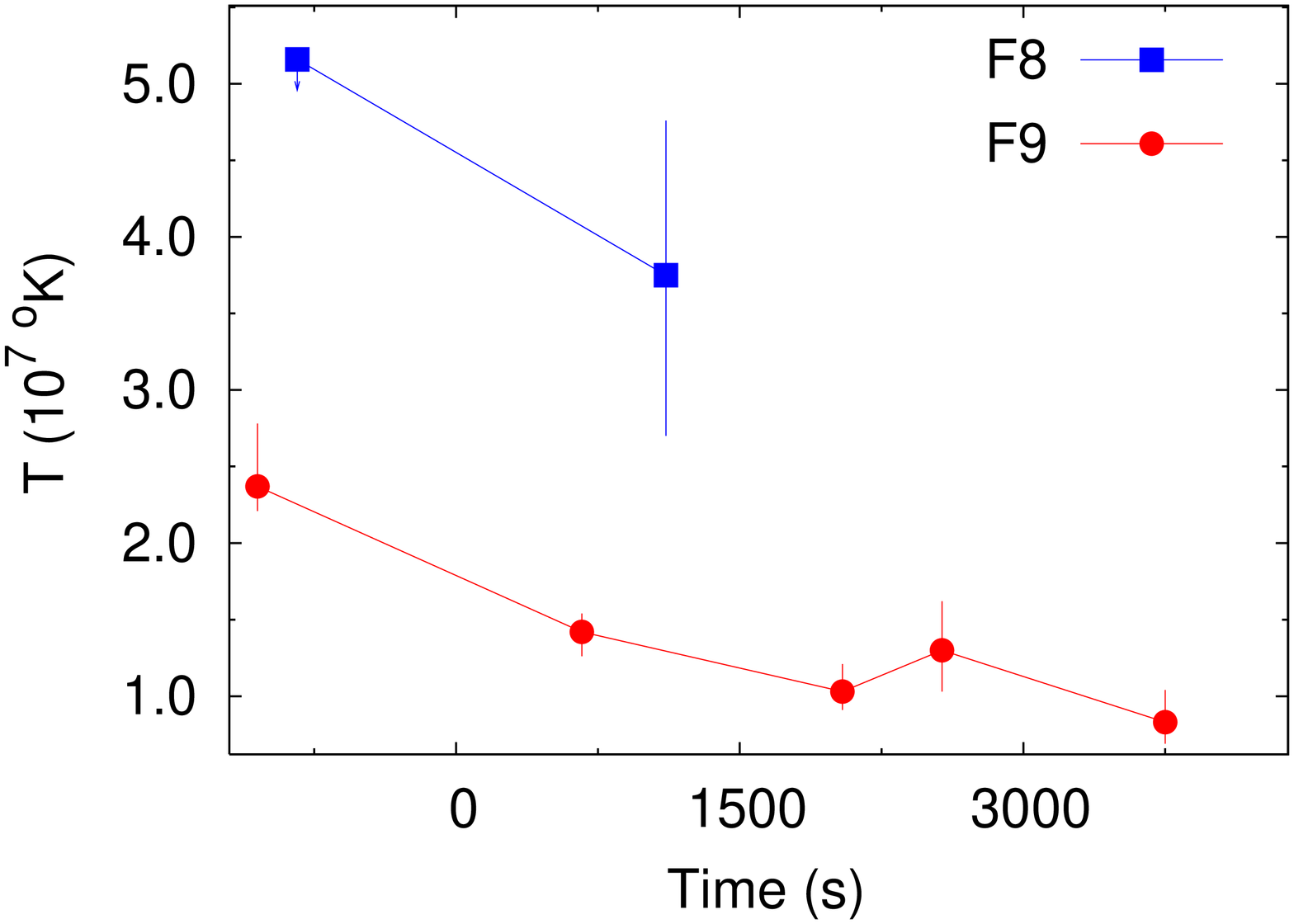} }
\subfigure[]{ \includegraphics[height=8.0cm, width=6.5cm,angle=-90]{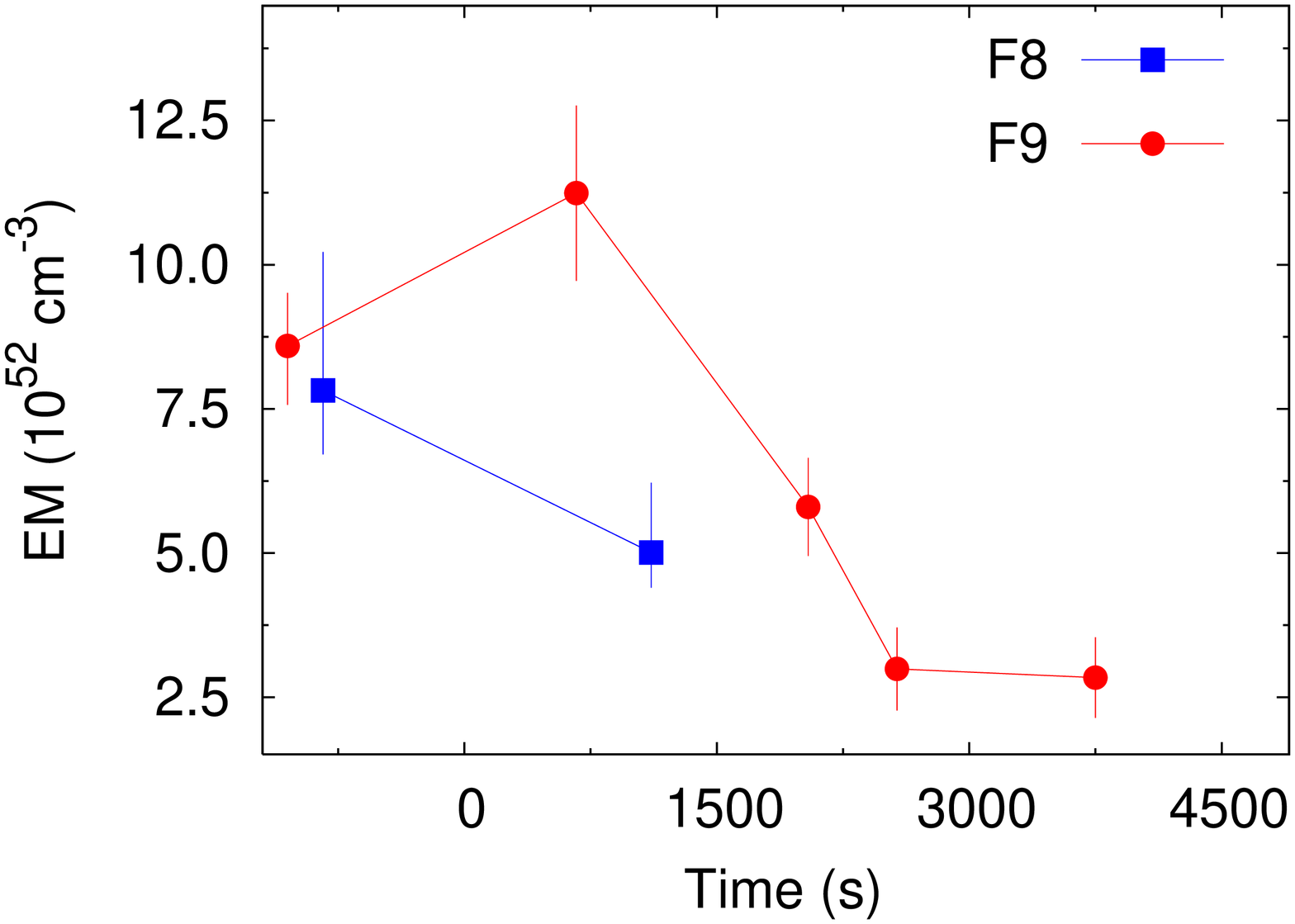} }
\caption{Plots of (a) temperature and (b) emission measure
 during the flares F8 and F9
of the star IM Vir. Here arrow shows upper limit on data point.}
\label{fig:imZTEM}
\end{figure}

\begin{figure}
\subfigure[]{\includegraphics[height=8.0cm, width=6.5cm,angle=-90]{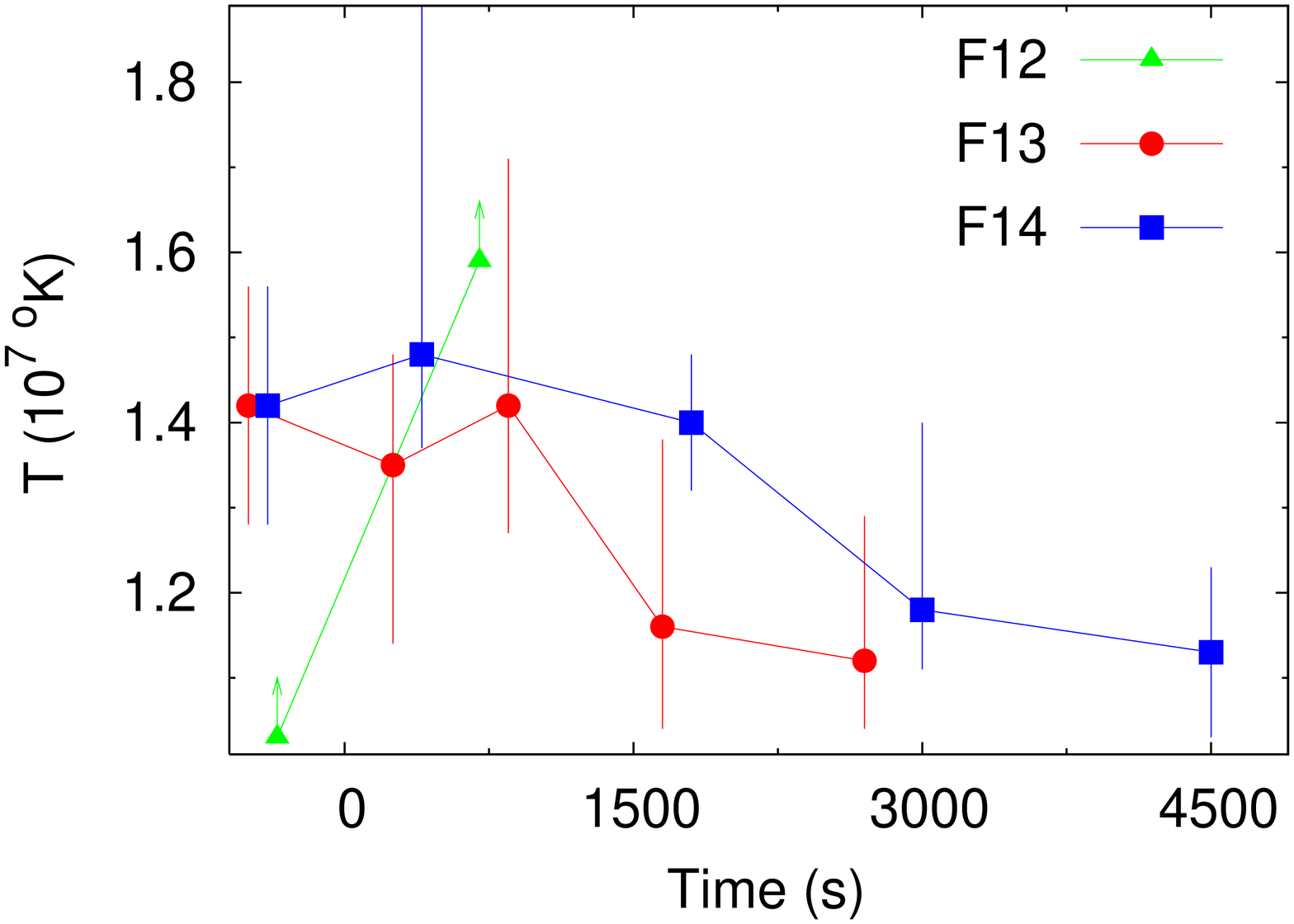} }
\subfigure[]{\includegraphics[height=8.0cm, width=6.5cm,angle=-90]{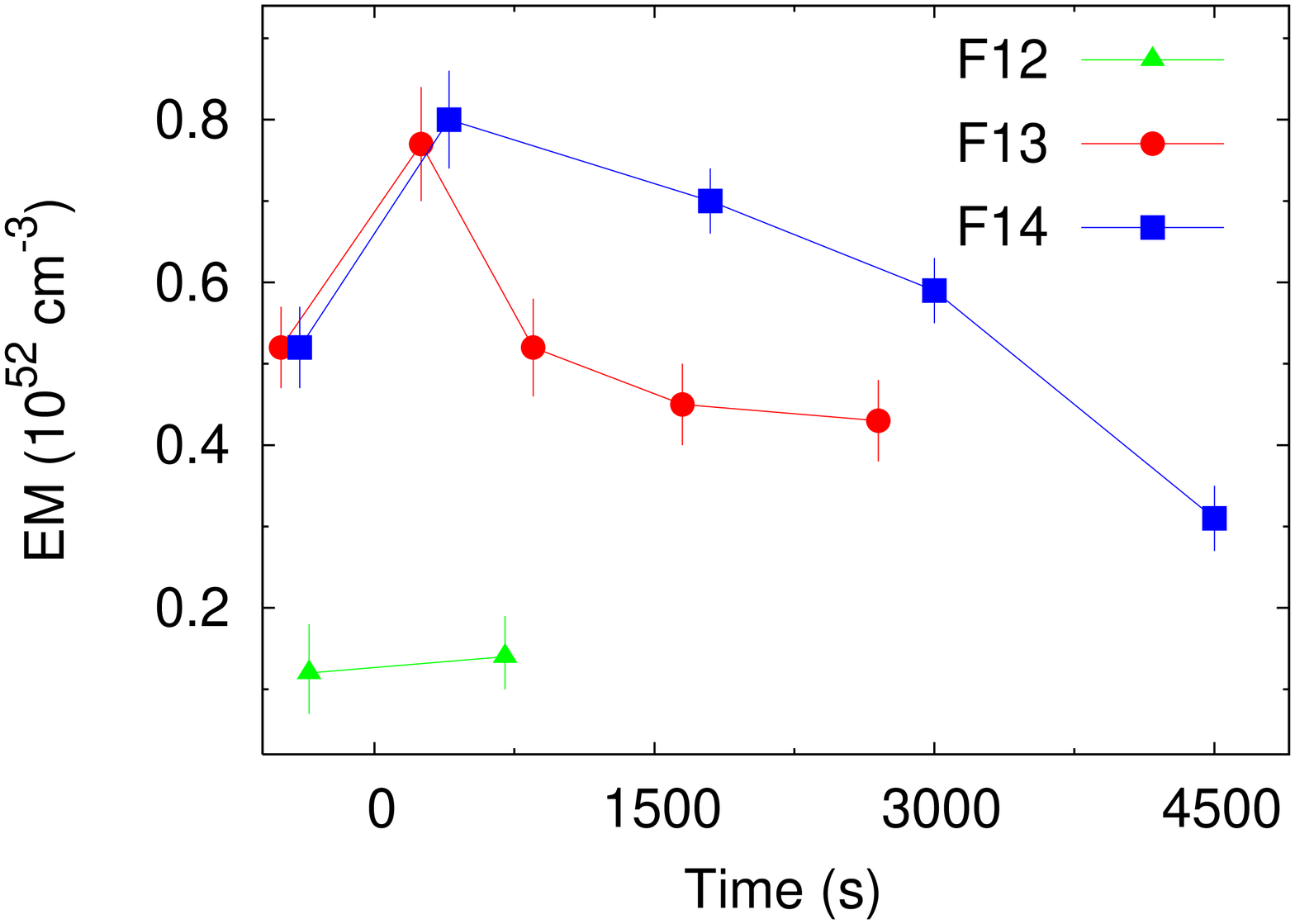} }
\caption{Plots of (a) temperature and (b) emission measure  during the flares F12, F13
and F14  of the star CC Eri. Here, arrow shows lower limit on the data point}
\label{fig:ccZTEM}
\end{figure}

\begin{figure}
\subfigure[]{ \includegraphics[height=8.0cm, width=6.5cm,angle=-90]{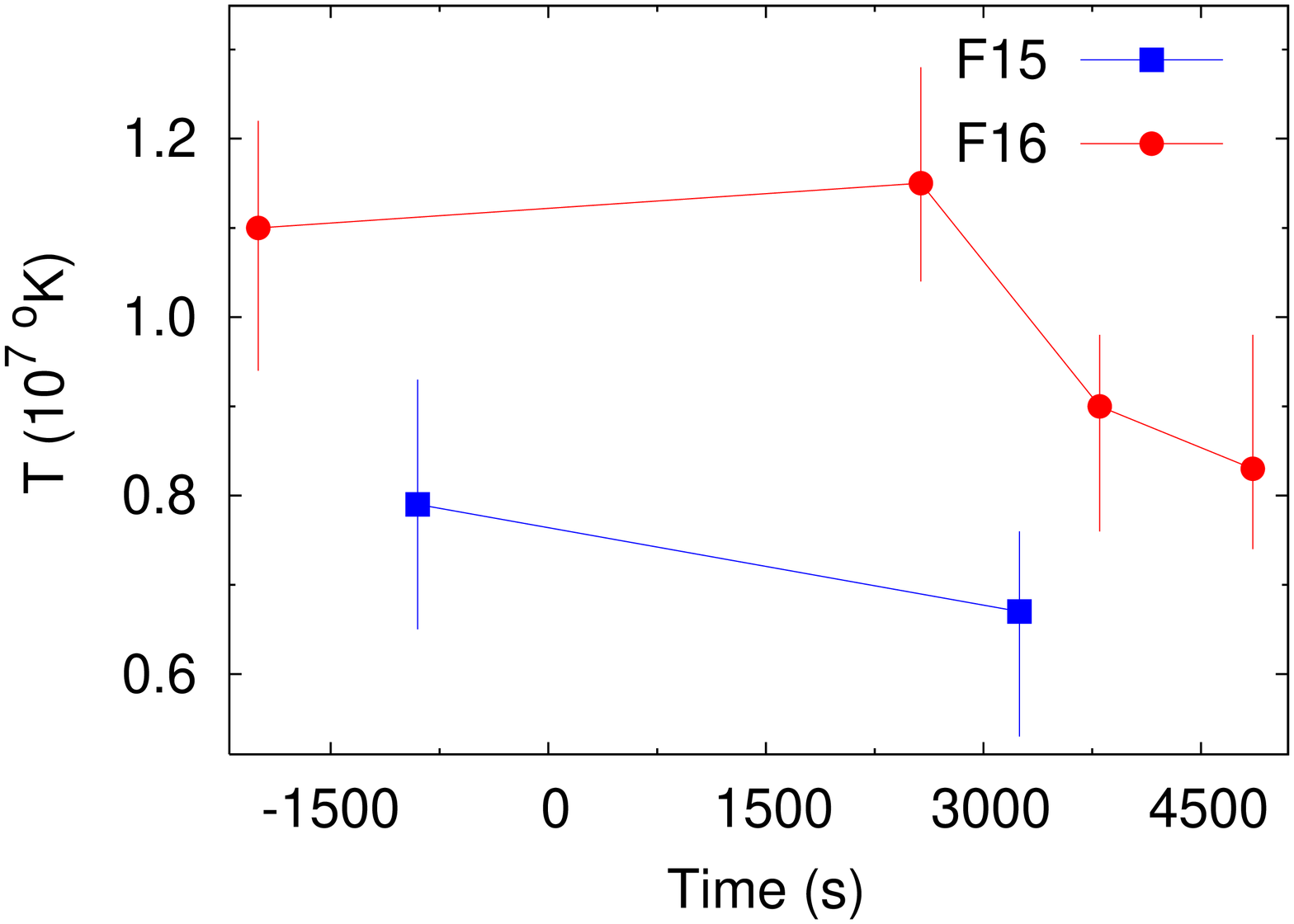} }
\subfigure[]{ \includegraphics[height=8.0cm, width=6.5cm,angle=-90]{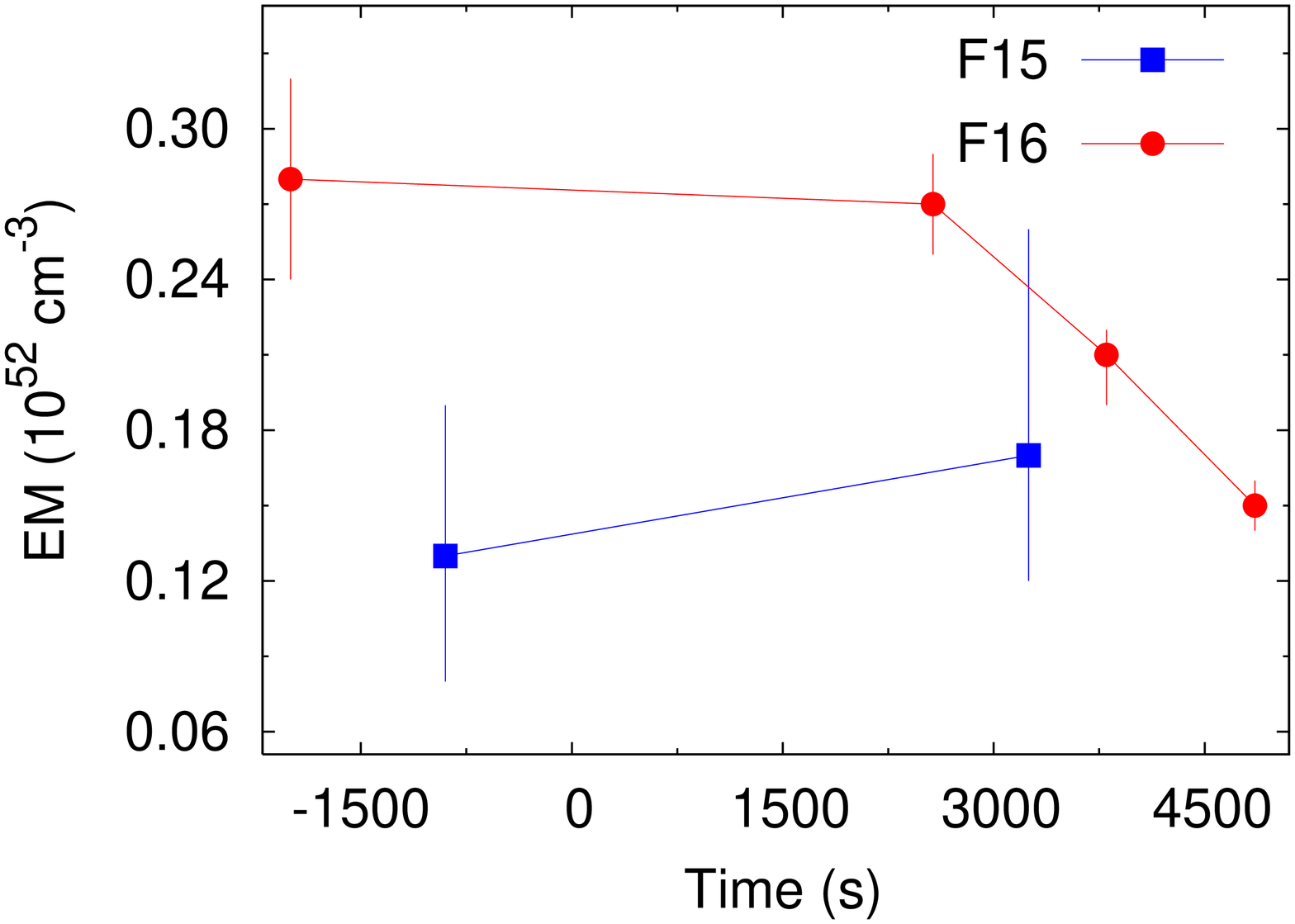} }
\caption{Plots of (a) temperature and (b) emission measures
during the flares F15 and  F16
of the star EP Eri.}
\label{fig:epZTEM}
\end{figure}

 In order to trace the spectral changes  during
the flares,  we have analysed the spectra of the different
time intervals shown in Figure \ref{fig:mospnlc}.
To study the flare emission only, we have performed 1-T spectral fits of the
data, with the quiescent emission taken into account by including its
best-fit 2-T model as a frozen background contribution. This is equivalent
to consider the flare emission subtracted of the quiescent level, allows us
to derive one ``effective'' temperature and one emission measure of the
flaring plasma.
The abundances were kept fix to that of the quiescent emission.
The best fit spectral parameters for each flare segments are listed in
Table 7. Figures \ref{fig:v368ZTEM} - \ref{fig:epZTEM} show the
temporal evolution of the temperature  and
corresponding emission measure of the flares detected in the targeted stars,
where '0' time corresponds to the flare peak time.

\subsubsection{V368 Cep}
\label{sec:V368}
\begin{figure}
\subfigure[]{ \includegraphics[height=8.0cm, width=6.5cm,angle=-90]{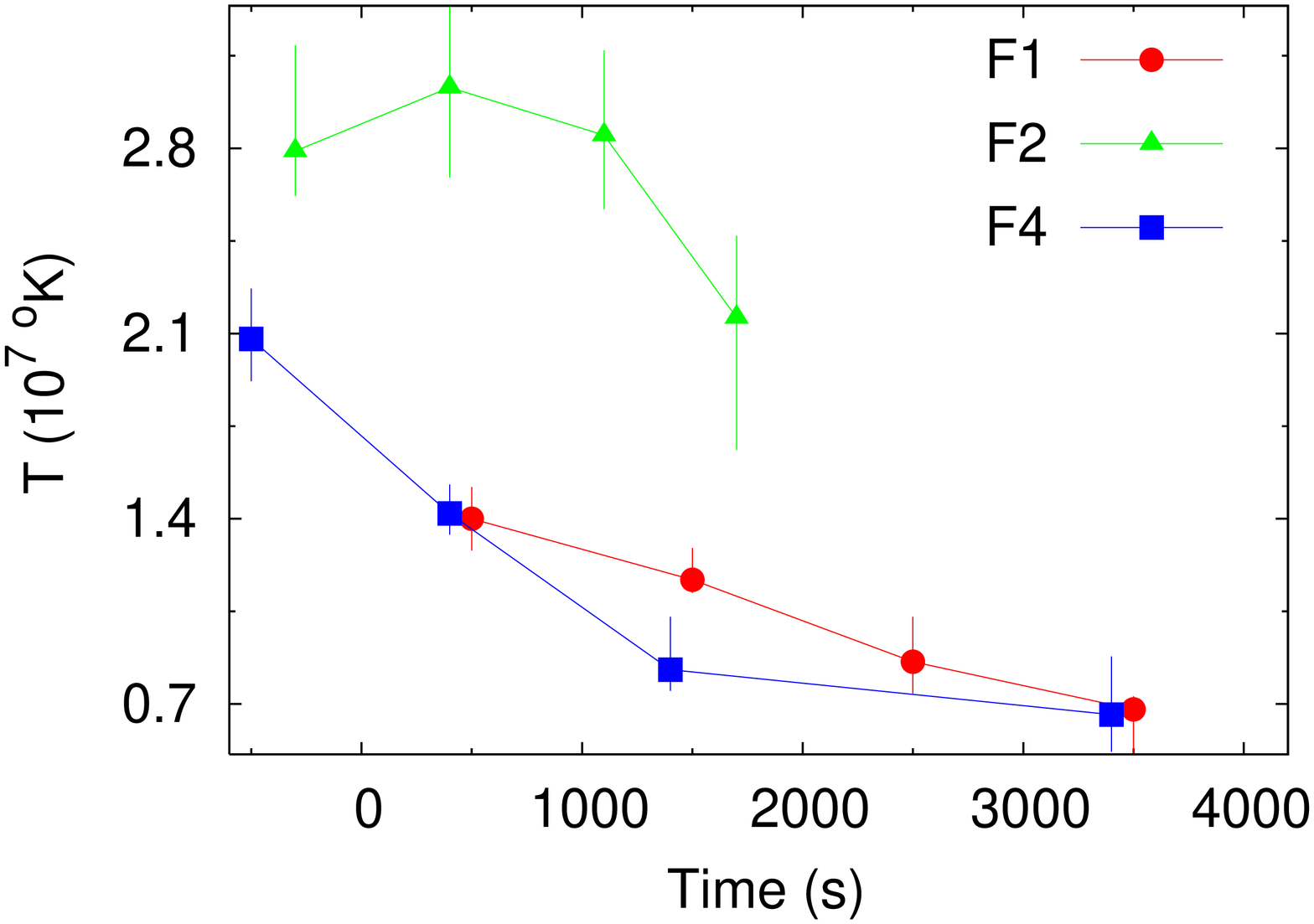} }
\subfigure[]{ \includegraphics[height=8.0cm, width=6.5cm,angle=-90]{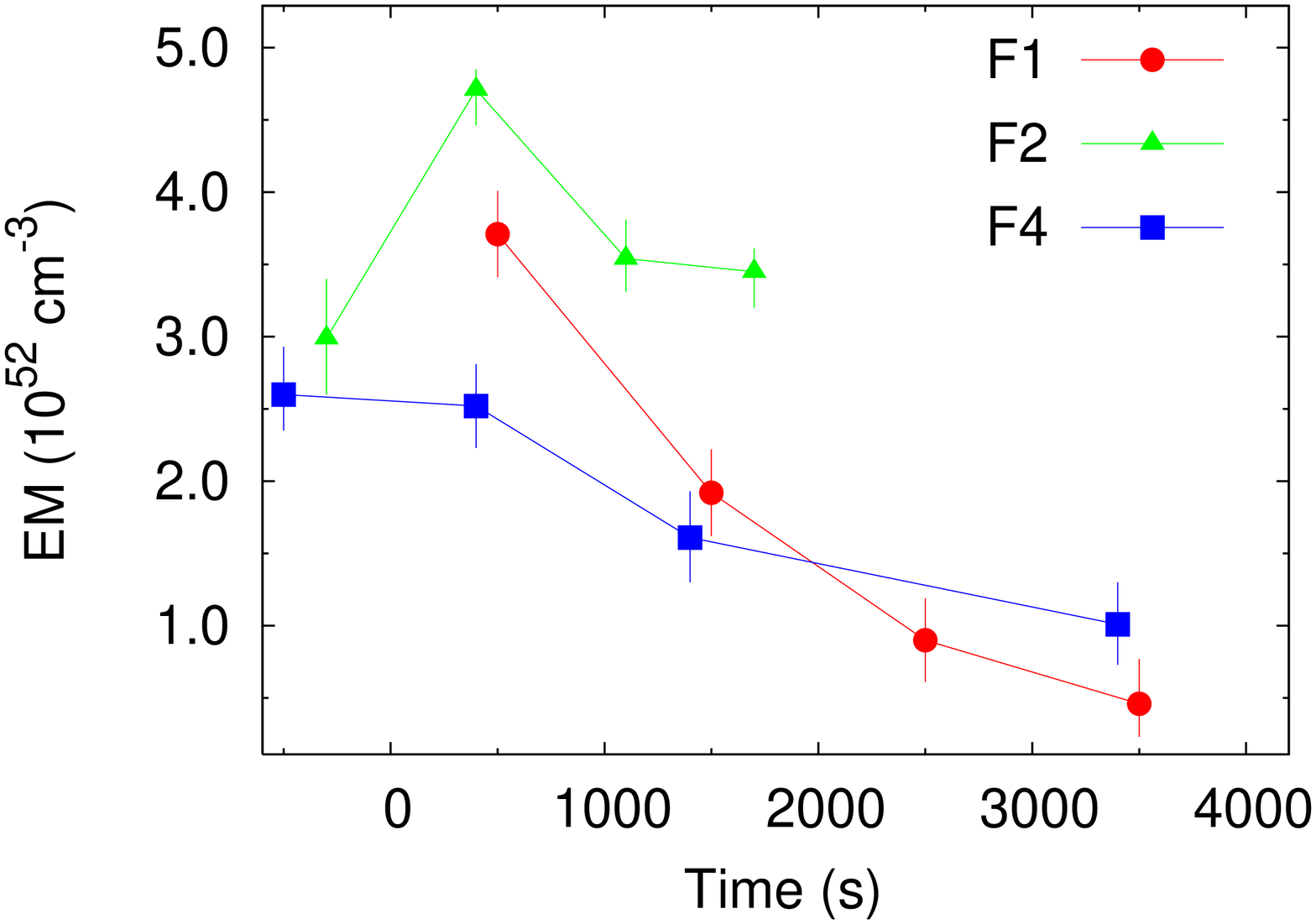} }
\caption{Plots of (a) temperature and (b) emission measure
 during the flares F1, flare F2,
and F4 of the star V368 Cep.  Here, '0' time corresponds to flare
peak and negative scale of time indicates the flare during its rise phase.}
\label{fig:v368ZTEM}
\end{figure}

Spectral data of flares F1, F2 and F4 were binned into four segments. However,
for the flare F3  data could be collected over only
 one segment.  Both the temperature and the corresponding
emission measure were found to decrease from the flare peak to the quiescent
state (see also Table 7). As shown in Figure \ref{fig:v368ZTEM}(a), the highest temperature
for the flare F1 was  14 MK and decreased along the decay path to  6.8 MK. A similar
trend was also found in the emission measure, where it increased by a
factor of seven (see Figure \ref{fig:v368ZTEM} (b)).
The temperature and emission measure for the flare F2
were peaked simultaneously during the decay phase D5.
However, for the flare
F4 both parameters were peaked during the rise phase R2.
The maximum luminosity was found during the flares F2 and
peaked at  $1.18\times10^{30}$ \lum, which is $\sim 1.7$ times more than
that of the quiescent state.

\subsubsection{XI Boo}
\begin{figure}
\subfigure[]{ \includegraphics[height=8.0cm, width=6.5cm,angle=-90]{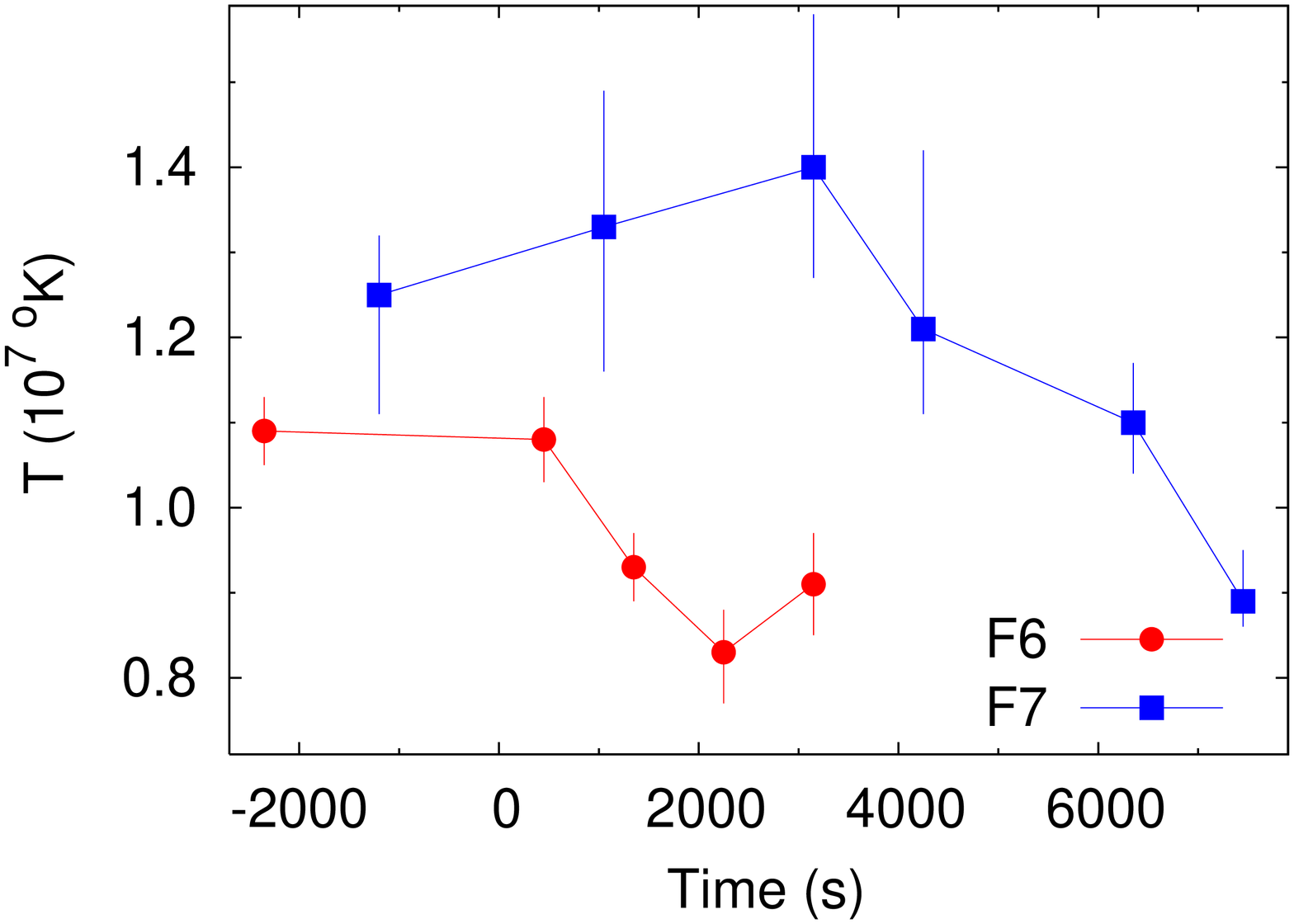} }
\subfigure[]{ \includegraphics[height=8.0cm, width=6.5cm,angle=-90]{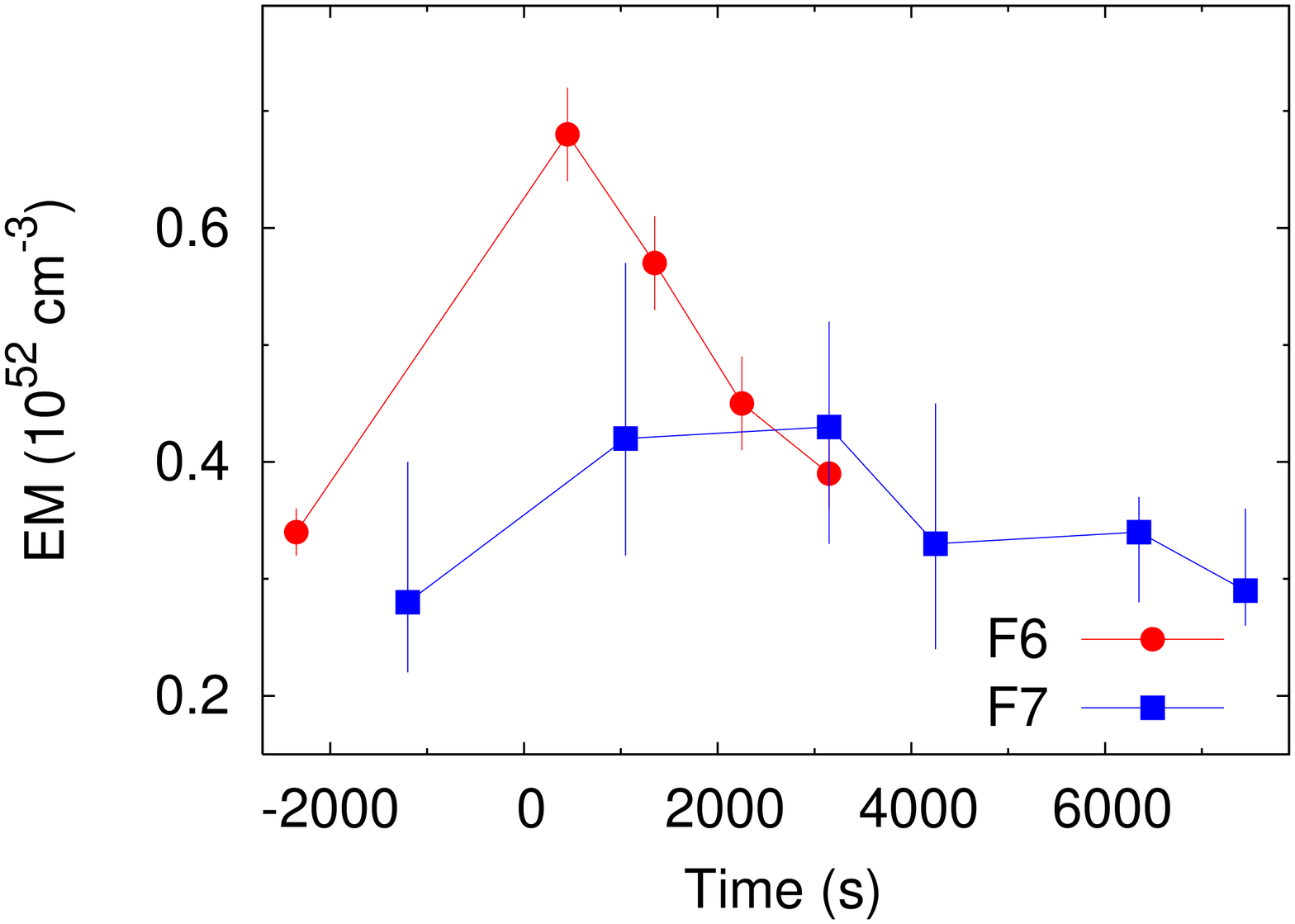} }
\caption{Plots of (a) temperature and (b) emission measure of
dominant component of 2T fit during the flares F6  and F7
of the star XI Boo.}
\label{fig:xiZTEM}
\end{figure}

To see the flare evolution, flare F6 and F7  have binned into five and six intervals, respectively.
For the flare segments R2, D5, D6, D7 and D8, the 1T APEC fit gave high value of
$\chi^2_\nu$ (=1.5 for R2, 1.75 for D5, 1.61 for D6, 1.55 for D7 and 1.34 for D8).
The spectral fit to the data for these time segments improves the $\chi^2$
by using the 2T plasma model (see Table 7). The temperature and the corresponding
emission measure of the cool component for these time segments were found to be constant
at $0.6\pm0.1$ keV and $4.0\pm1.0 \times 10^{51}$ cm$^{-3}$, respectively.
The spectral parameters for hot component of the best fit 2T plasma model
are reported in Table 7.
The evolution of the temperature of hot component and the corresponding emission measure are shown
in Figures \ref{fig:xiZTEM} (a) and  \ref{fig:xiZTEM} (b), respectively.
For the flares F6 and F7, both the temperature and corresponding emission measure were
reached to the maximum value  during the decay phase.
 As mentioned in Table 6 and 7, during the flares F6 and F7,
XI Boo was $\sim 2$ times more X-ray luminous than that of the
quiescent state.

\subsubsection{IM Vir}
The spectral data of the flares F8 and F9 were binned in two and four different time segments, respectively
(see Figure \ref{fig:mospnlc}(c)) and spectra of each segment were fitted with 1T APEC model.
The plots of  temperature  and the corresponding emission
measure are shown
in Figures \ref{fig:imZTEM} (a), and (b), respectively.
For the flare F9, the maximum fit temperature,
was reached during the rising phase R2, somewhat earlier than the
maximum of its emission measure. The emission measure was increased by a factor
of four during the flare F9. The luminosity at the peak of flare F9  was
found to  be  $\sim 2$ times
more than that of the quiescent state.

\subsubsection{V471 Tau}
\begin{figure}
\subfigure[]{ \includegraphics[height=8.0cm, width=6.5cm,angle=-90]{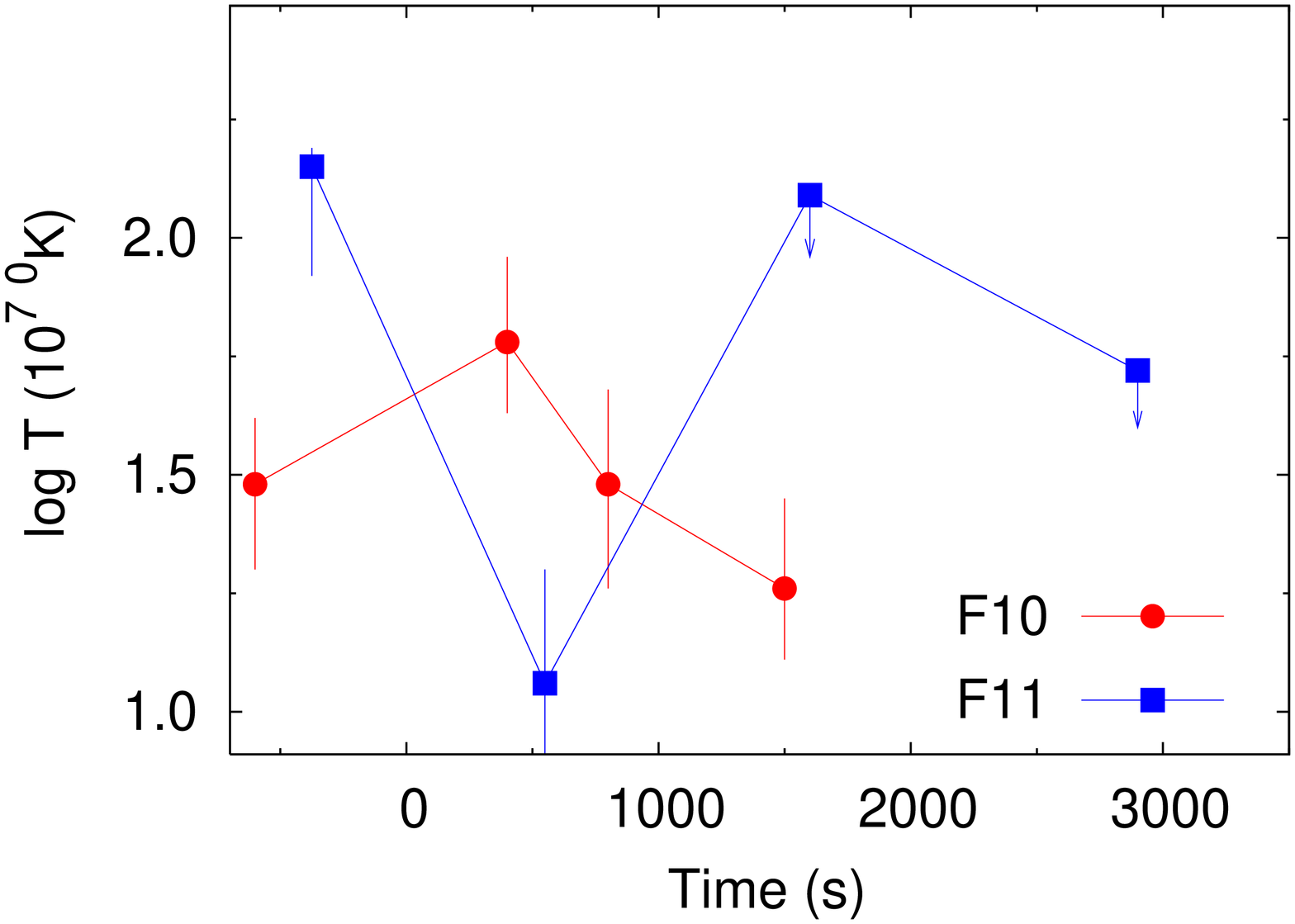} }
\subfigure[]{ \includegraphics[height=8.0cm, width=6.5cm,angle=-90]{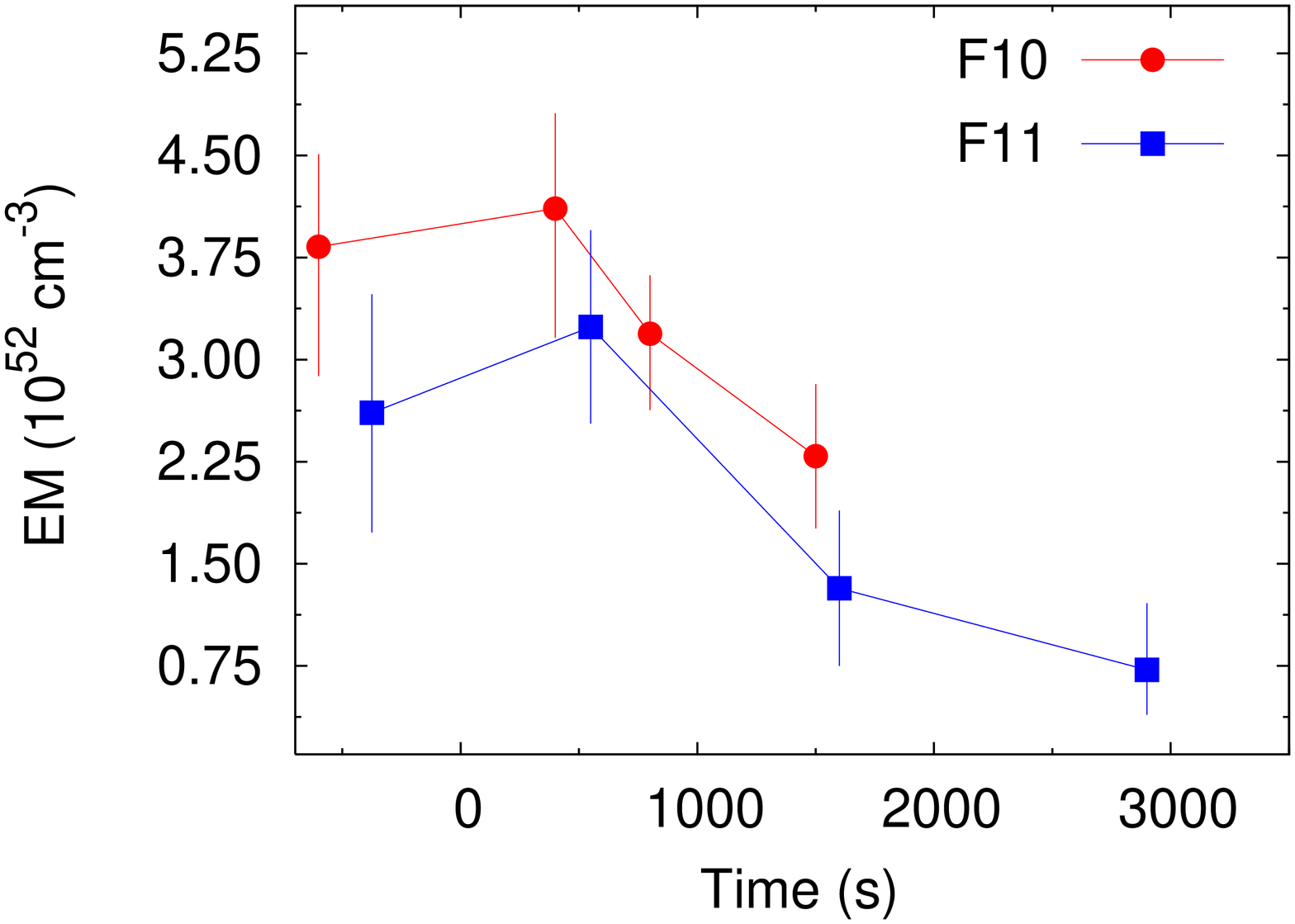} }
\caption{Plots of (a) temperature and (b) emission measure
during the flares F10  and F11
of the star V471 Tau.  Here, arrow shows the upper limit on data point.}
\label{fig:v471ZTEM}
\end{figure}

Two flares detected in the star V471 Tau were examined by analysing four
different time intervals during the flares. Figures.
\ref{fig:v471ZTEM} (a) and (b) show the plot of temperature
and the corresponding emission measure during the flares F10
and F11.  For the flare F10 the temperature reached  its
maximum value of $17.8$ MK and decreased along  the decay
phase (see Table 7).  The corresponding emission measure was also
peaked  during the decay phase of the flare F10 (see Figure
\ref{fig:v471ZTEM}(b) and Table 7).  For the flare F11,
both the temperature and the emission measure follow similar
trend to the flare light curve.
The X-ray luminosity was increased by a factor of $\sim 1.6$ for the
flare F10 and $\sim 1.4$ for the flare F11.

\subsubsection{CC Eri}
The evolution of temperature and the corresponding emission measure for
three flares F12, F13 and F14 are shown in Figures \ref{fig:ccZTEM} (a) and (b),
respectively.
Small variations were found in the  temperature and the emission
measure during
the flares F13 and F14.  Both the temperatures and the corresponding emission
measures were found to decrease during the decay
phase for all flares (see also Table 7). The temperature and the corresponding
emission measure for all the flares were peaked during the decay phase. However, it appears that
temperature is peaked before the emission measure for the flare F13. The peak
luminosity during flares F13 and F14 was found to be $6.0 \times 10^{28}$  \lum ~more
than that of the quiescent state.

\subsubsection{EP Eri}
Spectra of different time segments of the flares are fitted with 1T plasma models.
Figures \ref{fig:epZTEM} (a) and (b) show the evolution  of
temperature and the corresponding emission
measure for the flares F15 and F16. For the flare F16, both the temperature and
the corresponding emission measure were peaked during the rise phase R2,  remained
constant upto the decay phase D2 and then decreased along the flare decay
without any appreciable changes in the luminosity. The luminosity during both
flares was found to be similar to that of the quiescent state (see Table 6 and Table 7).

\subsection{Loop modeling of the X-ray flares}
\label{sec:model}
Flares can not be resolved spatially on a star. However, by  an analogy
with solar flares and using flare loop models, it is possible to infer
the physical size and morphology of the loop structures involved in a stellar
flare.
The most widely used but approximate methods for analyzing stellar flares are:
i) Two-ribbon flare method (Kopp \& Poletto 1984), ii)  Quasi-Static
Cooling method (van den Oord \& Mewe 1989), iii)  Pure radiation
cooling method ({\rm Pallavicini et al. 1990}), iv) Rise and decay time method
(Hawley et al. 1995), and v) Hydrodynamic method (Reale et al. 1997).
The Two-ribbon flare model assumes that the flare decay is entirely
driven by heating released by magnetic reconnection of higher and higher
loops and neglects completely the effect of plasma cooling.  The other
four methods are instead based on the cooling of plasma  confined in a single
flaring loop.  In the
hypothesis of flares occurring inside closed coronal structures,
the decay time of the X-ray emission roughly scales as the
plasma cooling time. In turn, the cooling time scales with the
length of the structure which confines the plasma: the longer
the decay, the larger is the structure (e.g. Haisch et al.1983). A loop
thermodynamic decay time has been derived (van den Oord \&
Mewe 1989; Serio et al. 1991) as:

\begin{equation}
\tau_{th} = \frac{120 L_9}{\sqrt{T_7}}
\label{eq:l1}
\end{equation}

where $L_9$ and $T_7$ are the loop half-length and the maximum
temperature (T$_{max}$) of the flaring plasma, in units of $10^9$ cm and $10^7$ K,
respectively. The timescale above is derived under the
hypothesis of impulsive heat released at the beginning of a flare.

The hydrodynamic model includes both plasma cooling and
the effect of heating during flare decay.
Reale et al. (1997) presented a method to infer the geometrical size
and other relevant physical parameters of the flaring loops, based on the
decay time and on evolution of temperature and the emission measure (EM)
during the flare decay. The empirical formula for the estimation of the
unresolved flaring loop length has been derived as:

\begin{equation}
L_{9} = \frac{\tau_{d}\sqrt{T_{7}}}{120 f(\zeta)}   ~~~~~~~~ f(\zeta) \geq 1
\label{eq:l2}
\end{equation}

\noindent
where $\tau_{d}$ is the decay time
of the light curve, and $f(\zeta)$ is a non-dimensional correction factor larger
than one. $\zeta$ is  the slope of the decay path in the density-temperature
diagram (Sylwester et al. 1993) and is maximum ($\sim 2$) if heating is
negligible and minimum ($\sim 0.5$) if heating dominates the decay. The
correction factor $f(\zeta)$  and loop maximum temperature are calibrated
for  the XMM-Newton EPIC spectral response and are given as (Reale 2007):

\begin{equation}
f(\zeta) = \frac{0.51}{\zeta - 0.35} + 1.36 ~~~~~~~~~~~\rm{ (for ~0.35 < \zeta \leq 1.6)}
\label{eq:fxi}
\end{equation}

\begin{equation}
T_{max} = 0.13 T_{obs}^{1.16}
\label{eq:tmax}
\end{equation}

Other physical properties of the flaring plasma can be inferred from the
analysis of flare data. The analysis of X-ray spectra provides
values of the temperature and emission measure.
From the emission measure (EM) and the plasma density ($n_e$), the
volume of the flaring loop is estimated as (Reale 2002):

\begin{equation}
V = \frac{EM}{n^2_e}
\label{eq:density}
\end{equation}

\noindent
where V is the  loop volume. It has been
shown  that  in the equilibrium condition the loop scaling laws hold, linking maximum
pressure, temperature, loop length  and heating rate per unit volume as
(Rosner, Tucker \& Vaiana 1978, Kuin \& Martens 1982):

\begin{equation}
T_{max}  = 1.4 \times 10^3 (pL)^{1/3}
\label{eq:pressure}
\end{equation}

\begin{equation}
E_H \approx 10^{-6} T_{max}^{3.5} L^{-2}  ~~~ergs ~s^{-1} cm^{-3}
\label{eq:EH}
\end{equation}

\noindent
The minimum magnetic field necessary to confine the flaring plasma can be
simply estimated as

\begin{equation}
B = \sqrt{8 \pi p}
\label{eq:B}
\end{equation}

Because of limited statistics, we have modeled only ten flares.
The results of the model parameters are listed in Table 8. In this
table,
column 1 is name of the star and the corresponding flare, column 2 is
slope of density-temperature diagram ($\zeta$), column 3 contains
loop maximum temperature based on spectral fit and equation (5),
 semiloop length is  determined using the equation (3) and given in column
 4, pressure in the loop at the flare peak is estimated using the
loop scaling law (see equation (7); column 5), column 6 contains maximum
electron density at the flare loop (we have assumed a totally ionized
hydrogen plasma i.e. $p = 2n_ekT_{max}$; since pressure derived from equation (7) is a maximum value, therefore, a upper limit on density is estimated), volume of the flaring plasma
is estimated from maximum emission measure and from the estimated
maximum electron density, and therefore, is treated as
lower limit (equation (6); column 7), column 8
is heating rate per unit volume and estimated using equation (8),
the minimum magnetic filed necessary to confine the flaring plasma
at loop apex $B$ (see equation \ref{eq:B}) is given in column 9,
the loop aspect ratio $\beta$ (=$r/L$, $r$ is radius of the loop
 cross section, assuming circular) estimated from the
loop scaling law $V = 2\pi \beta^2 L^3$ and number of loops
filling the flare volume estimated assuming that $\beta = 0.1$
for a single loop are given in column 9 and 10, respectively.
The model parameters for the different flares from each star are
presented and discussed below.

\subsubsection{Flares F1, F2 and F4 of V368 Cep}
\begin{figure}
\centering
\subfigure[]{\includegraphics[height=6.5cm, width=6.5cm,angle=-90]{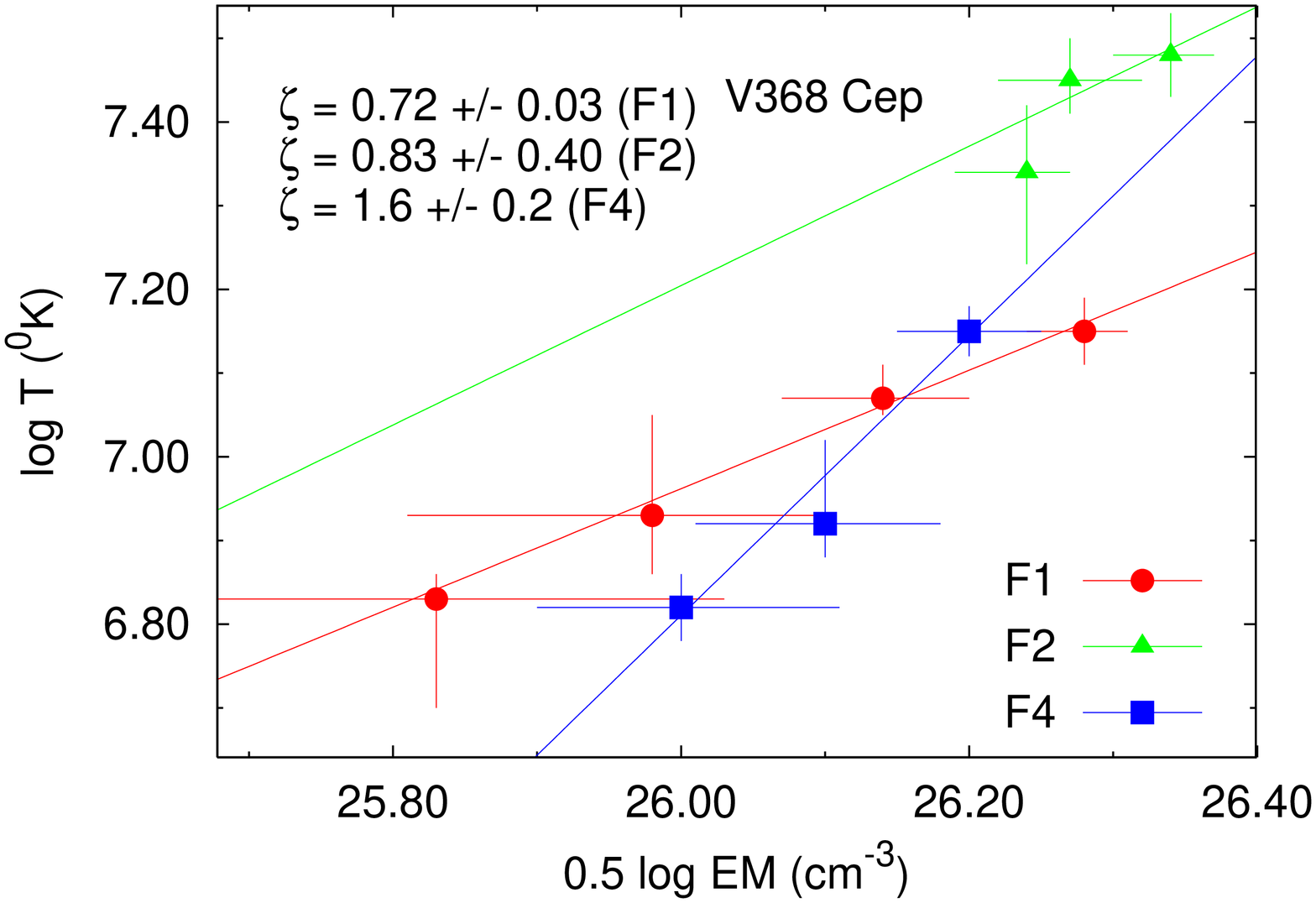}}
\subfigure[]{\includegraphics[height=6.5cm, width=6.5cm,angle=-90]{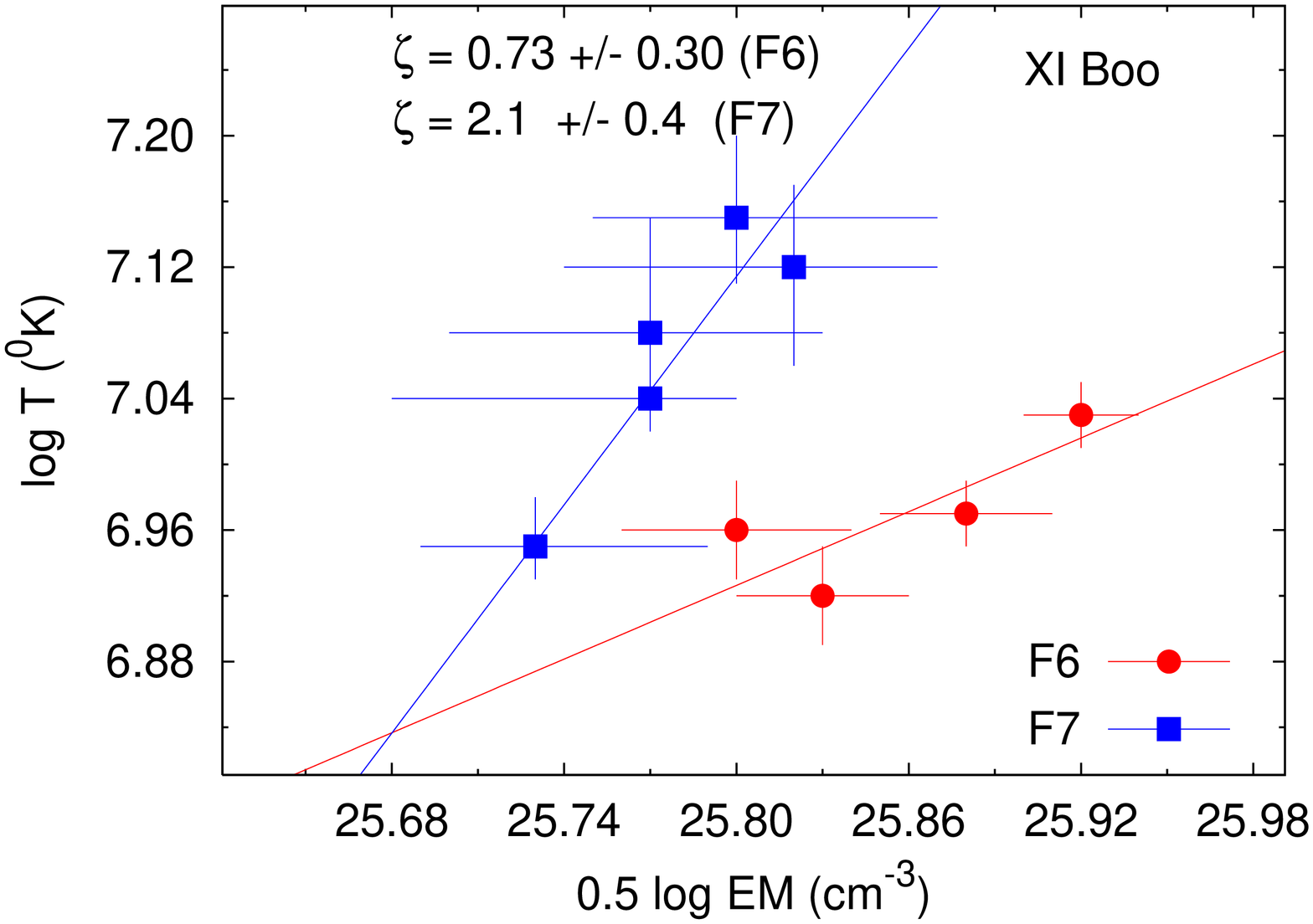}}
\subfigure[]{\includegraphics[height=6.5cm, width=6.5cm,angle=-90]{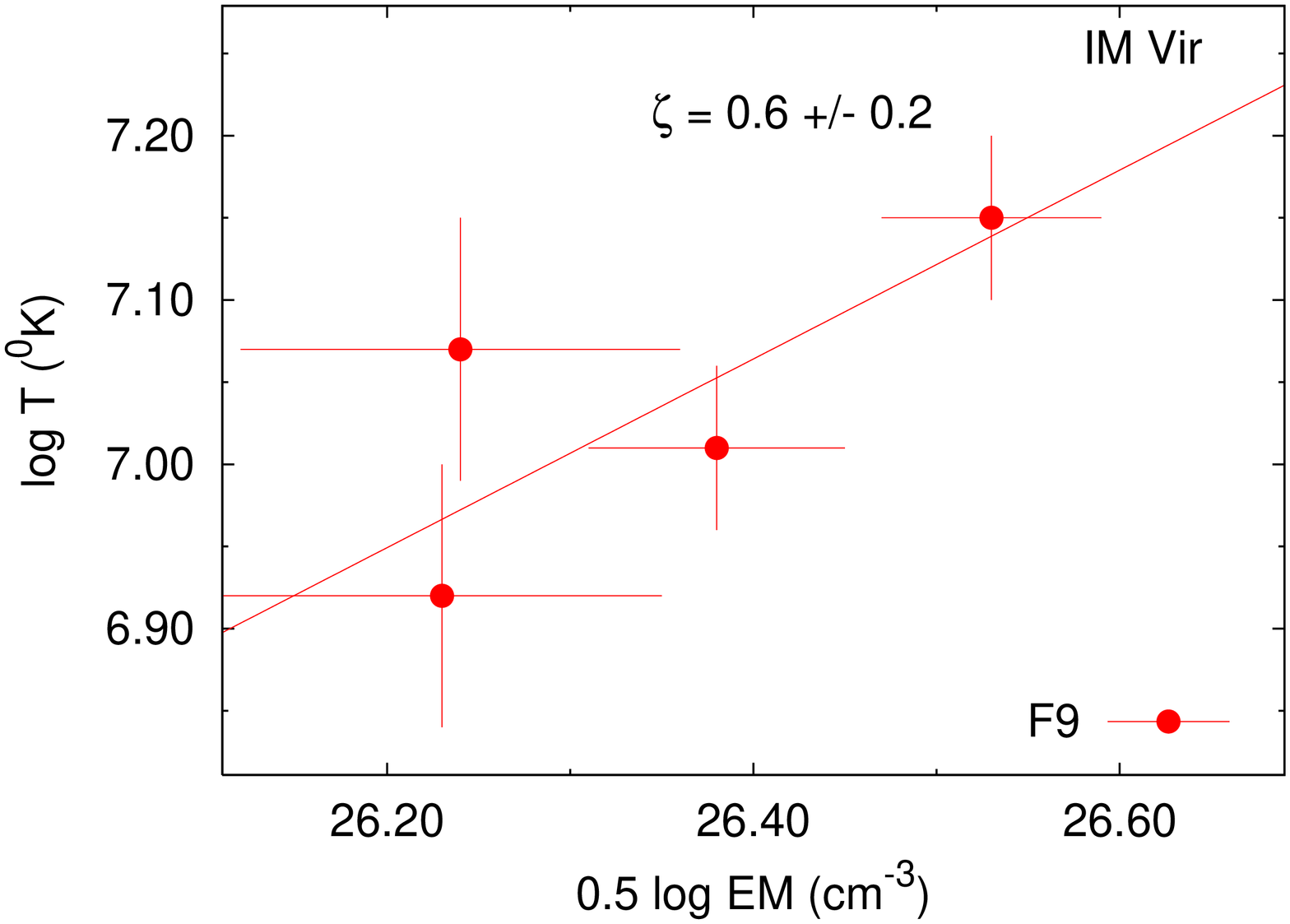}}
\subfigure[]{\includegraphics[height=6.5cm, width=6.5cm,angle=-90]{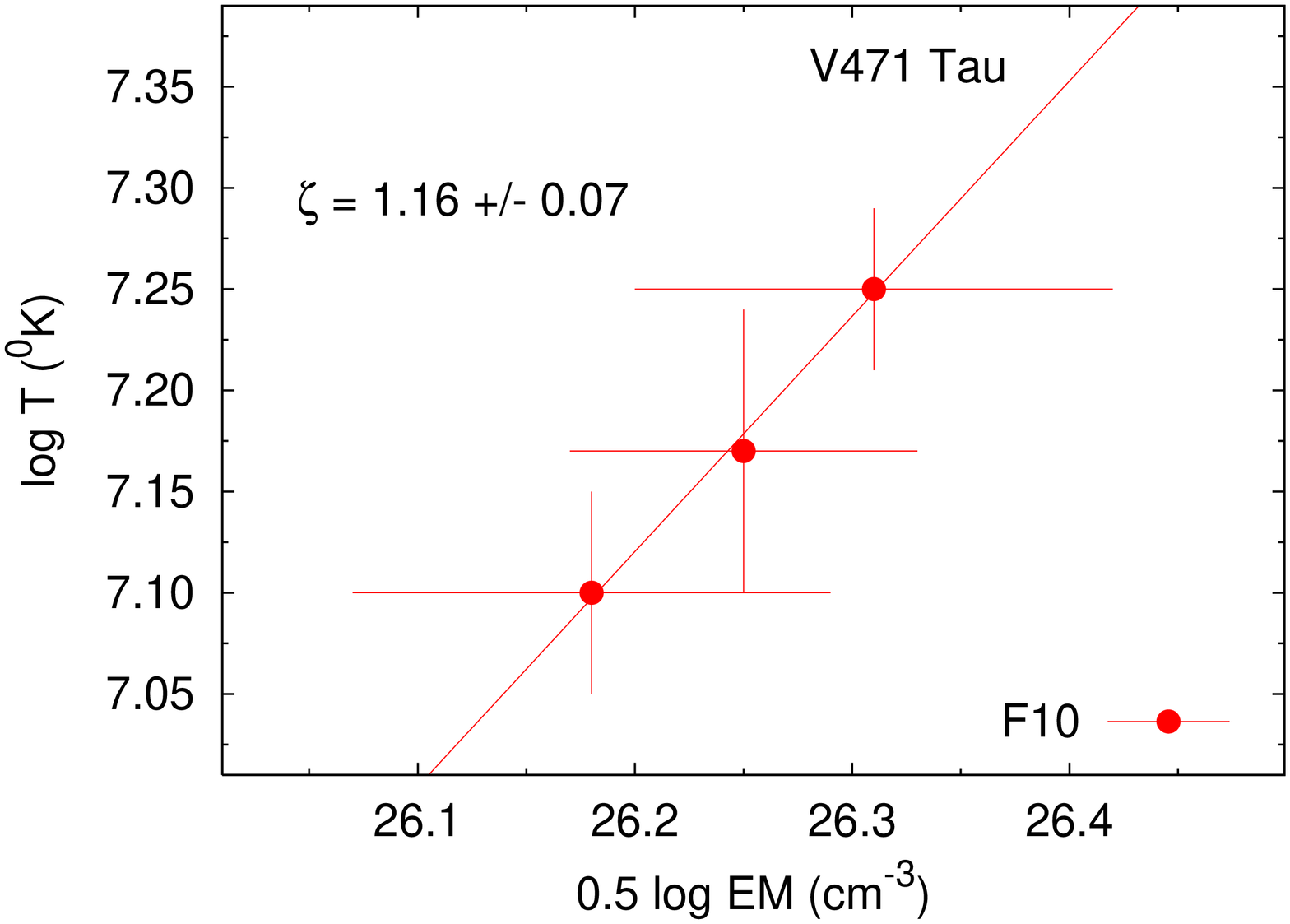}}
\subfigure[]{\includegraphics[height=6.5cm, width=6.5cm,angle=-90]{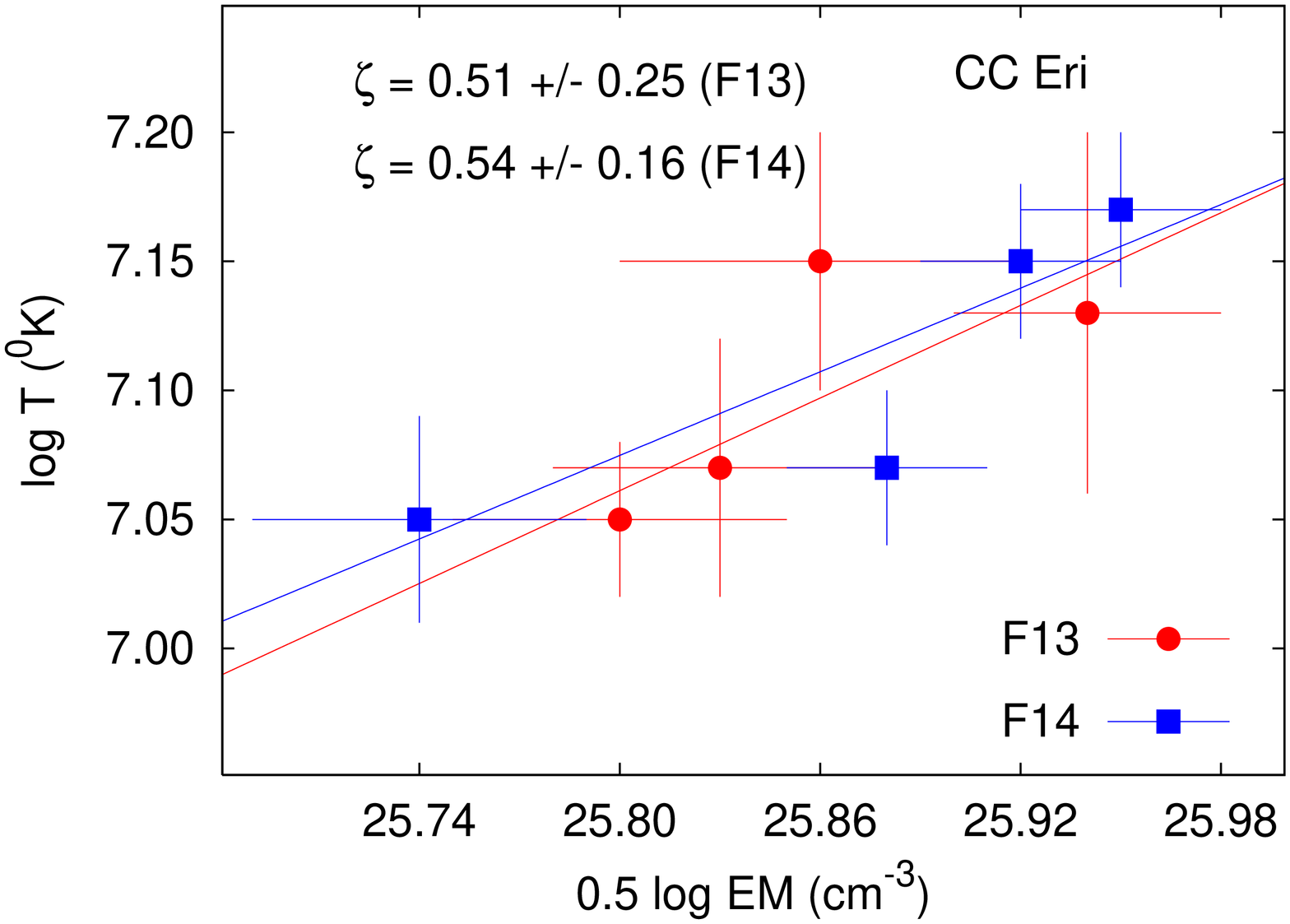}}
\subfigure[]{\includegraphics[height=6.5cm, width=6.5cm,angle=-90]{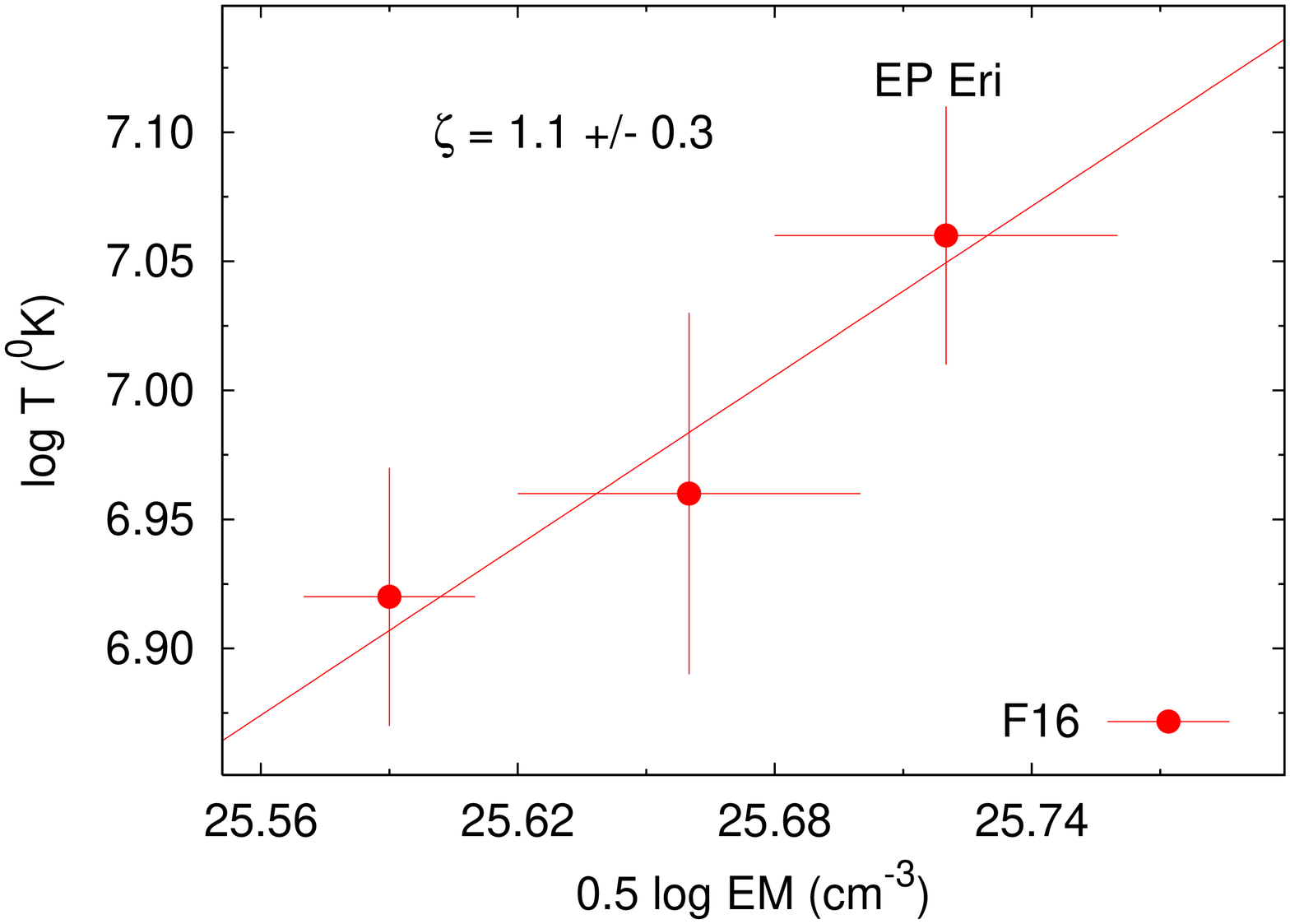}}
\caption{ The density-temperature diagram, where EM$^{1/2}$ has been used as a proxy of density.
Symbols for the each flares are given at the bottom right corner.
Straight lines represent the best linear fit to the corresponding data. $\zeta$ is
 slope of density-temperature diagram.}
\label{fig:nt}
\end{figure}

The average temperatures of the loops, usually
lower than the real loop maximum temperatures, are found from the
spectral analysis of the data as given in \S \ref{sec:flare}.
According to equation (\ref{eq:tmax}), the loop maximum temperatures for the flare
F1, F2 and F4 are found to be $25.5\pm2.5$, $62\pm8$ and $40.3\pm1.8$ MK, respectively. Figure \ref{fig:nt} (a)
shows the density-temperature (n-T) diagram,
where EM$^{1/2}$ has been used as a proxy of density. In Figure \ref{fig:nt}
(a) the flares F1 ,F2 and F4 are represented by solid circles , solid triangles and solid squares, respectively.
The solid lines represent the best linear fit to the corresponding data,
 providing  the slope  $\zeta$.
These values of
$\zeta$ indicate  the presence of a sustained
heating during the decay of the flares F1 and F2. However, sustained heating is negligible during the decay of the flare F4.
As discussed in \S \ref{sec:model}, the e-folding time of the light curve and the  slope
of n-T path can be used to estimate the loop half-length.
According to equation (3)  the semiloop lengths for the flares F1, F2 and F4 are
determined to be $6.0\pm0.6 \times 10^9  ~(=0.1R_\star)$, $19.8\pm8.6
~(= 0.4R_\star)$ and $11.7\pm1.3 \times 10^9 ~(= 0.2R_\star)$ cm,
respectively. These loop lengths are much smaller than the pressure
scale height\footnote{defined as $h_p = 2kT/\mu g \sim 5000 \times T_{max}/(g/g_\odot)$,
where T is plasma temperature in the loop, $\mu$ is the molecular weight of the
plasmas and g is the surface gravity of the star.}  $h_p  > 1.2\times10^{11}$ cm.
Maximum pressure in the loop at the  flare peak is estimated to be
$>10^3$ dyne cm$^{-2}$ for the flares F1, F2 and F4,
respectively (see equation (7)).
Assuming that the hydrogen plasma is totally ionized ($p = 2n_ekT_{max}$),
the  maximum plasma density in the loop at the flare peak is estimated
in the order of $10^{11}$ cm$^{-3}$
for the flares F1, F2 and F4, respectively.
We computed a volume of $\sim 1.78\times 10^{30}$ cm$^{-3}$ for the flare F1
using the observed peak EM of $3.7\times10^{52}$ cm$^{-3}$.
Similarly, loop volume  is determined to be $7.2\times10^{29}$ cm$^3$ for flare F2
and $7.4 \times10^{29}$ cm$^3$  for the flare F4.
A hint for the heat pulse intensity comes from the flare maximum temperature. By applying
the loop scaling laws and loop maximum temperature (see equation (\ref{eq:EH})) the heat
pulse intensity for the flares F1, F2 and F4 would be the order of 2.36, 4.75
and 3.06 ergs cm$^{-3}$ s$^{-1}$, respectively.
From the pressure of the flare plasma the magnetic field required
to confine the plasma should be  more than the 160 Gauss for all the flares.

\subsubsection{Flares F6 and F7 of XI Boo}
The evolution of flare F6 and F7 in n - T plane is shown in
Figure \ref{fig:nt} (b), together with a least square fit to the decay phase.
The resulting best-fit slopes ($\zeta$) for the decaying phase of flare F6 and F7
indicate
that the flare F6 is driven by the time scale of the heating process, whereas
sustained heating is negligible during the decay of the  flare F7. The intrinsic flare
peak temperatures for flare F6 and F7 are, applying equation (5) to the observed maximum temperature
, $18.9\pm0.8$ and $25.3\pm2.7$ MK, respectively.
For the flare F6 the  semiloop length is estimated to
be  less than the pressure
scale height, $h_p = 10^{11}$ cm. Maximum pressure and density in the loop at
the peak of the flare F6 are estimated to be $130\pm 6$ dyne cm$^{-2}$ and $2.5\pm1.0
\times 10^{10}$ cm$^{-3}$, respectively. Using equation \ref{eq:density},
volume of the flaring plasma is estimated to be $1.1\times10^{31}$ cm$^3$.
A minimum magnetic field of $\sim  58$ Gauss is required to confine
the flaring plasma at the peak of the flare.
Although, the value of $\zeta$
for the flare F7 is outside the domain of the validity of the
method (see Real 2007), the estimated loop parameters obtained
from the hydrodynamic loop modeling are listed in Table 8.
The estimated semiloop length  for the flare F7
is smaller than the pressure scale height (=$1.3\times10^{11}$ cm).

\subsubsection{Flare F9 of IM Vir}
Using  equation (\ref{eq:tmax}) the maximum loop temperature for the
flare F9 is found to be $46.6\pm3.7$ MK.
Figure \ref{fig:nt} (c) shows the n-T diagram for the flare F9.
From the best linear fit, we found  the slope of the n-T diagram
implies that the heating  during the flare decay
was as strong as in the flares F1, F2 and F6.
The resulting semiloop length
is found to be
 ten times smaller than the pressure
scale height ($h_p = 2.3\times10^{11}$ cm),
 which implies a loop in hydrostatic
equilibrium with a plasma pressure of $2020\pm1100$  dyne cm$^{-2}$ and
 maximum heating rate of $\approx  2.07 $ ergs cm$^{-3}$ s$^{-1}$.
The maximum  plasma density and loop volume  for the flare are estimated
to be $1.6\pm0.8 \times10^{11}$ cm$^{-3}$ and $4.55\times10^{30}$ cm$^{3}$, respectively.
The minimum confining magnetic field is estimated to be  $\sim 226$ Gauss.

\subsubsection{Flare F10 of V471 Tau}
The maximum temperature in the loop at the flare peak was found to be $33.5\pm3.3$ MK for the flare F10.
(see Figure \ref{fig:nt}(d) for the n-T diagram of this flare).
The slope of the decay path of n-T diagram
indicates the presence of sustained heating.
Using equation (3) the semiloop length  of the flare F10 is calculated to be
$\sim 0.4R_\star$ and is  $\sim 5$ times smaller than the pressure scale height $h_p = 1.2\times10^{11}$ cm.
Using these parameters and the scaling laws for the loops, the maximum pressure, the maximum density
and the minimum magnetic field confining the flaring plasma in the
loop at the flare peak, the volume of flaring plasma and the heating rate per unit volume for
F10 are estimated to be $p= 630\pm200$ dyne cm$^{-2}$, $n_e=6.8\pm1.0\times10^{10}$ cm$^{-3}$ and $B=126$ Gauss,
$8.9\times10^{30}$ cm$^{3}$ and 0.46 ergs s$^{-1}$ cm$^{-3}$.

\subsubsection{Flares F13 and F14 of CC Eri}
Using  equation (\ref{eq:tmax}) the maximum
temperature in the loop responsible for the flares F13 and F14 are found to be $25.7\pm2.9$ and $27.0\pm2.1$ MK,
respectively.
Figure \ref{fig:nt} (e) shows the n-T diagram for these flares. The slopes in
n-T diagram for the flares F13 and F14
indicate the presence of strong sustained heating, and showing that
the observed decay is driven by the time-evolution of the heating process.
The value of $\zeta$  at the low extreme of the error bar
for the flare F13 is compatible with the lower
asymptotic value for which equation (4) can be applied.
Therefore, an upper limit for semi-length of the loop can be derived
by using the value of $\zeta$ at the high extreme of the error bar.
The resulting semi-length for the flares F13 and F14 are $<1.4 \times 10^{10} (=0.3/0.5 R_\star) $ and
$2.0\pm1.0 \times 10^{10} (=0.4/0.7 R_\star)$ cm, respectively.
These values of the loop length are much smaller than the pressure scale
height $h_p = 1.1\times 10^{11}$ for flare F13 and $h_p = 7.3\times10^{10}$
cm for flare F14.
A magnetic field more than 95 Gauss is required to confine the flaring plasma
at the top of both flares. A detailed analysis
of these flares is also given in Crespo-Chacon et al. (2007). The estimated
loop parameters for both flares are found well within $1\sigma$ level to that of
Crespo-Chacon et al. (2007).

\subsubsection{Flare F16 of EP Eri}
The trend in n-T diagram  suggests  the
presence of significant heat during the decay of the flare F16
(See Figure \ref{fig:nt} (f) and Table 8). The temperature at
the peak of the flare is determined to be $20.2\pm2.2$ MK.
The derived semiloop length $L =  2.4\pm0.5 \times 10^{10}$ cm ($=0.4R_\star$) is
less than the pressure scale height $h_p = 9 \times 10^{10}$ cm.
Using the same approach as discussed above, the maximum pressure
and the maximum density are derived at $p = 130\pm50$ dyne cm$^{-2}$ and $n_e =
2.3\pm0.5 \times 10^{10}$ cm$^{-3}$. Using the maximum emission
measure, the loop volume is estimated to be
$5.38\times10^{30}$ cm$^{3}$. The confining magnetic filed is estimated as
$B > 57$ Gauss.

\section{Discussion and Conclusions}
\label{sec:discussion}
We have carried out an analysis of XMM-Newton observations
of six G-K dwarfs. Light curves revealed flaring in all the stars
on time scale ranging from seconds to kiloseconds, with
various peak strengths. A total of 17 flares were
identified in these  G-K dwarfs. The decay time of these flares ranges
from 0.5  to 10 ks.  A similar range of flare decay times
in the pre-main-sequence stars of Pleiades cluster was found by Stelzer, Neuh{\"a}user \& Hambaryan (2000).
Among the 17 flares presented here, only four flares (F6 and F7  of XI Boo, F9 of IM Vir and
F14 of CC Eri) appear to be long decay flare(\td ~$\geq$ 1 hr). Infact, the flare F7 of
XI Boo was one of the longest duration flares observed with \td ~
$\sim 10$ ks.  Such long decay  flares have so far been reported
in CF Tuc (\td ~= 79.2 ks; Kr\"{u}ster \& Schmitt 1996), EV Lac (\td ~=  37.8 ks; Schmitt
1994), Algol (\td ~= 30.2 ks; Ottmann \& Schmitt 1996),  AD Leo
(\td ~= 7.9 ks; {\rm Favata, Micela \& Reale 2000a}) and some pre-main sequence stars ({\rm Stelzer, Neuh{\"a}user \& Hambaryan 2000, Favata et al. 2005}).
This flare classification is purely based on the decay time.
It has been shown that even much longer flares can occur
in a single loop (Favata et al. 2005), while relatively short flares with heating dominated
decay are probably arcade flares. For the flares F6, F9 and F14, the decay
path is driven by the sustained heating, therefore, these flares can not
be classified as long decay flares. However, the sustained heating during the decay of
the flare F7 is negligible. Thus, these flares can be classified as arcade.
The flare decay time for the remaining 12 flares are similar to  that of the solar compact
flares. The morphological differences in these different types of flares
indicate the different processes of energy released.  In compact flares energy
is probably released only
during an impulsive phase, whereas in the two-ribbon flares a prolonged energy release
is apparently required to explain their long decay time (Pallavicini, Serio \& Vaiana 1977; Pallavicini et al. 1988;
Priest 1981; Poletto, Pallavicini \& Kopp 1988).
The rise time of these  flares has been found to be  less than 1 ks,
which is similar to the rise time of impulsive flares observed in the
M dwarfs ({\rm Pallavicini et al. 1990}).
However, for some flares (F6, F7 and F9) the rise time was found to be more than 1 ks.
 We found that the decay times of the flares observed here are more in the soft
band than in the medium band or the  hard band.
Some flares, e.g., F2, actually showed a hard peak during the decay.
But even then the \td ~in soft band being more than in the hard band.
This could probably be due to  the softening of the spectrum during
the decay due to the plasma cooling i.e. emission gradually exists
form high energy band and enters more deeply in the soft energy band
or the higher energy electrons streaming down from
the coronal heights to the chromosphere that can heat the plasma
to higher temperature, lose their energy faster than the lower
energy electrons.
Most of the flares observed in our sample were found to have more
peak flux in the soft band than in the medium and hard band.

The total energy released during the X-ray flares observed in the sample of
G-K dwarfs is observed to be in the range of $2.3\times10^{32}$ to $6.1\times10^{34}$ ergs.
This shows that these flares are ten times more
energetic than the flares in the M-type dwarfs ({\rm Pallavicini
et al. 1990}), and $10^2$ to $10^4$ times more energetic than the solar
flares (Moore et al. 1980, Wu et al. 1986). However, they are not as energetic
as a large flare observed in Algol (E$_{tot} = 10^{35}$ ergs;
White et al. 1986), but are as energetic as the flares observed in
G9 dwarf ZS 76 (E$_{tot}$ = $1.4 \times 10^{34}$ ergs; Pillitteri et al. 2005).
The large flare, F7, of XI Boo was the largest in terms of both the
decay time and the ratio of the peak luminosity to the quiescent luminosity (a
factor of 2.2). However, given the low quiescent  emission (\lx ~=
$5.1\times10^{28}$ \lum) from the XI Boo, this flare was not particularly
prominent in terms of either peak X-ray luminosity (\lx ~= $1.1\times10^{29}$
\lum) or the total
X-ray energy (E$_{tot}$ = $6.4\times10^{33}$ ergs). Other large flare F9 of IM Vir
shows a large peak X-ray luminosity ($1.6\times10^{30}$ \lum)
as well as the total energy released (E$_{tot}$ = $6.12\times10^{34}$ ergs).

\begin{figure}
\includegraphics[height=10.0cm, width=10cm,angle=-90]{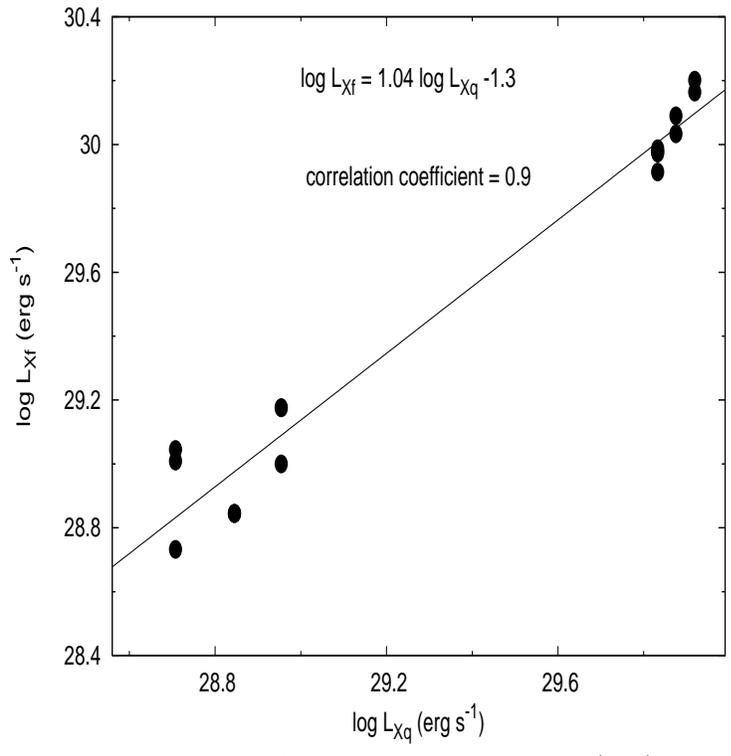}
\caption{Plot of peak flare luminosity (L$_{Xf}$) versus stellar quiescent luminosity (L$_{Xq}$)}
\label{fig:lxlq}
\end{figure}

Inspite of the fact that the
peak flare luminosity may vary by a factor of 2 (see, e.g.,  IM Vir in Table 7)
, there is a good correlation between the peak flare luminosities (L$_{Xf}$) and the quiescent state
stellar X-ray luminosities (L$_{Xq}$). The form of the correlation is
log L$_{Xf}$ =  1.04 log L$_{Xq}$ + 1.3, with the correlation coefficient, r, of
0.9 (see Figure \ref{fig:lxlq}).  A similar correlation was found by  {\rm Pallavicini et al. (1990)}
for the M dwarfs. This indicates that there is a  direct connection between
the flaring and the quiescent X-ray emission, which is probably due to their
common emission mechanism for X-rays.

Flares F2-F3-F4, F6-F7 and F10-F11 show the similar structure i.e. before
ending the first flare the next flare starts.
Similar loop systems have been
observed to the flare on the Sun (e.g. so-called Bastille-day
flare; Aschwanden \& Alexander 2001), and in a stellar analogue
 dMe star Proxima Centauri (Reale et al. 2004).
In these two events, a double ignition in nearby loops was observed or
suggested, and the delay between the ignitions appears to scale with the
loop sizes.  Similar type of double exponential
decay was also found in the young stars
ZS 76 (Pillitteri et al. 2005) and Cygnus OB2 (Albacete Colombo et al. 2007).
This implies that
in these dwarfs a steady active region undergoes a strong magnetic
reconnection event resulting in an
intense flare, and is followed by an arcade of reconnected
loops that slowly decay.

We have performed time resolved spectroscopy of these flares.
The coronal spectrum during
the flare can be represented with a 1T model, with quiescent state
taken into account as a frozen  background contribution.
During the flares F1, F9 and F11 the emission measure
was increased by a factor of five or more. Such large
variation in emission measures were also seen during the
flares detected in AB Dor (Maggio et al. 2000).  Both
the temperature and emission measure show the well-defined trends
i.e. the changes in the temperature and the emission measure  are correlated
with the variations observed in the light curves during the flares.
Similar trends were also seen during the flares detected in the pre-main sequence stars
of the Orion nebula cluster (Favata et al. 2005) and star-forming complex
L1551 in Taurus (Giardino et al. 2006).
Reale, Peres \& Orlando (2001)
found  that the height of emission measure distributions are variable during the different
phases of the solar flares, while its width and the peak temperature of the distribution undergo much smaller changes.  The peak flare temperature
is found in the range of 20 - 60 MK for all the flares.
These values are intermediate between those of flares observed on active
stars like Algol ($T_{max} \sim $ 100-150 MK, Favata et al. 2000b) and
AB Dor ($T_{max} \sim$ 140-170 MK, Maggio et al. 2000) and those
found in flares on the dMe star AD Leo (range 20-50 MK,
Reale \& Micela 1998; Favata et al. 2000a), and  on solar-type
Pleiades (20-40 MK, Briggs \& Pye 2003), and young stellar
objects ($T_{max} \sim$ 80-270 MK; Favata et al. 2001,2005).
For most of flares both emission measure and temperature
peaked simultaneously. However,
for few flares (F9, F11, F13), it appears  that the temperature
was evolved before the emission measure.
Similar delay is often observed both in solar flare (Sylwester et al. 1993)
and in  flares from the stars Algol(van Den Oord \& Mewe 1989),
EV Lac (Favata et al. 2000a),
AB Dor (Maggio et al. 2000), and YY Gem (Stelzer et al. 2002).
This is
probably due to an impulsive flare event , in which loop does not
reach equilibrium conditions, the density begins to decay later than
the temperature.

We have modeled the flares using the
hydrodynamic  model (see \S \ref{sec:model};
Reale et al. 1997) based on the decay phase of the flare.
The derived semiloop lengths for all flares are found to be in the range
of  $0.6 - 8.0 \times 10^{10}$ cm.
Alternatively, Reale (2007)
derive the semiloop length from the rise phase and peak phase of the flare
 as:
\begin{equation}
L_9^\prime \approx  3 \psi^2 T_{max}^{1/2} t_M
\label{eq:loop_rise}
\end{equation}

\noindent
where $L_9^\prime$ is semiloop length in the unit of $10^9$ cm,
$\psi = T_{max}/T_M$, $T_{max}$ is maximum temperature
in the unit of $10^7$ K, $T_M$ is temperature at density maximum and
$t_M$ is time in the unit of $10^3$ s at which density maximum occurs.
The time of the maximum emission measure is a good proxy
for $T_M$.  In case where both temperature and emission measure were
simultaneously peaked during the rise phase, only rise time was used
in the estimation of the semiloop length.
The semiloop lengths estimated
from this approach for the flares F4 ($L_9^\prime = 13.6\pm3.2$ cm), F6($L_9^\prime = 16.8\pm1.8$ cm),
F13($L_9^\prime = 6.6\pm1.4$ cm), F14($L_9^\prime = 11.4\pm3.1$ cm) and
F16($L_9^\prime = 20.8\pm5.8$ cm) are found to be
consistent with that of estimated from the decay phase analysis (see also Table 8).
The estimated loop length, $L_9^\prime$, for the flares
F2 ($9\pm2$ cm), F7($36\pm8$ cm) and F10 ($7\pm2$) are found to be less than that of determined from decay phase.
However, for the flare F9 of IM Vir the estimated loop length ($\L_9^\prime = 41\pm13$) is two times
more than that of estimated from decay phase. The inconsistency in the estimation of
the loop length from the two approaches is probably due to the involvement of
other different coronal loops during the decay of the flare or the heat pulse
triggering the flare is not a top-hat function.
This new analysis allows us
to derive the loop lengths of the flares for which either time resolved spectral
information is not available or decay phase is not observed.
These flares are F11 of V471 Tau, and F15 and F17 of EP Eri.
The estimated semiloop lengths for these flares F11, F15 and F17 are $6.2 \times 10^{10}$,
$2.1\times10^{10}$ cm and $3.8\times 10^9$ cm, respectively.
Therefore, the magnetic structures confining  the plasma for all the observed flares
in G-K dwarfs are smaller than the  star themselves, and
are not as large as observed  in a single giant HR 9024 ($L=R_\star/2$; Testa et al. 2007)
and some pre-main-sequence analogues ($L >> R_\star$; Favata et al. 2005, Giardino et al.
2006).

For all the flares, the estimated maximum electron density under assumption of a totally ionised
hydrogen plasma is found in the order of $10^{10-11}$ cm$^{-3}$. This is compatible with
the values expected for the plasma in coronal condition (Landini et al. 1986).
To satisfy the energy balance relation for the flaring as a whole, the maximum
X-ray luminosity must be lower than the total energy rate ($H = E_H.V$; see \S
\ref{sec:model}) at the flare peak.
The rest of the input energy is used for thermal conduction, kinetic energy and
radiation at lower frequencies.  For the flares F2 and F4 of V368 Cep
the maximum X-ray luminosity observed is about 34\%  of $H$.  Similarly, for the flares F7 of
XI Boo and F10 of V471 Tau the maximum X-ray luminosity is about 30\% of $H$.
However, for other flares (see Table 7) only 16 - 22 \% of $H$ is observed as a peak
luminosity.
These values are in agreement with those reported
for the solar flares, where the soft X-ray radiation only accounts for upto 20\% of total energy
(Wu et al. 1986). In comparison, the fraction of X-ray
radiation to the total energy has been found to be 15\% and 35\% for the M dwarfs Proxima Centauri and EV Lac, respectively (Reale et al.
2004; Favata et al.  2000c).
Applying loop scaling law $V = 2\pi \beta^2 L^3$, and if the detected flares are produced
by a single loop, their aspect ratio ($\beta$) were estimated in the range of 0.1 to 0.3 for the
flares F2, F4, F7, F9, F13, F14 and F16 (see Table 8). Similar cross section was also observed for the
solar coronal loops, for which typical values of $\beta$ are in the range of 0.1 -0.3. If we assume
$\beta =0.1$, these flares occurred in the arcades, which are composed of 2 to 11 loops (see also Table 8).
However, for the flares F6 and F10 the value of $\beta$ (assuming single loop) is
estimated to be 0.51 and 0.37, respectively. In comparison to the
solar coronal loops, the cross section for these flare loops are very large.
It appears that the flares F6 and F10 occurred in the arcades that contain 26 and 14 loops, respectively.

Therefore, we conclude that the observed flares in G-K
dwarfs are similar to the solar arcade flares, which are as strong as M dwarfs and are much smaller than the
flare observed in dMe star, giants and pre-main-sequence analogous.

\section*{Acknowledgments}
We thank the reviewer of the paper for very useful comments and suggestions.
This research has made use
of data obtained from HEASARC, provided by the NASA
Goddard Space Flight Center.

\newpage
\begin{deluxetable}{llccl}
\tablecolumns{5}
\tablewidth{-0pt}
\tablecaption{\leftskip 0mm{ General properties of stars in the sample}}
\label{tab:parameter}
\tablehead{
\colhead{Name}   & \colhead{Spectral Type }  & \colhead{V}    &\colhead{Period}&\colhead{Distance} \\
                 &                           & \colhead{(mag)}&\colhead{(days)}&\colhead{(pc)} }
\startdata
V368 CEP   &    G9V         &  7.54 & 2.74      & 19.7 \\
XI Boo     &    G8V/K4V     &  4.55 & 6.2       &  6.6 \\
IM VIR     &    G5V/KV      &  9.00 & 1.3086    & 60.0 \\
V471 Tau   &    K2V/dW      &  9.71 & 0.52      & 46.7 \\
CC Eri     &    K7Ve/M4V    &  8.76 & 1.56      & 11.5 \\
EP ERI     &    K2V         &  6.00 & 6.85      & 10.3 \\
\enddata
\end{deluxetable}
\clearpage

\begin{deluxetable}{llccc}
\tablecolumns{5}
\tablewidth{-0pt}
\tablecaption{\leftskip0mm{Log of observations with the XMM-Newton}}
\label{tab:log}
\tablehead{
\colhead{Star} & \colhead{Instrument} &\colhead{Start time} & \colhead{Exposure time}& \colhead{offset(')} \\
\colhead{Name} & \colhead{(mode,filter)}&\colhead{date (UT)} & \colhead{(s)} &}
\startdata
V368 CEP  &MOS1(SW,thick)   &2003-12-27(19:23:43)& 29762& 0.045 \\
          &MOS2(SW,thick)   &2003-12-27(19:23:43)& 29768&       \\
          &PN(SW, thick)    &2003-12-27(19:28:55)& 29570&       \\
XI Boo*   &MOS1(T, medium)  &2001-01-19(11:25:06)& 58397& 0.150\\
          &MOS2(SW, medium) &2001-01-19(11:25:06)& 58997&      \\
IM VIR    &MOS1(FF, medium) &2004-07-15(04:49:45)& 53488& 9.093\\
          &MOS2(FF, medium) &2004-07-15(04:49:45)& 53487&      \\
          &PN(FF, medium)   &2004-07-15(05:32:47)& 51077&      \\
V471 Tau  &MOS1(LW, medium) &2004-08-01(06:52:10)& 60678& 1.052\\
          &MOS2(LW, medium) &2004-08-01(06:52:10)& 60682&      \\
          &PN(SW, medium)   &2004-08-01(06:57:35)& 60471&      \\
CC Eri    &MOS1(PFW, thick) &2003-08-08(08:40:36)& 39463& 0.065\\
          &MOS2(T, thick)   &2003-08-08(08:40:27)& 39212&      \\
          &PN(PFW, thick)   &2003-08-08(09:21:36)& 36703&      \\
EP Eri    &MOS1(SW, thick)  &2004-01-20(17:53:18)& 50013& 0.015 \\
          &MOS2(SW, thick)  &2004-01-20(17:53:19)& 50013&       \\
          &PN(SW, thick)    &2004-01-20(17:58:29)& 50013&       \\
\enddata
\tablenotetext{*}{XI Boo was observed for 895 s by EPIC PN detector}
\tablenotetext{~}{{\it Note:} FF: Full frame, FPW: FullPrime window
PPW: PrimePartial Window, SW: Small window, T: timing, LW: Large window}
\vspace{-0.5cm}
\end{deluxetable}
\clearpage

\begin{table}
\caption{ Time interval of 17 flares observed in six G-K dwarfs.}
\begin{tabular}{lcccccc}
\hline \hline
 Object    & Flare &  Time-interval& Duration&Peak flare flux$^{1}$ & Quiescent state flux$^1$\\
           &       &  (ks after the observation start)& (ks)   & $10^{-11}$ \lum ~cm$^{-2}$&$10^{-11}$ \lum ~cm$^{-2}$ \\
\hline
 V368 Cep  &  F1   &    ..      -    4.0  &  4.0 & 2.28 &1.30\\
           &  F2   &   19.2     -   22.5  &  3.3 & 2.39 &..\\
           &  F3   &   22.5     -   23.9  &  1.4 & 2.36 &..\\
           &  F4   &   23.9     -   27.8  &  3.9 & 1.90 &..\\
   XI Boo  &  F5   &    8.1     -   12.7  &  4.6 & 2.87 &2.14\\
           &  F6   &   12.7     -   21.1  &  8.4 & 5.66 &..\\
           &  F7   &   21.1     -   32.4  & 11.3 & 5.57 &..\\
   IM Vir  &  F8   &    4.5     -    9.3  &  4.8 & 0.20 &0.13\\
           &  F9   &    9.3     -   18.9  &  9.6 & 0.25 &..\\
   V471 Tau&  F10  &   14.4     -   18.5  &  4.1 & 0.39 &0.24\\
           &  F11  &   18.5     -   23.0  &  4.5 & 0.31 &..\\
   CC Eri  &  F12  &    5.8     -    7.5  &  1.7 & 2.38 &1.83\\
           &  F13  &    8.3     -   12.5  &  4.2 & 3.45 &..\\
           &  F14  &   27.1     -   36.5  &  9.4 & 3.43 &..\\
   EP Eri  &  F15  &    4.5     -    8.5  &  4.0 & 0.62 &0.48\\
           &  F16  &   35.1     -   44.2  &  8.9 & 0.65 &..\\
           &  F17  &   47.4     -    ..   &  ..  & 0.66 &..\\
\hline
\end{tabular}
\vbox{
$^1$ count rates are converted into the flux using WebPIMMS (http://heasarc.gsfc.nasa.gov/Tools/w3pimms.html)
}
\end{table}

\begin{deluxetable}{lcccccccccccc}
\tablecolumns{13}
\tablewidth{-0pt}
\tabletypesize{\scriptsize}
\setlength{\tabcolsep}{0.03in}
\tablecaption{\leftskip0mm{Parameters obtained from fitting of equation 1 to the flare light
curves obtained with MOS and PN detectors  in the energy band of 0.3-10.0 keV}}
\label{tab:tab3}
\tablehead{
Object   & FN&\multicolumn{4}{c}{MOS} && \multicolumn{4}{c}{PN}  & \lx$^a$ & E$_{tot}$\\
\cline{3-6} \cline{8-11}
  Name   & &A$_0$ &\td &\tr & q          && A$_0$ &\td &\tr& q & ($10^{29}$& $(10^{33}$ ergs)\\
         & &(\cts)& (s)& (s)&(\cts)&&(\cts) &(s)&(s)&(\cts) &\lum)&}
\startdata
V368 Cep &F1& $ 7.0\pm0.2$  &$1213\pm74 $  & \nodata      & 3.6 && $ 9.0 \pm0.2 $ & $1226\pm86$   &\nodata      & 5.1 &31.3& 12.5  \\
         &F2& $ 6.8\pm0.1$  &$1658\pm141$  & $315\pm43$   &     && $ 9.1 \pm0.3 $ & $2387\pm341$  & $347\pm53 $ &     &39.6& 13.1  \\
         &F3& $ 6.5\pm0.2$  &$ 451\pm49 $  & $479\pm44$   &     && $ 8.7 \pm0.2 $ & $496 \pm36 $  & $515\pm79 $ &     &~9.5& 9.5   \\
         &F4& $ 4.5\pm0.3$  &$1117\pm110$  & $649\pm181$  &     && $ 7.0 \pm0.2 $ & $1232\pm133$  & $681\pm91 $ &     &29.9& 11.7  \\
XI Boo   &F5& $3.42\pm0.12$ &$2829\pm950 $ & $879\pm213$  & 2.66&&    \nodata     &    \nodata    &   \nodata   &     &~0.5& 0.23  \\
         &F6& $ 6.9\pm0.1$  &$4393\pm333 $ & $3169\pm259$ &     &&    \nodata     &    \nodata    &   \nodata   &     &~4.3& 3.61  \\
         &F7& $6.78\pm0.06$ &$9885\pm362 $ & $3351\pm244$ &     &&    \nodata     &    \nodata    &   \nodata   &     &~5.7& 6.44  \\
IM Vir   &F8& $0.67\pm0.04$ &$1949\pm578 $ & $205\pm130$  &0.357 && $ 1.02\pm0.05$ & $2025\pm603 $ & $311\pm121$ &0.64&27.9& 13.4  \\
         &F9& $0.77\pm0.02$ &$3474\pm680 $ & $1323\pm277$ &     && $ 1.27\pm0.04$ & $3450\pm289 $ & $2041\pm401$&     &63.7&61.2   \\
V471 Tau &F10& $1.44\pm0.04$ &$2874\pm605 $ & $595\pm204$  &0.96 && $ 2.12\pm0.05$ & $2135\pm286 $ & $580\pm109$&1.28 &42.4&17.4   \\
         &F11& $1.31\pm0.03$ &$1272\pm301 $ & $ 864\pm189$ &     && $ 1.70\pm0.02$ & $1847\pm519 $ & $827\pm307$&     &38.3&17.2   \\
CC Eri   &F12& $ 4.9\pm0.3$  &$656\pm152 $  & $256\pm100$  &3.86 && $ 4.6 \pm0.2 $ & $ 781\pm98  $ & $335\pm52$ &2.35 &~2.0& 0.34  \\
         &F13& $ 6.9\pm0.2$  &$3027\pm441$  & $493\pm50$   &     && $ 6.4 \pm0.3 $ & $2844\pm312 $ & $722\pm61$ &     & 6.7& 2.82  \\
         &F14& $ 7.9\pm0.1$  &$4800\pm188$  & $577\pm91$   &     && $ 7.4 \pm0.2 $ & $5992\pm231$  & $915\pm51$ &     & 6.7& 6.26  \\
EP Eri   &F15& $1.59\pm0.04$ &$ 939\pm340$  & $897\pm439$  &1.20 && $ 2.48\pm0.10$ & $1094\pm293$  & $967\pm201$& 1.90&~1.4& 0.56  \\
         &F16& $1.72\pm0.04$ &$4476\pm 867$ & $2800\pm1103$&     && $ 2.54\pm0.04$ & $4052\pm 514$ & $3200\pm 780$&   &~3.1& 2.76  \\
         &F17& $1.68\pm0.03$ &    \nodata   & $820\pm136$  &     && $ 2.51\pm0.03$ &               &$1014\pm597$&     &~0.8& \\
\enddata
\tablenotetext{~}{{\it Note:} FN is flare name, A$_0$ is count rate at flare peak,
\td ~is flare decay time, \tr ~is flare rise time, q is quiescent state count rate, \lx ~is
total X-ray luminosity during the flare and E$_{tot}$ is total X-ray energy emitted during the flare}
\tablenotetext{a}{\lx ~for each flare is integrated using the values given in Table 6 based on spectral fits}
\end{deluxetable}

\begin{deluxetable}{lccccccccccccccc}
\tablecolumns{16}
\tablewidth{-0pt}
\rotate
\tablecaption{\leftskip-4cm{Parameters obtained from the fitting of flare
light curves observed with PN detector in the soft, medium and the hard bands.}}
\tabletypesize{\scriptsize}
\label{tab:3banddecay}
\tablehead{
Object    & FN    & \multicolumn{4}{c}{ Soft (0.3-0.8 keV)}        && \multicolumn{4}{c}{Medium (0.8-1.6 keV)}       && \multicolumn{4}{c}{Hard (1.6-10.0 keV)}   \\
\cline{3-6} \cline{8-11} \cline{13-16}
Name      &    &   \td  &  \tr  &  $A_0$   &q     &&  \td    & \tr   &  $A_0$   &q    && \td  & \tr &$A_0$ &q     \\
          &    & (s)    &  (s)  & (\cts)   &(\cts)&& (s)    &  (s)  & (\cts)   &(\cts)&&  (s)    &  (s)  & (\cts)   &(\cts)}
\startdata
V368 Cep  & F1 &$1485\pm171$ &\nodata     &$3.96\pm0.09$& 2.69&&$1105\pm110$ &\nodata     &$3.54\pm0.09$&2.15&&$840\pm97  $& \nodata   &$0.74\pm0.03$& 0.31\\
          & F2 &$2749\pm747$ &$374\pm79  $&$3.90\pm0.07$& ... &&$1949\pm318$ &$249\pm43  $&$3.86\pm0.07$&... &&$1546\pm212$&$186\pm37  $&$1.35\pm0.05$& ...\\
	  & F3 &$318\pm47  $ &$545\pm90  $&$4.21\pm0.10$& ... &&$ 610\pm113$ &$707\pm184 $&$3.45\pm0.10$&... &&$324\pm45  $&$283\pm56  $&$0.81\pm0.07$&...  \\
	  & F4 &$1258\pm264$ &$545\pm90  $&$3.46\pm0.10$& ... &&$1184\pm156$ &$792\pm201 $&$3.02\pm0.07$&... &&$1130\pm330$&$709\pm224 $&$0.52\pm0.04$&...  \\
XI Boo    & F6 &$5552\pm748$ &$2821\pm235$&$2.62\pm0.04$&1.40 &&$4665\pm340$ &$3429\pm291$&$3.74\pm0.07$&1.22&&$2120\pm191$&$          $&$0.47\pm0.02$&0.05 \\
	  & F7 &$10155\pm786$&$4120\pm643$&$2.66\pm0.03$&...  &&$10599\pm543$&$3320\pm296$&$3.86\pm0.04$&... &&$6772\pm492$&$2683\pm556$&$0.46\pm0.01$&...  \\
IM Vir    & F9 &$3810\pm455$ &$1194\pm296$&$0.56\pm0.02$&0.32 &&$3226\pm427$ &$848\pm112 $&$0.56\pm0.02$&0.27&&$ 790\pm130$&$350\pm171 $&$0.21\pm0.01$&0.06 \\
V471 Tau  & F10&$2945\pm530$ &$851\pm146 $&$0.98\pm0.03$&0.65 &&$2465\pm454$ &$676\pm179 $&$0.98\pm0.03$&0.53&&$1409\pm549$&$806\pm251 $&$0.18\pm0.02$&0.07 \\
          & F11&$1962\pm267$ &$1203\pm272$&$0.87\pm0.03$&...  &&$1254\pm258$ &$1149\pm209$&$0.77\pm0.03$&... &&$278\pm121 $&$247\pm106 $&$0.16\pm0.02$&...  \\
CC Eri    & F12&$591\pm149 $ &$383\pm99  $&$1.62\pm0.08$&1.27 &&$641\pm90  $ &$503\pm151 $&$2.22\pm0.08$&1.27&&$452\pm138 $&$460\pm123 $&$0.33\pm0.04$&0.07 \\
          & F13&$3182\pm501$ &$800\pm115 $&$2.99\pm0.09$&...  &&$2429\pm232$ &$712\pm92  $&$4.23\pm0.05$&... &&$2245\pm551$&$692\pm153 $&$1.03\pm0.10$&...  \\
	  & F14&$6445\pm334$ &$1552\pm138$&$3.40\pm0.06$&...  &&$5765\pm360$ &$1262\pm194$&$3.87\pm0.10$&... &&$3934\pm252$&$ 513\pm93 $&$1.14\pm0.05$&...  \\
EP Eri    & F15&$1223\pm256$ &$609\pm194 $&$1.51\pm0.03$&1.21 &&$715\pm255 $ &$601\pm201 $&$0.91\pm0.07$&0.68&&$...       $&$ ...      $&$...        $&...  \\
\enddata
\tablenotetext{~}{{\it Note:} FN is flare Name, $A_0$ is count at flare peak,
\td is ~flare decay time, \tr is ~flare rise time, q is the quiescent state count
rates.}
\end{deluxetable}

\begin{deluxetable}{lcccrcrcccr}
\tablecolumns{11}
\tablewidth{-0pt}
\tabletypesize{\normalsize}
\setlength{\tabcolsep}{0.05in}
\tablecaption{\leftskip0mm{Spectral parameters derived for the quiescent emission of each program star from the analysis of PN spectra (MOS spectra for
XI Boo) accumulated during the time interval 'Q' as shown in Fig. \ref{fig:mospnlc}, using APEC 2T model.}}
\label{tab:tab4}
\tablehead{
Object   & N$_H$ & Z & kT$_1$& EM$_1$~& kT$_2$& EM$_2$~&\lx & $\chi_\nu^2$ (DOF) }
\startdata
V368 Cep & 1.9$_{-0.5}^{+0.4}$ & 0.14$_{-0.01}^{+0.01}$ & 0.35$_{-0.01}^{+0.01}$ &  7.4$_{-0.6}^{+0.5}$  & 0.90$_{-0.01}^{+0.01}$ &  5.9$_{-0.3}^{+0.2}$ & 6.8 & 1.71(429)\\
XI Boo   &$<0.95$              & $ < 0.23$              & 0.20$_{-0.04}^{+0.08}$ & 1.51$_{-0.51}^{+0.70}$& 0.57$_{-0.01}^{+0.01}$ & 1.2$_{-0.1 }^{+0.1 }$& 0.51 & 1.68(72)\\
IM Vir   &2.8$_{-1.0}^{+0.9}$  & 0.12$_{-0.02}^{+0.02}$ & 0.49$_{-0.04}^{+0.05}$ &  7.4$_{-1.1}^{+1.2}$  & 0.97$_{-0.04}^{+0.05}$ &  8.3$_{-2.3}^{+1.0}$ & 8.3 & 1.13(280)\\
V471 Tau &2.4$_{-0.6}^{+0.6}$  & 0.16$_{-0.01}^{+0.02}$ & 0.41$_{-0.02}^{+0.02}$ &  6.2$_{-0.6}^{+0.7}$  & 0.90$_{-0.02}^{+0.02}$ &  6.5$_{-0.4}^{+0.4}$ &  7.5 & 1.16(341)\\
CC Eri   & 1.3$_{-0.7}^{+0.8}$ & 0.18$_{-0.02}^{+0.02}$ & 0.31$_{-0.01}^{+0.01}$ &  0.6$_{-0.1}^{+0.2}$  & 0.83$_{-0.07}^{+0.07}$ &  0.5$_{-0.1}^{+0.2}$ &  0.9 & 1.05(302)\\
EP Eri   & $<0.4$              & 0.21$_{-0.01}^{+0.01}$ & 0.36$_{-0.01}^{+0.01}$ & 0.8$_{-0.2}^{+0.1}$   & 0.69$_{-0.03}^{+0.02}$ &  0.3$_{-0.1}^{+0.2}$ &  0.7 & 1.44(291)\\
\enddata

\tablenotetext{~}{{\it Note:}
$N_H$ is in $10^{20}$ cm$^{-2}$
temperatures (kT) are in keV, emission measures (EM) are in $10^{52}$
cm$^{-3}$, and X-ray luminosity (\lx) is in $10^{29}$ \lum.
$\chi_\nu^2$ is the minimum reduced $\chi^2$ and DOF stands for degrees
of freedom.
}
\end{deluxetable}

\begin{deluxetable}{lcccccc}
\tablecolumns{7}
\tablewidth{-0pt}
\tabletypesize{\normalsize}
\setlength{\tabcolsep}{0.05in}
\tablecaption{\leftskip0mm{Spectral parameters for each temporal segment of the
flare in the targeted stars. The temporal segments of the
flare are marked by
vertical lines in Figure \ref{fig:mospnlc}.}}
\label{tab:tab6}
\tablehead{
Object  & LS & FS & kT& EM~&\lx & $\chi_\nu^2$ (DOF) }
\startdata
V368 Cep & F1 &  D1 &  1.20$_{-0.10}^{+0.11}$ & 3.7$_{-0.3}^{+0.3}$    & 9.4  & 0.97(213)\\
         &    &  D2 &  1.01$_{-0.04}^{+0.10}$ & 1.9$_{-0.3}^{+0.3}$    & 8.0  & 1.11(197)\\
         &    &  D3 &  0.74$_{-0.11}^{+0.23}$ & 0.9$_{-0.3}^{+0.3}$    & 7.2  & 0.94(186)\\
         &    &  D4 &  0.58$_{-0.15}^{+0.04}$ & 0.5$_{-0.2}^{+0.4}$    & 6.7  & 0.82(170)\\
         & F2 &  R1 &  2.40$_{-0.15}^{+0.35}$ & 3.0$_{-0.4}^{+0.4}$    & 9.7  & 1.19(157)\\
         &    &  D5 &  2.61$_{-0.29}^{+0.33}$ & 4.7$_{-0.2}^{+0.1}$    & 11.8 & 1.14(220)\\
         &    &  D6 &  2.46$_{-0.24}^{+0.28}$ & 3.5$_{-0.2}^{+0.3}$    & 9.5  & 1.02(163)\\
         &    &  D7 &  1.86$_{-0.43}^{+0.26}$ & 3.4$_{-0.2}^{+0.2}$    & 8.8  & 1.12(163)\\
         & F3 &  .. &  0.98$_{-0.07}^{+0.07}$ & 4.6$_{-0.5}^{+0.6}$    & 9.5  & 1.18(234)\\
         & F4 &  R2 &  1.79$_{-0.23}^{+0.16}$ & 2.6$_{-0.3}^{+0.3}$    & 8.2  & 1.06(255)\\
         &    &  D8 &  1.22$_{-0.07}^{+0.09}$ & 2.5$_{-0.3}^{+0.3}$    & 8.2  & 0.88(183)\\
         &    &  D9 &  0.71$_{-0.07}^{+0.18}$ & 1.6$_{-0.3}^{+0.3}$    & 7.1  & 1.04(174)\\
         &    &  D10&  0.57$_{-0.12}^{+0.19}$ & 1.0$_{-0.3}^{+0.3}$    & 6.4  & 0.96(220)\\
XI Boo   & F5 &  .. &  0.81$_{-0.08}^{+0.21}$ & 0.10$_{-0.02}^{+0.02}$ & 0.54 & 1.37(89)\\
         & F6 &  R1 &  0.94$_{-0.04}^{+0.04}$ & 0.34$_{-0.02}^{+0.02}$ & 0.74 & 1.17(103)\\
         &    &  D1 &  0.93$_{-0.04}^{+0.04}$ & 0.68$_{-0.04}^{+0.04}$ & 1.02 & 0.97(72)\\
         &    &  D2 &  0.80$_{-0.04}^{+0.04}$ & 0.57$_{-0.04}^{+0.04}$ & 0.93 & 0.90(65)\\
         &    &  D3 &  0.71$_{-0.05}^{+0.04}$ & 0.45$_{-0.04}^{+0.04}$ & 0.83 & 0.87(61)\\
         &    &  D4 &  0.78$_{-0.05}^{+0.05}$ & 0.39$_{-0.03}^{+0.03}$ & 0.78 & 1.21(60)\\
         & F7 & *R2 &  1.08$_{-0.12}^{+0.06}$ & 0.28$_{-0.06}^{+0.12}$ & 0.93 & 1.27(93)\\
         &    & *D5 &  1.15$_{-0.15}^{+0.14}$ & 0.42$_{-0.09}^{+0.15}$ & 1.11 & 1.20(85)\\
         &    & *D6 &  1.21$_{-0.11}^{+0.15}$ & 0.43$_{-0.10}^{+0.08}$ & 1.07 & 0.93(82)\\
         &    & *D7 &  1.04$_{-0.08}^{+0.18}$ & 0.33$_{-0.09}^{+0.10}$ & 0.99 & 0.95(81)\\
         &    & *D8 &  0.95$_{-0.05}^{+0.06}$ & 0.34$_{-0.06}^{+0.03}$ & 0.88 & 1.09(78)\\
         &    &  D9 &  0.77$_{-0.03}^{+0.05}$ & 0.29$_{-0.03}^{+0.07}$ & 0.73 & 1.07(73)\\
IM Vir   & F8 &  R1 &  $ <5.16$               & 7.8$_{-1.1}^{+2.4}$    & 14.6 & 1.10(60)\\
         &    &  D1 &  3.23$_{-0.91}^{+1.56}$ & 5.0$_{-0.6}^{+1.2}$    & 13.3 & 1.23(88)\\
         & F9 &  R2 &  2.04$_{-0.14}^{+0.35}$ & 8.6$_{-1.0}^{+0.9}$    & 15.9 & 1.08(103)\\
         &    &  D2 &  1.23$_{-0.14}^{+0.10}$ & 11.2$_{-1.5}^{+1.5}$   & 15.8 & 0.86(36)\\
         &    &  D3 &  0.89$_{-0.10}^{+0.15}$ & 5.8$_{-0.8}^{+0.8}$    & 11.8 & 1.22(77)\\
         &    &  D4 &  1.02$_{-0.13}^{+0.27}$ & 3.0$_{-0.7}^{+0.7}$    & 10.3 & 1.12(81)\\
         &    &  D5 &  0.72$_{-0.12}^{+0.18}$ & 2.8$_{-0.7}^{+0.7}$    & 9.90 & 0.79(31)\\
V471 Tau & F10&  R1 &  1.23$_{-0.12}^{+0.12}$ & 4.4$_{-0.8}^{+0.8}$    & 10.6 & 0.85(49)\\
         &    &  D1 &  1.29$_{-0.17}^{+0.14}$ & 5.6$_{-0.7}^{+0.7}$    & 12.3 & 1.18(74)\\
         &    &  D2 &  1.24$_{-0.14}^{+0.09}$ & 3.5$_{-0.6}^{+0.6}$    & 10.3 & 1.03(64)\\
         &    &  D3 &  1.08$_{-0.12}^{+0.07}$ & 2.3$_{-0.5}^{+0.5}$    & 9.2  & 0.80(80)\\
         & F11&  R2 &  1.86$_{-0.90}^{+2.13}$ & 2.6$_{-0.9}^{+0.9}$    & 10.8 & 1.14(43)\\
         &    &  D4 &  0.91$_{-0.15}^{+0.21}$ & 3.2$_{-0.7}^{+0.7}$    & 9.7  & 0.89(50)\\
         &    &  D5 &  $ < 2.79$              & 1.3$_{-0.6}^{+0.6}$    & 9.5  & 1.35(56)\\
         &    &  D6 &  $ <1.48$               & 0.7$_{-0.3}^{+0.5}$    & 8.3  & 0.79(72)\\
CC Eri   & F12&  R1 &      $>1.03           $ & 0.12$_{-0.05}^{+0.06}$ & 1.0  & 0.85(74)\\
         &    &  D1 &  2.37$_{-0.77}^{+1.45}$ & 0.14$_{-0.04}^{+0.05}$ & 1.0  & 1.24(135)\\
         & F13&  R2 &  1.23$_{-0.12}^{+0.12}$ & 0.52$_{-0.05}^{+0.05}$ & 1.3  & 1.14(129)\\
         &    &  D2 &  1.17$_{-0.18}^{+0.11}$ & 0.77$_{-0.07}^{+0.07}$ & 1.5  & 1.24(82)\\
         &    &  D3 &  1.23$_{-0.13}^{+0.25}$ & 0.52$_{-0.06}^{+0.06}$ & 1.3  & 1.26(86)\\
         &    &  D4 &  1.00$_{-0.10}^{+0.19}$ & 0.45$_{-0.05}^{+0.05}$ & 1.3  & 0.87(101)\\
         &    &  D5 &  0.97$_{-0.07}^{+0.15}$ & 0.43$_{-0.05}^{+0.05}$ & 1.2  & 0.91(126)\\
         & F14&  R3 &  1.23$_{-0.12}^{+0.12}$ & 0.52$_{-0.05}^{+0.05}$ & 1.3  & 1.14(129)\\
         &    &  D6 &  1.28$_{-0.10}^{+0.35}$ & 0.80$_{-0.06}^{+0.06}$ & 1.5  & 1.32(120)\\
         &    &  D7 &  1.21$_{-0.07}^{+0.07}$ & 0.70$_{-0.04}^{+0.04}$ & 1.4  & 1.20(218)\\
         &    &  D8 &  1.02$_{-0.06}^{+0.19}$ & 0.59$_{-0.04}^{+0.04}$ & 1.3  & 1.27(209)\\
         &    &  D9 &  0.98$_{-0.09}^{+0.09}$ & 0.31$_{-0.04}^{+0.04}$ & 1.2  & 1.17(217)\\
EP Eri   & F15&  R1 &  0.68$_{-0.20}^{+0.21}$ & 0.13$_{-0.05}^{+0.06}$ & 0.7  & 1.24(74 )\\
         &    &  D1 &  0.57$_{-0.20}^{+0.16}$ & 0.17$_{-0.05}^{+0.09}$ & 0.7  & 1.31(126)\\
         & F16&  R2 &  0.95$_{-0.14}^{+0.10}$ & 0.28$_{-0.04}^{+0.04}$ & 0.8  & 1.14(185)\\
         &    &  D2 &  0.99$_{-0.10}^{+0.11}$ & 0.27$_{-0.02}^{+0.02}$ & 0.8  & 1.10(187)\\
         &    &  D3 &  0.78$_{-0.12}^{+0.07}$ & 0.21$_{-0.02}^{+0.01}$ & 0.8  & 1.16(195)\\
         &    &  D4 &  0.72$_{-0.08}^{+0.13}$ & 0.15$_{-0.01}^{+0.01}$ & 0.7  & 1.01(140)\\
         & F17&  .. &  0.81$_{-0.08}^{+0.08}$ & 0.27$_{-0.04}^{+0.04}$ & 0.8  & 1.05(190)\\
\enddata
\tablenotetext{~}{{\it Note:} LS stands for light curve segments, FS stands for
``Flare Segments'', $N_H$ is in units of $10^{20}$ cm$^{-2}$,
temperatures (kT) are in keV, emission measures (EM) are in $10^{52}$
cm$^{-3}$, and X-ray luminosity (\lx) is in $10^{29}$ \lum.
$\chi_\nu^2$ is the minimum reduced $\chi^2$ and DOF stands for degrees
of freedom.\\
*Spectral parameters of hot component in APEC 2T fit.
}
\end{deluxetable}

\begin{deluxetable}{rcccccccccc}
\tablecolumns{11}
\tablewidth{-0pt}
\tabletypesize{\scriptsize}
\tablecaption{\leftskip0mm{ Parameters derived for flares.
}}
\label{tab:tab4}
\tablehead{
Object(FN)  &$\zeta  $     &  $  T_{max}$  & $  L       $   &   $p$          & $ n_e^*$        & $V^*$      & $ E_h$ &$B$ &$\beta$& $NL$ \\
(1)         & (2)          &     (3)       &  (4)           &    (5)         &  (6)          & (7)      & (8)    &(9) &  (10)  & (11) }
\startdata
 V368Cep(F1)&$0.72\pm0.03$ & $2.55\pm0.25$ & $ 6.0 \pm0.6 $ & $1.02\pm 0.32$ & $1.44\pm0.20$ & $  1.78$ & $ 2.36$&160 & ..   &  ..  \\
 V368Cep(F2)&$0.80\pm0.40$ & $6.20\pm0.80$ & $19.8 \pm8.6 $ & $4.37\pm 2.55$ & $2.56\pm1.16$ & $  0.72$ & $ 4.75$&331 & 0.12 &$\sim 2$ \\
 V368Cep(F4)&$1.60\pm0.20$ & $4.03\pm0.18$ & $11.7 \pm1.3 $ & $2.05\pm 0.36$ & $1.84\pm0.32$ & $  0.74$ & $ 3.06$&227 & 0.27 &$\sim 8$\\
   XIBoo(F6)&$0.73\pm0.30$ & $1.89\pm0.08$ & $18.6 \pm7.5 $ & $0.13\pm 0.06$ & $0.25\pm0.10$ & $ 10.54$ & $ 0.08$& 58 & 0.51 &$\sim 26$\\
   XIBoo(F7)&$2.1 \pm0.4 $ & $2.53\pm0.27$ & $79.4 \pm6.1 $ & $0.07\pm 0.02$ & $0.11\pm0.02$ & $ 37.81$ & $ 0.01$& 43 & 0.11 &$\sim 1$\\
   IMVir(F9)&$0.60\pm0.20$ & $4.66\pm0.37$ & $18.3 \pm8.9 $ & $2.02\pm 1.10$ & $1.57\pm0.78$ & $  4.55$ & $ 2.07$&226 & 0.34 &$\sim 12$\\
V471TauF(10)&$1.16\pm0.07$ & $3.35\pm0.33$ & $21.7 \pm2.5 $ & $0.63\pm 0.20$ & $0.68\pm0.10$ & $  8.88$ & $ 0.46$&126 & 0.37 &$\sim 14$\\
  CCEriF(13)&$0.51\pm0.25$ & $2.57\pm0.29$ & $ < 14       $ & $    >0.43   $ & $    >0.60  $ & $ <2.15$ & $>0.41$&103 & 0.33 &$\sim 11$\\
  CCEriF(14)&$0.54\pm0.15$ & $2.70\pm0.21$ & $20.0 \pm10.0$ & $0.35\pm 0.20$ & $0.47\pm0.25$ & $  3.60$ & $ 0.26$& 95 & 0.26 &$\sim 7$\\
  EPEriF(16)&$1.10\pm0.30$ & $2.02\pm0.22$ & $23.5 \pm4.5 $ & $0.13\pm 0.05$ & $0.23\pm0.05$ & $  5.38$ & $ 0.07$& 57 & 0.26 &$\sim 7$\\
\enddata
\tablenotetext{~}{(1) FN is flare name}
\tablenotetext{~}{(2) $\zeta$ is slope of decay path in density-temperature diagram,}
\tablenotetext{~}{(3) $T_{max}$  is the maximum temperature in the loop at the flare peak in  unit of $10^7$ K based on spectral fit and equation (5),}
\tablenotetext{~}{(4) $L$  is half-length of the flaring loop in  unit of $10^9$ cm and is determined using equation (3),}
\tablenotetext{~}{(5) $p$ is the maximum pressure in the loop at the flare peak in units of $10^3$ dyne ~cm$^{-2}$ (see equation (7)),}
\tablenotetext{~}{(6) $n_e$ is the maximum electron density in the loop at the flare peak in unit of $10^{11}$ cm$^{-3}$, assuming that
hydrogen plasma is totally ionised ($p = 2n_ekT_{max}$),}
\tablenotetext{~}{(7) $V$  is volume of flaring plasma in unit of $10^{30}$ cm$^{3}$ and estimated using the equation (6),}
\tablenotetext{~}{(8) $E_h$  is heating rate per unit volume at the flare peak in the units ergs s$^{-1}$ cm$^{-3}$ estimate from the loop
scaling laws (see equation (8)).}
\tablenotetext{~}{(9) $B$ is minimum magnetic field necessary for confinement in Gauss.}
\tablenotetext{~}{(10) $\beta$ is loop aspect ratio ($r/L$).}
\tablenotetext{~}{(11) $NL$ is number of loops involve in the flaring plasma.}
\tablenotetext{~}{(*) The  pressure as derived from
equation (7) is a maximum value, appropriate for the equilibrium condition,
which can be reached only for very long-lasting heating. The density and
volume values derived thereby are therefore be treated as upper and
lower limits, respectively.}
\end{deluxetable}

\begin{figure}
\includegraphics[height=13.5cm, width=7.5cm,angle=-90]{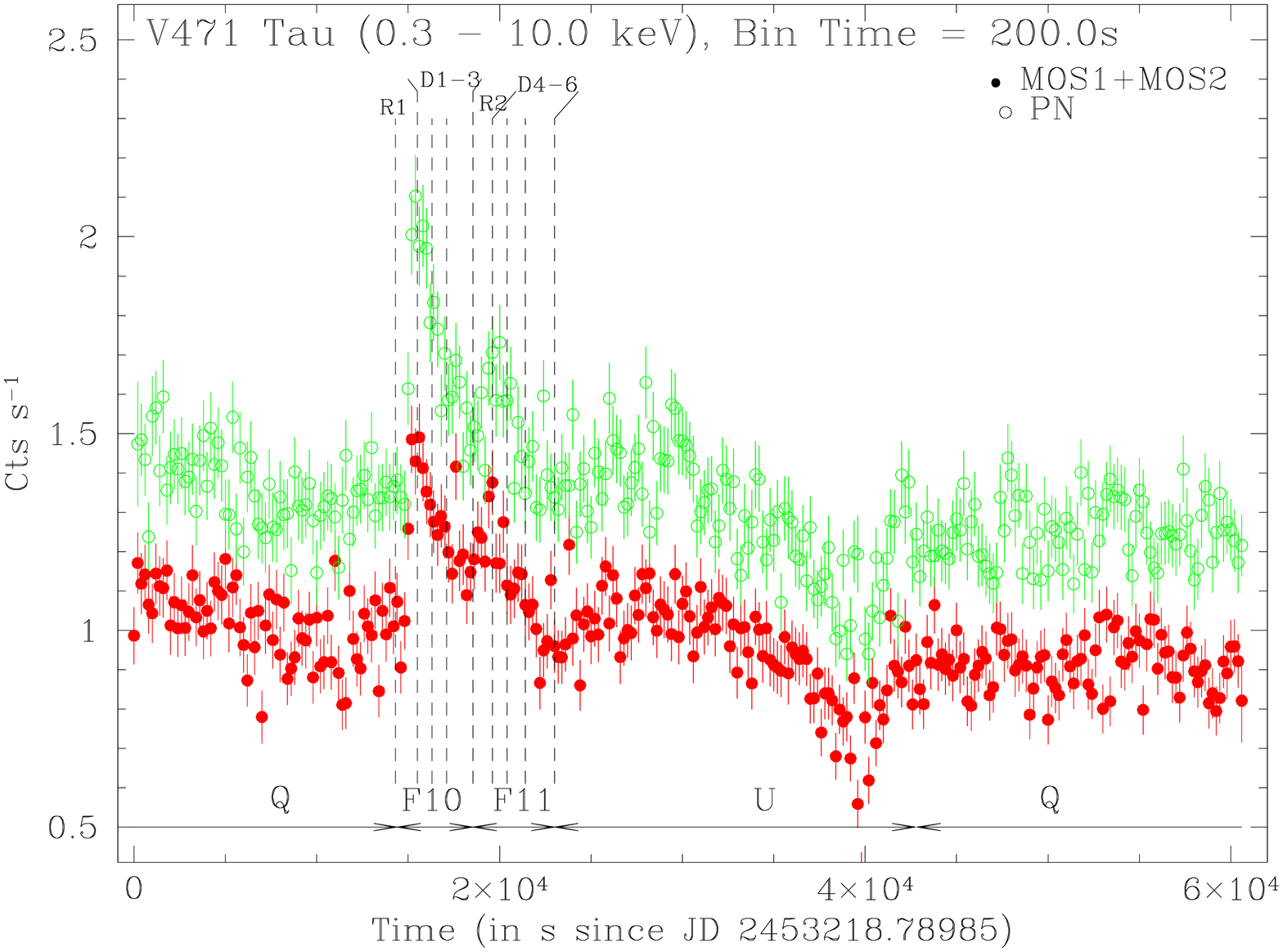}
\includegraphics[height=13.5cm, width=7.5cm,angle=-90]{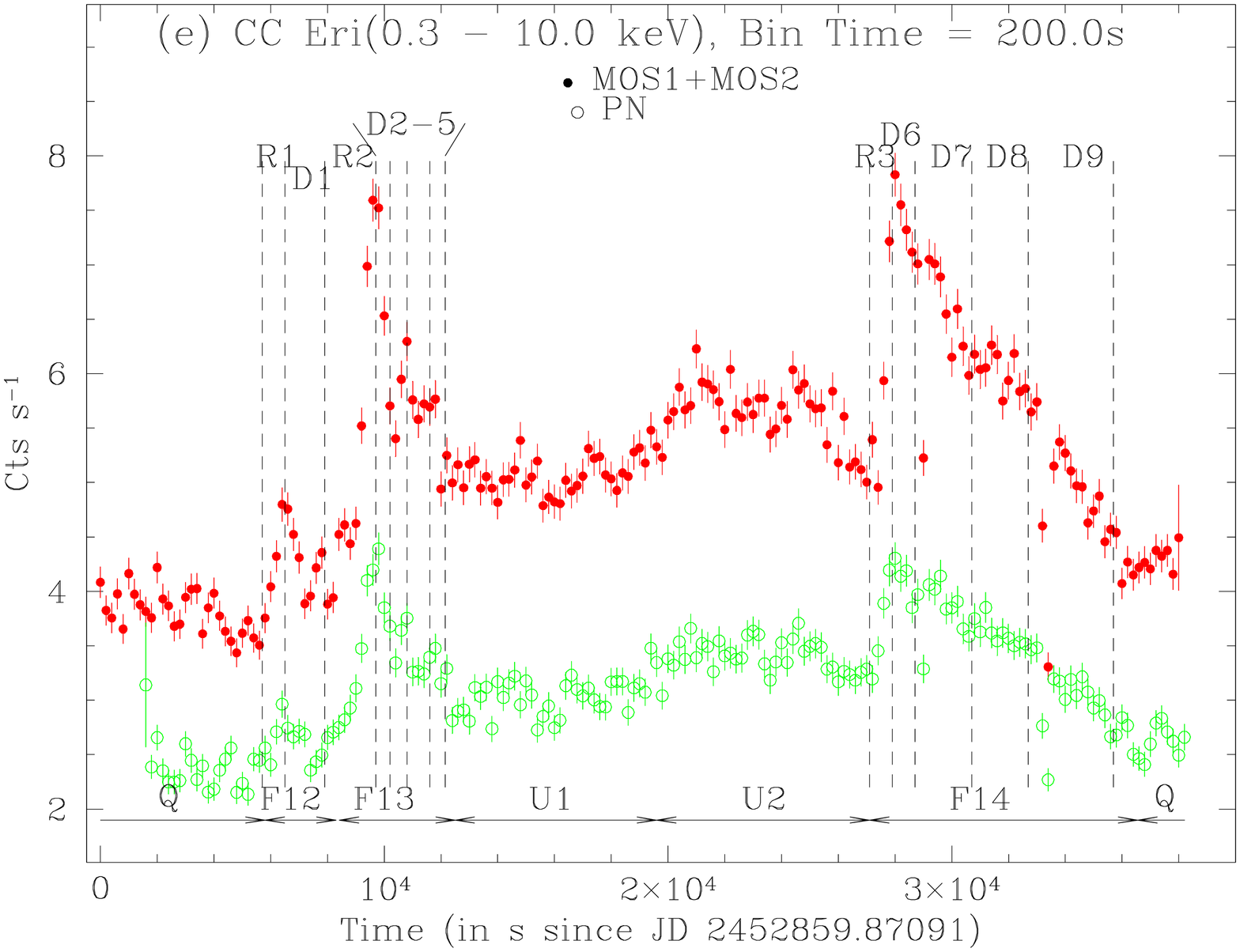}
\includegraphics[height=13.5cm, width=7.5cm,angle=-90]{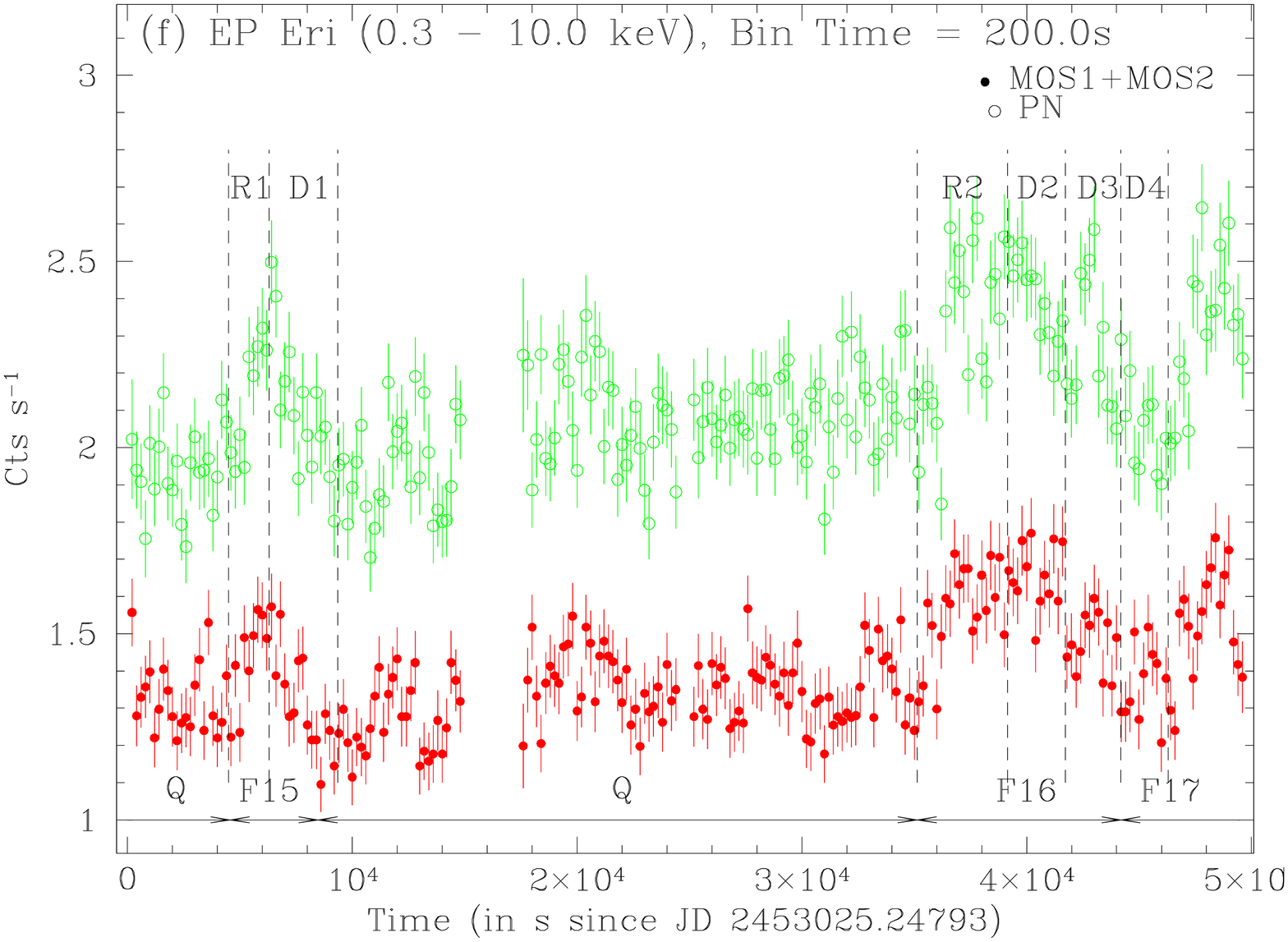}
\vbox{Fig. \ref{fig:mospnlc} Continued}
\end{figure}

\begin{figure}
\hbox{
\includegraphics[height=8.5cm, width=9.0cm,angle=-90]{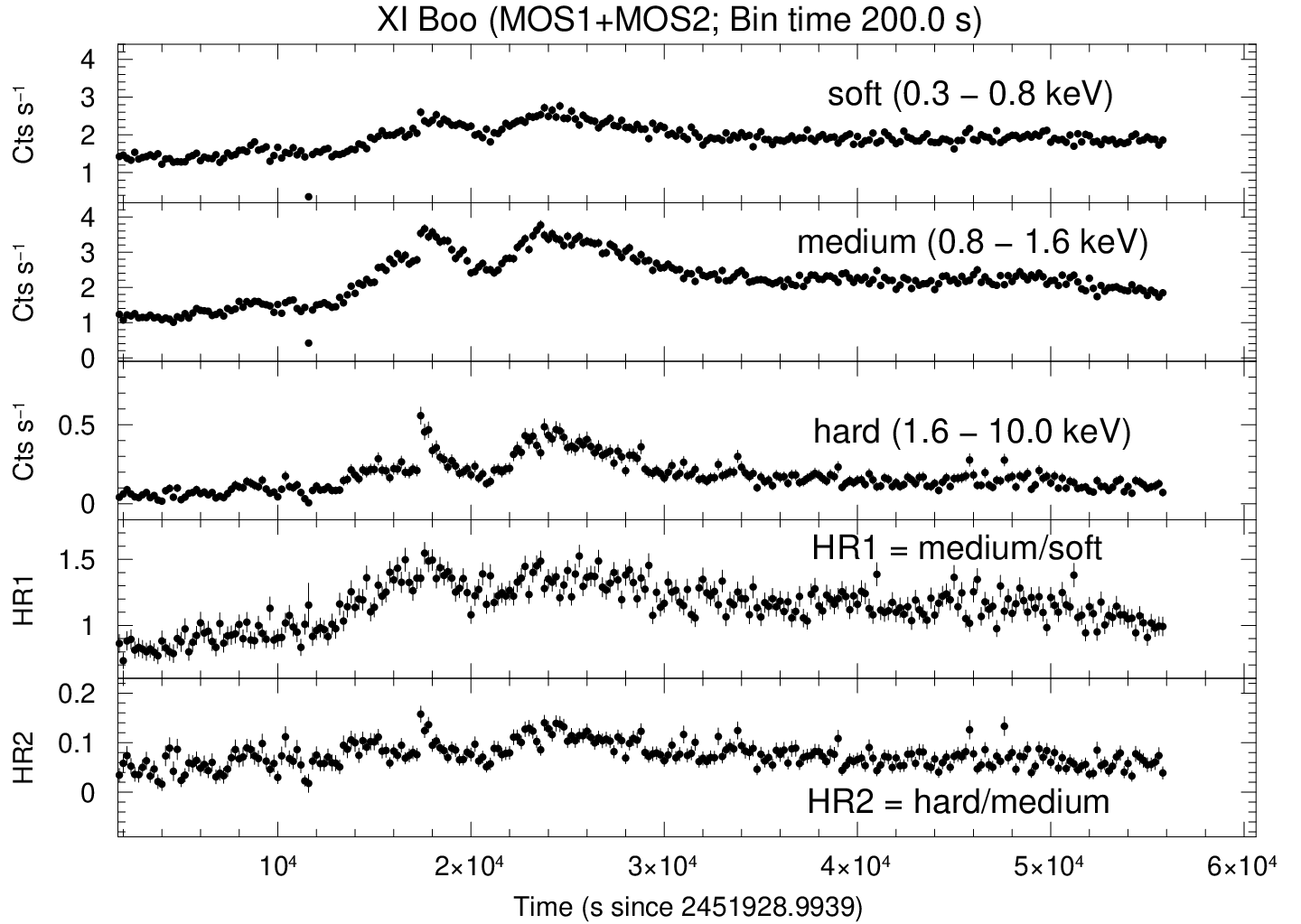}
}
\hbox{ Fig. \ref{fig:lchr}(b) -- Similar to Fig. \ref{fig:lchr} (a), but for XI Boo
(only MOS)}
\end{figure}

\begin{figure}
\hbox{
\includegraphics[height=8.5cm, width=9.5cm,angle=-90]{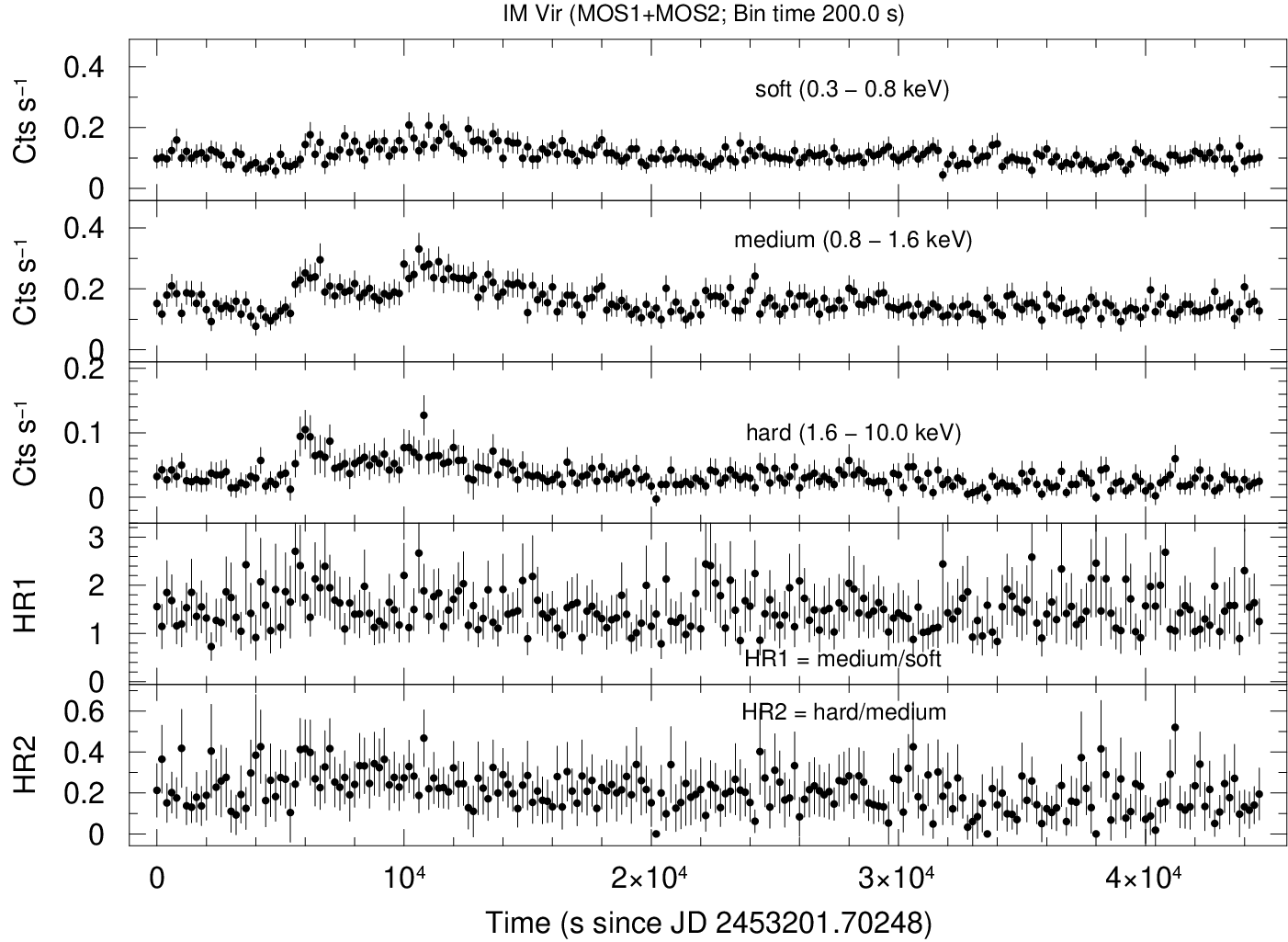}
\includegraphics[height=8.5cm, width=9.5cm,angle=-90]{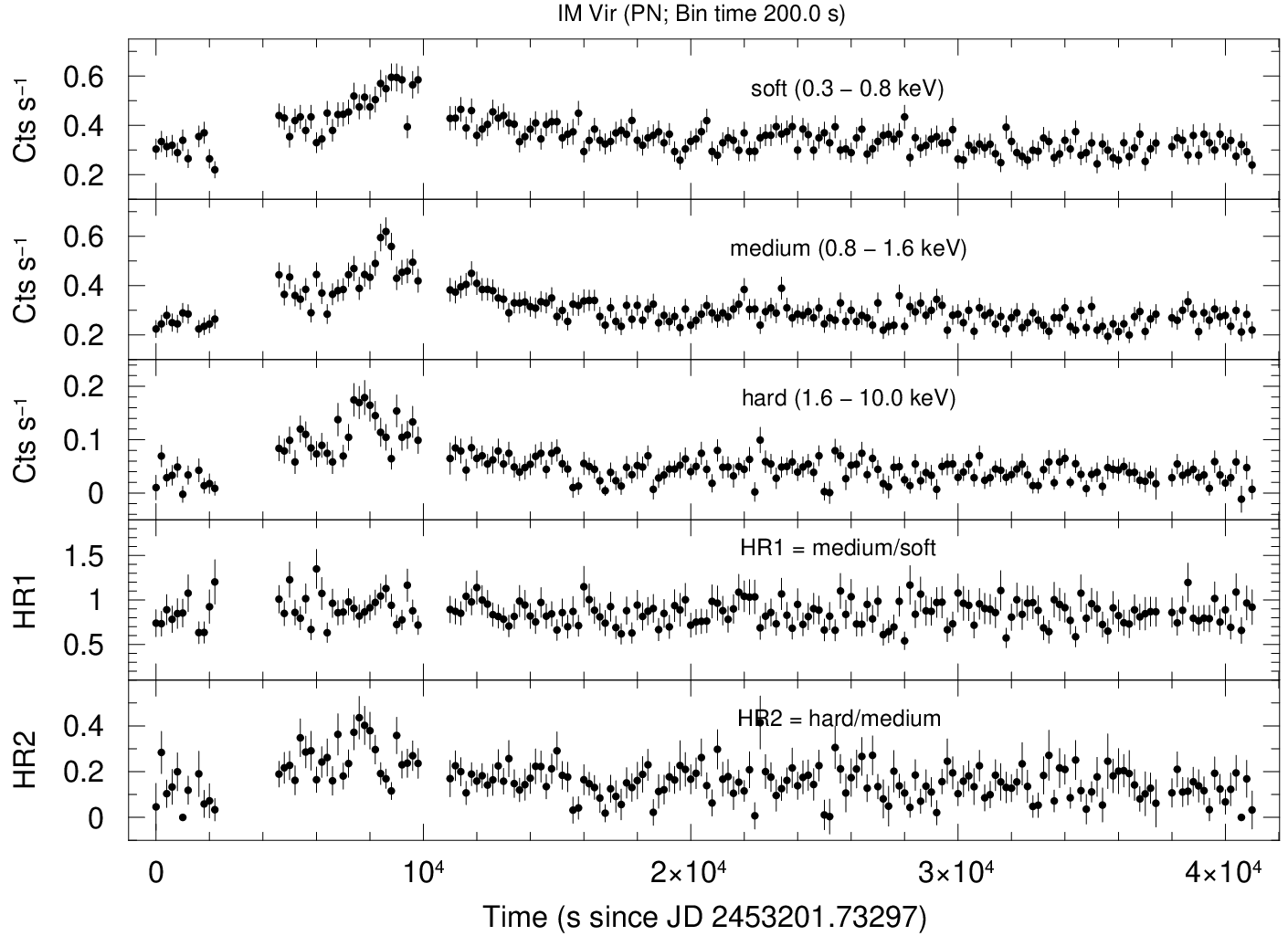}
}
\hbox{
Fig. \ref{fig:lchr}(c) -- Similar to Fig. \ref{fig:lchr} (a), but for IM Vir}
\end{figure}

\begin{figure}
\hbox{
\includegraphics[height=8.5cm, width=9.5cm,angle=-90]{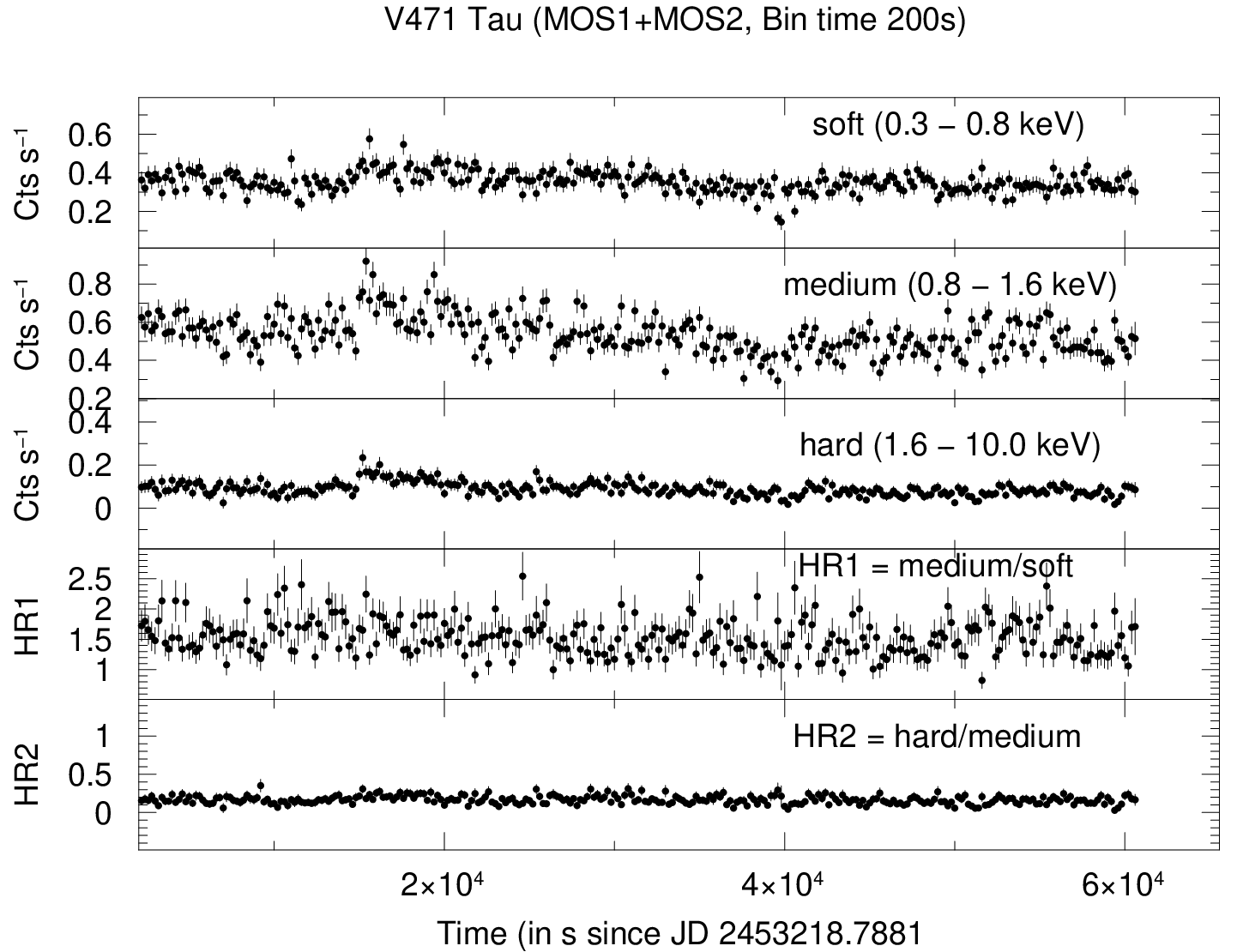}
\includegraphics[height=8.5cm, width=9.5cm,angle=-90]{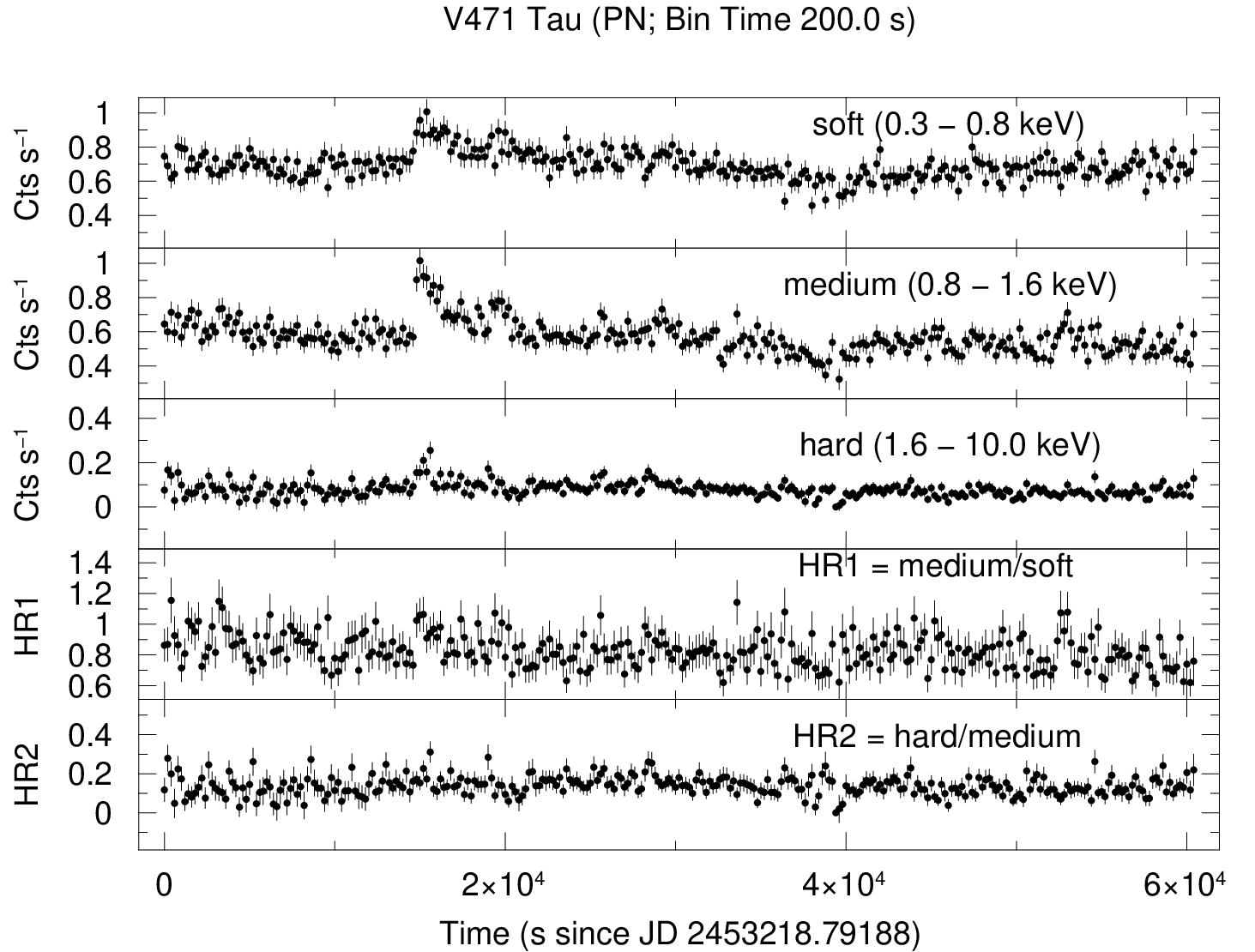}
}
\hbox{
Fig. \ref{fig:lchr}(d) -- Similar to Fig. \ref{fig:lchr} (a), but for V471 Tau}
\end{figure}

\begin{figure}
\hbox{
\includegraphics[height=8.5cm, width=9.5cm,angle=-90]{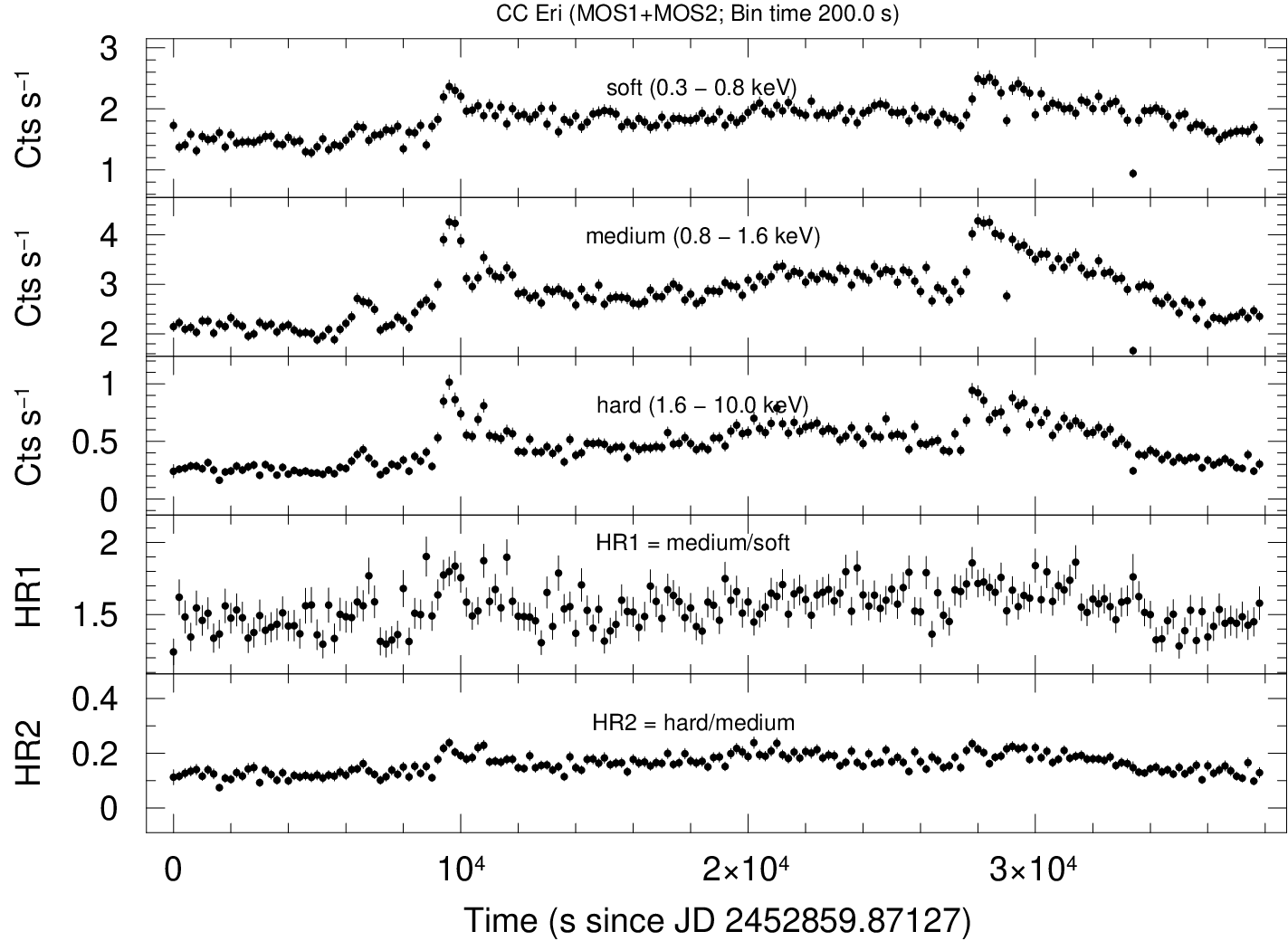}
\includegraphics[height=8.5cm, width=9.5cm,angle=-90]{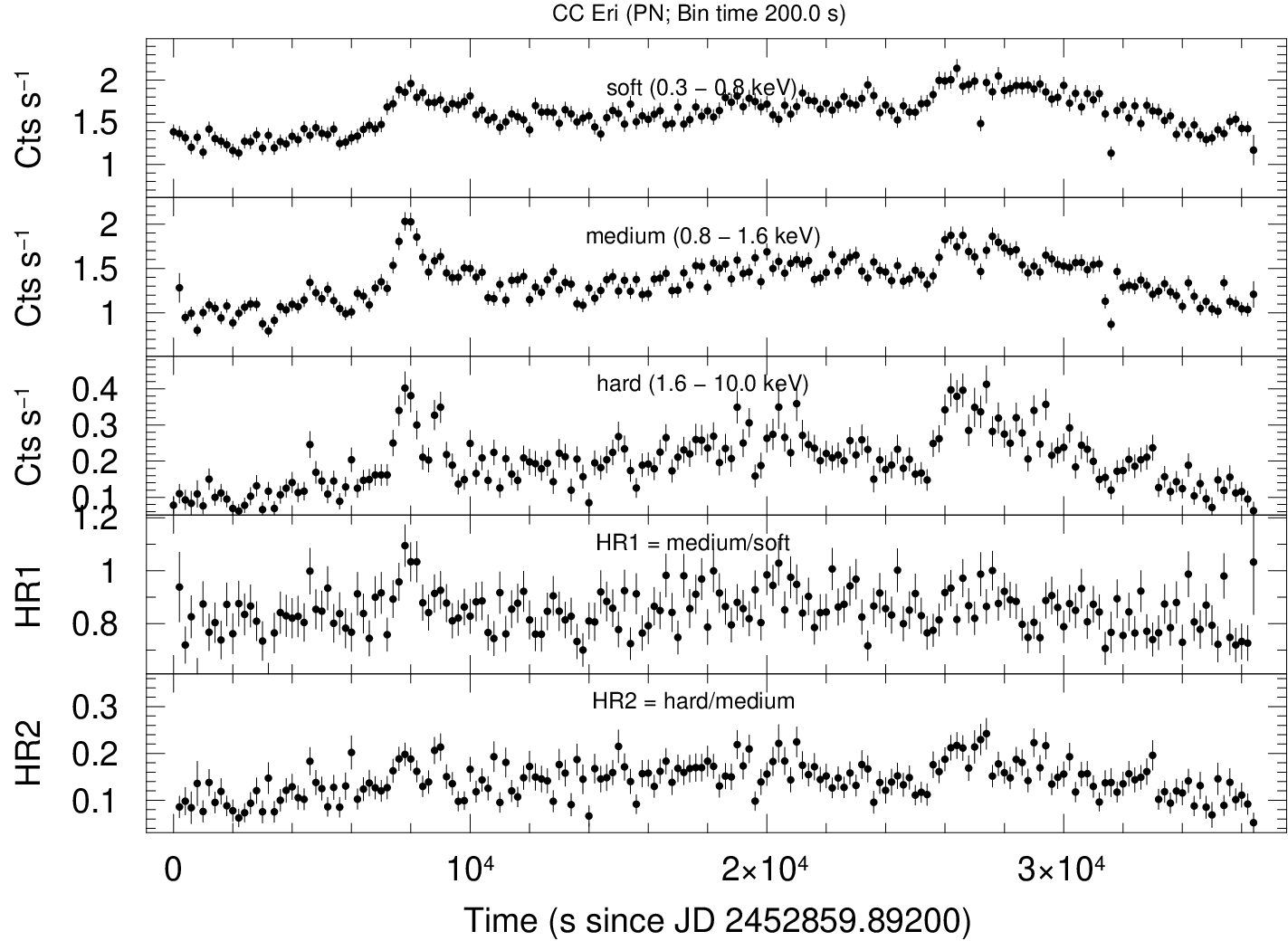}
}
\hbox{Fig. \ref{fig:lchr}(e) -- Similar to Fig. \ref{fig:lchr} (a), but for CC Eri}
\end{figure}

\begin{figure}
\hbox{
\includegraphics[height=8.5cm, width=9.5cm,angle=-90]{EPEri_mos_b200.ps}
\includegraphics[height=8.5cm, width=9.5cm,angle=-90]{EPEri_pn_b200.ps}
}
\hbox{ Fig. \ref{fig:lchr}(f) -- Similar to Fig. \ref{fig:lchr} (a), but for EP Eri}
\end{figure}


\clearpage

\begin{figure*}
\centering
\vbox{
\includegraphics[height=11.0cm, width=7.5cm,angle=-90]{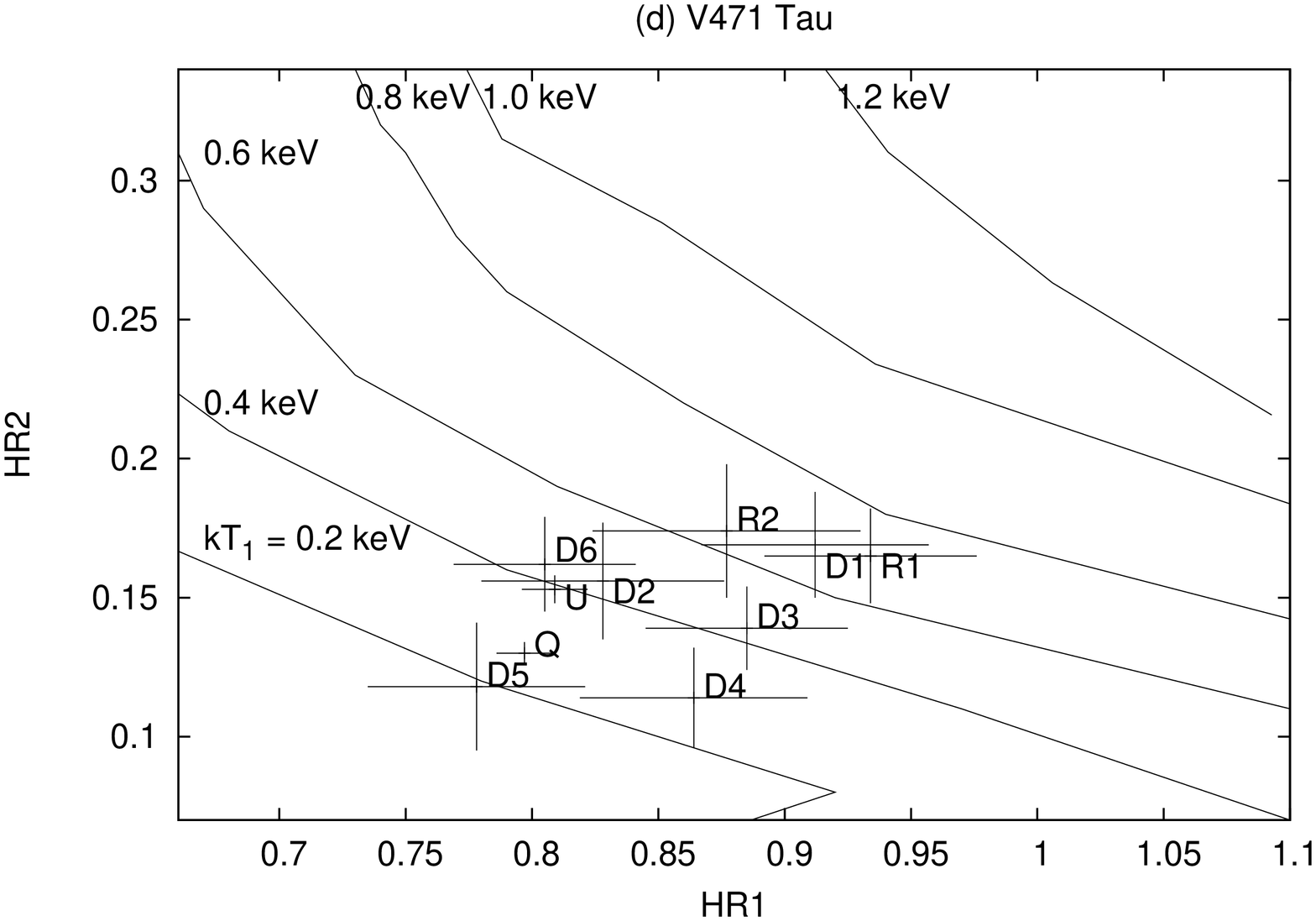}
\includegraphics[height=11.0cm, width=7.5cm,angle=-90]{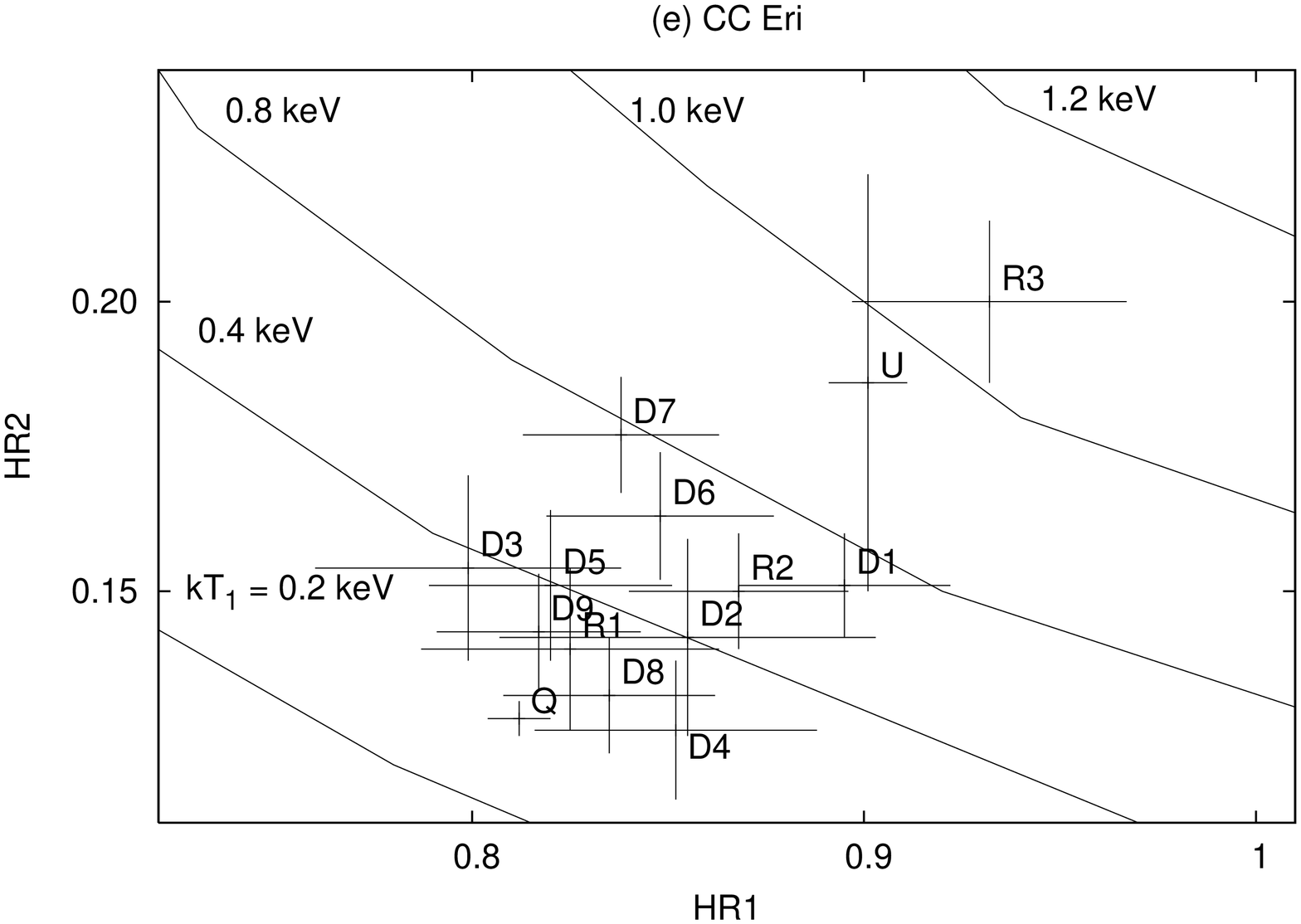}
\includegraphics[height=11.0cm, width=7.5cm,angle=-90]{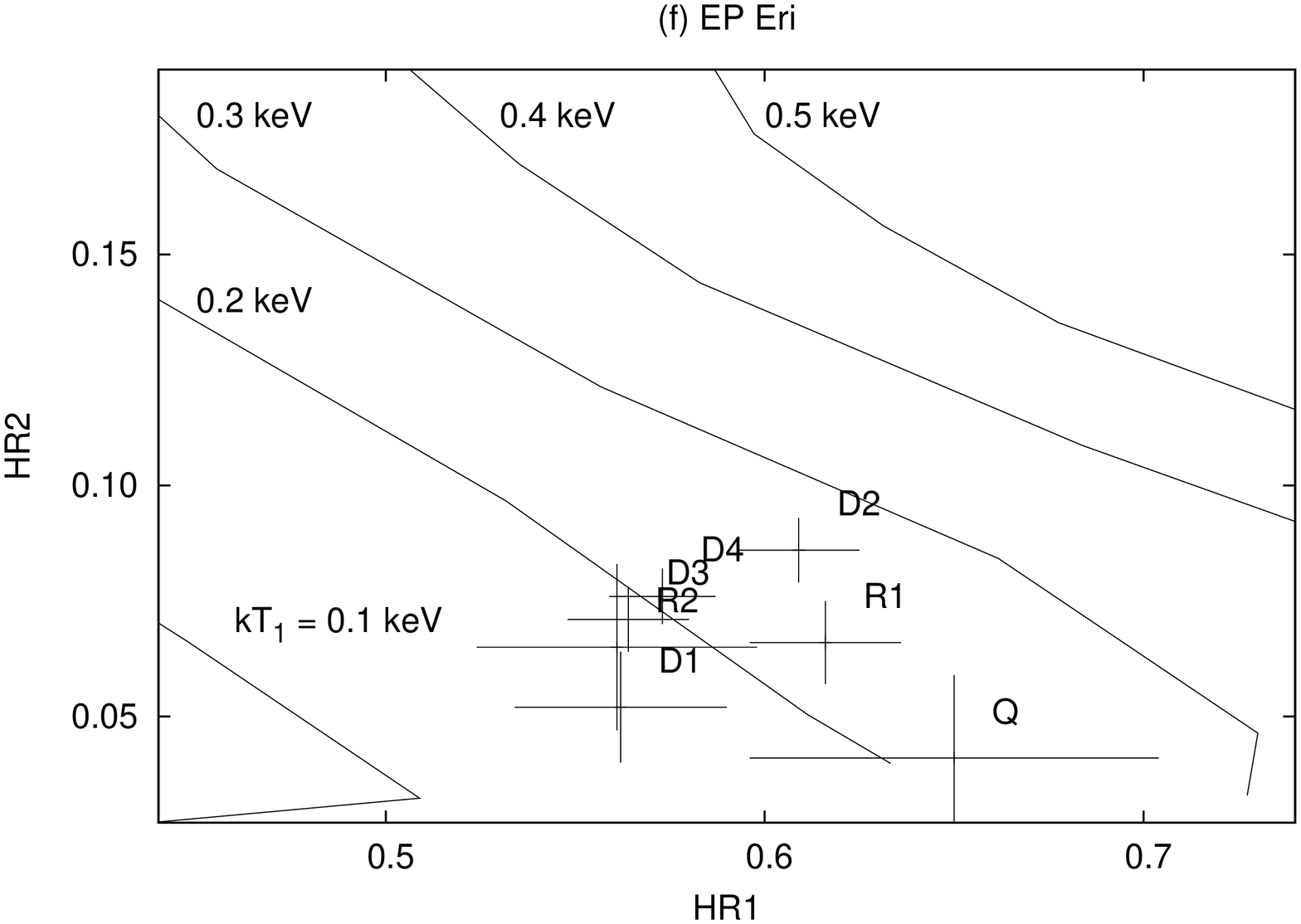}
}
\vbox{Fig. \ref{fig:hr} Continued}
\end{figure*}


\clearpage

\end{document}